\documentclass[fleqn,usenatbib]{mnras}

\usepackage{float}
\usepackage{caption}

\usepackage[T1]{fontenc}

\DeclareRobustCommand{\VAN}[3]{#2}
\let\VANthebibliography\thebibliography
\def\thebibliography{\DeclareRobustCommand{\VAN}[3]{##3}\VANthebibliography}

\usepackage{nccmath}
\usepackage{empheq} 

\usepackage{graphicx}	
\usepackage{amsmath}	
\usepackage{amssymb}	
\usepackage[dvipsnames]{xcolor}
\definecolor{myBlue}{rgb}{0,0.5,1}
\usepackage[normalem]{ulem} 

\usepackage{xspace}

\newcommand{\Windrate}{\dot\Sigma_{\rm w}}
\usepackage{newtxtext,newtxmath}

\title[Thermal winds and outbursts evolution of LMXBs]{The effect of thermal winds on the outbursts evolution of LMXB systems}

\author[A. L. Avakyan et al.]{
A. L. Avakyan,$^{1}$\thanks{E-mail: artur.avakyan@astro.uni-tuebingen.de}
G. V. Lipunova,$^{2,4}$
K. L. Malanchev$^{3,4}$
\\
$^{1}$ Institut f{\"u}r Astronomie und Astrophysik T{\"u}bingen, Universit{\"a}t T{\"u}bingen, Sand 1, 72076 T{\"u}bingen, Germany \\
$^{2}$ Max-Planck-Institut f\"ur Radioastronomie, Auf dem H\"ugel 69, 53121 Bonn, Germany \\
$^{3}$ Department of Astronomy, University of Illinois Urbana-Champaign, 1002 W Green Street, Urbana, IL 61801, USA \\
$^{4}$ Sternberg Astronomical Institute, Lomonosov Moscow State University,  Universitetskiy Prospekt, 13, 119992 Moscow, Russia
}

\date{Accepted XXX. Received YYY; in original form ZZZ}

\pubyear{2023}

\def\lup{{4U\,1543$-$47}}

\begin{document}
\label{firstpage}
\pagerange{\pageref{firstpage}--\pageref{lastpage}}
\maketitle

\begin{abstract}
Theoretical models of accretion discs and observational data indicate that the X-ray emission from the inner parts of an accretion disc can irradiate its outer regions and induce a thermal wind, which carries away the mass and angular momentum from the disc.
Our aim is to investigate the influence of the thermal wind on the outburst light curves of black hole X-ray binary systems. 
We carry out numerical simulations of a non-stationary disc accretion with wind using upgraded open code {\sc freddi}. We assume that the wind launches only from the ionised part of the disc and may turn off if the latter shrinks fast enough.
Our estimates of the viscosity parameter $\alpha$ are shifted downward compared to a scenario without a wind. Generally, correction of $\alpha$ depends on the spectral hardness of central X-rays and the disc outer radius, but unlikely to exceed a factor of 10 in the case of a black hole low-mass X-ray binary (BH\,LMXB).
We fit 2002 outburst of {BH\,LMXB}~\lup{} taking into account the thermal wind. 
The mass loss in the thermal wind is of order of the accretion rate on the central object at the peak of the outburst.
New estimate of the viscosity parameter $\alpha$ for the accretion disc in this system is about two times lower than the previous one. Additionally, we calculate evolution of the number of hydrogen atoms towards \lup{} due to the thermal wind from the hot disc.
\end{abstract}



\section{Introduction}\label{intro}

X-ray Binary (XRB) is a binary system featuring an accreting compact object, which can be either a neutron star or a black hole (BH). Up to now, about several hundreds of XRBs are known in the Galaxy~\citep{AvakyanLMXBcat, NeumannHMXBcat, FotinHMXBcat}\footnote{\url{http://astro.uni-tuebingen.de/~xrbcat/}}$^{,}$\footnote{\url{https://binary-revolution.github.io/}}. XRB population is typically categorised into two groups: low- and high-mass X-ray binaries (LMXBs and HMXBs, respectively), based on the properties of the counterpart. In contrast to the HMXB case, the donor companion in a LMXB is a low-mass star, typically a main-sequence star or a white dwarf. In a LMXB, the matter coming from the donor star usually forms an accretion disc around the compact object. As the material falls onto the compact object through the disc, it releases a large amount of energy in the form of X-rays. The majority of LMXBs are X-ray transient sources, which occasionally show outburst behaviour.

During an outburst of a LMXB, almost all the energy is radiated as X-rays, so the `X-ray Novae' are some of the brightest objects in the X-ray sky. X-ray outbursts last much longer and repeat less frequently than outbursts in systems with accreting white dwarfs. Apparently, this can be related to central X-rays illuminating the outer disc during outbursts~\citep{MeyerHofm,KingRit,DIM}. Currently, few dozens of LMXB systems with BHs are known in the Galaxy~\citep[see, for example,][]{AvakyanLMXBcat,WATCH,cherep}, from which bright X-ray outbursts were detected, implying episodes of high accretion rate in these systems.

In this work we presume that during an outburst an outflow from an accretion disc takes place. The presence of such a wind in the LMXBs is supported by modern observations indicating the expansion of ionised matter. Outflows from accretion discs are observed as blue-shifted X-ray absorption lines~\citep[e.g.,][]{Ueda, Trigo}, as emission lines broadening in the optical range~\citep[e.g.,][]{Munoz, Casares, Charles}, and P-Cygni profiles~\citep{2016Natur.534...75M,2016ApJ...821L...9M,2016A&A...590A.114M,2023MNRAS.526L.149F}. In most cases, X-ray absorption lines of highly ionised Fe are observed in systems with an inclination of more than $\sim 50^{\circ}$. Therefore, the absorbing plasma should have a higher density closer to the equatorial plane, which suggests that it is a substance flowing out of the disc~\citep{HigProg}.

The nature of such winds and their physical characteristics are an open question. Namely, in addition to the model of thermal heating of matter by central radiation in optically thin regions of the disc~\citep[][hereinafter B83, S86 and W96, respectively]{Begelman1983, Shields, Woods}, other mechanisms are considered: 
hydromagnetic winds~\citep{MagnetOrigin, Pelletier92, 2010ApJ...715..636F, 2017NatAs...1E..62F, 2023arXiv230210226S}, local radiation pressure at super-Eddington accretion rates~\citep{SHAKURA_SUNAEV,Proga2002}, and radiation pressure in spectral lines ('line driven' winds) \citep{Shlosman85, Suleimanov95, Feld99v2, Shlosman85, Murray95, ProgaKall04}.

The observational rates of mass loss in the wind for the intermediate mass X-ray binary~(IMXB) Her~X-1 are obtained by~\citet{Kosec}. They find that the estimates strongly depend on the assumed wind geometry. According to them, if the wind moves along the disc, then the rate of mass loss in the wind in Her~X-1 is approximately equal to the accretion rate onto the compact object. However, if the wind is spherically symmetric, then the estimated mass-loss rate increases by an order of magnitude. \citet{Ponti12} provide observational estimates of the outflow rate in several X-ray transients; it follows that the ratios of the mass-loss rate in the wind to the accretion rate onto compact objects are in the range from $1$ to $10$. In hydrodynamical numerical simulations of thermally
driven disc winds~\citep{Higgin17, Luketic, Higgin19}, this ratio is obtained to lie in the range from $2$ to $15$. Apart from observational spectral signatures of outflowing matter, an outflow takes away angular momentum and mass and thus affects the long-term dynamical disc evolution.

In the standard model of disc accretion~\citep{Shakura72,SHAKURA_SUNAEV}, the dimensionless parameter of the turbulent viscosity $ \alpha \lesssim 1 $ is introduced, which characterises the angular momentum transfer rate in the disc. Parameter $\alpha$ determines the viscous time of the accretion disc and and can be estimated for cataclysmic variables and X-ray novae~\citep{Smak1984_IV,dubus_et2001,Kotko-Lasota2012}. It was found that two values of $\alpha$ are needed to explain observed long-term evolution: $\alpha_{\rm hot}$, which determines outburst evolution, and $\alpha_{\rm cold}$, quiescent times. For particular X-ray novae outbursts parameter $\alpha_{\rm hot}$ was estimated by \citeauthor{Lipunova-Shakura2002} (\citeyear{Lipunova-Shakura2000}; using analytical time-dependent solution for $\alpha$-disc, see~\citealt{Lipunova-Shakura2000}), \citet{Suleimanov+2008,Tetarenko18_1}.

During an outburst, inner parts of the accretion disc irradiate its outer parts. This self-irradiation plays a crucial role in the evolution of the disc, as pointed out by~\citet{KingRit, dubus_et2001}. Consequently, if a disc is irradiated, $\alpha_{\rm hot}$ cannot be determined uniquely from the X-ray outburst decay alone without analysing the optical light curves~\citep{lipunova_malanchev2017,Lipunova+2022}.

Using a disc instability code, \citet{Dubus_19} have analysed the effects of thermal wind on light curves of X-ray novae. The authors used the analytical model of the thermal wind from W96 and assumed that self-irradiation of the disc is enhanced due to scattering in the optically thin wind. They argue that thermal wind (by affecting self-irradiation) plays an important role in the outburst dynamics. However, there is still inconsistency between analytic estimates of X-ray illumination from scattering in the wind and the observed data, even in combination with direct illumination~\citep{Tetarenko20}. \citet{Tetarenko18_1} suggest that winds influence estimates of $\alpha$ obtained from the light curves.

Our open code \textsc{freddi} calculates X-ray and optical light curves of soft X-ray transients outbursts, solving the evolution equation for a viscous $\alpha$-disc of variable size~\citep{SOFT, Avakyan19, Avakyan21}\footnote{\textsc{freddi} can be freely downloaded from the web page \url{https://github.com/hombit/freddi/}}. Main input parameters of \textsc{freddi} are the turbulent parameter $\alpha$, BH mass $m_{\rm x} \equiv M_{\rm x}/M_{\rm odot}$, and Kerr parameter $a_{\rm Kerr}$. Self-irradiation of the disc is taken into account and the degree of irradiation can vary with time. Calculated light curves or $\dot M(t)$ can be used to fit the observed ones to deliver an estimate of $\alpha$-parameter, see~\cite{lipunova_malanchev2017, Lipunova+2022}. For this, it is essential to perform a spectral analysis of data to try to obtain as accurately as possible the evolution of bolometric luminosity or accretion rate.

In this work, we report the results of implementing the Compton-heated wind to our code. For X-ray transients, the wind model is taken into account following~W96. We analyse how the action of the wind imitates large parameter $\alpha$ and show that a high value of $ \alpha $ parameter, potentially obtained from observed light curves, corresponds to a smaller $ \alpha $ if the wind is acting.

We also apply the model to 2002 outburst of X-ray transient~\lup{}~(V$^*$ IL Lup). The system is a BH LMXB that demonstrates outbursts about every decade~\citep{Kitam84, Harmon92, Miller02,Negoro+2021}. \citet{lipunova_malanchev2017}~(hereinafter LM17) have obtained the evolution of the central accretion rate $\dot{M}(t)$, using archival X-ray spectral observations by Proportional Counter Array aboard the Rossi X-ray Timing Explorer observatory~\citep[\textit{RXTE}/PCA,][]{RXTE}. In particular, it was demonstrated that a reflare around the 12th day after the maximum is an artefact of spectral evolution and underlying dependence of $\dot{M}(t)$ is monotonic during almost 30 days after the maximum, when the source was in the 'high/soft' state~\citep{2006csxs.book..157M} according to~\citet{2004ApJ...610..378P}.

For \lup, we compare the central accretion rate $\dot{M}_{\rm acc}(t)$ with the model taking the thermal wind into account. 
As a result, we obtain new estimate of the $\alpha$-parameter, comparing to LM17. Value of the self-irradiation parameter in \lup{} is determined from the $V$ optical light curve obtained by~\citet{Bux_Bai2004}.

The structure of the paper is as follows. In Section~\ref{wind_desc} the physical model of a thermal wind is briefly described. A viscous disc evolution model, which takes into account the wind, is reviewed in the same Section~\ref{wind_desc}. In Section~\ref{irrad_3}, we explain how we take into account the disc self-irradiation and irradiation of the companion star. Thermal wind model from W96 and S86 and numerical method of solving disc evolution equation are described in Section~\ref{windmod}. In Section~\ref{res_4u}, we present the results of disc modelling, as well as the fitting of the BH LMXB \lup{} outburst in 2002. The discussion and conclusions are given Section~\ref{discus} and Section~\ref{concl}, respectively. In appendix~\ref{appendix_num} we study the numerical-solution stability of the non-linear diffusion-type equation of the disc evolution. Appendix~\ref{appendix_ver} describes how we tested the work of our code~\textsc{freddi} by reproducing S86 results on luminosity oscillations. Appendix~\ref{appendix_val} overviews analytical solutions to some of the wind models that are used in the code to test it. Comparison of the two different self-irradiation models is given in appendix~\ref{appendix_cirr}.

\section{Thermal wind and the viscous disc dynamics}\label{wind_desc}
\subsection{Compton-heated wind}
\noindent

\begin{figure}
    \begin{center}
    \includegraphics[width=\columnwidth]{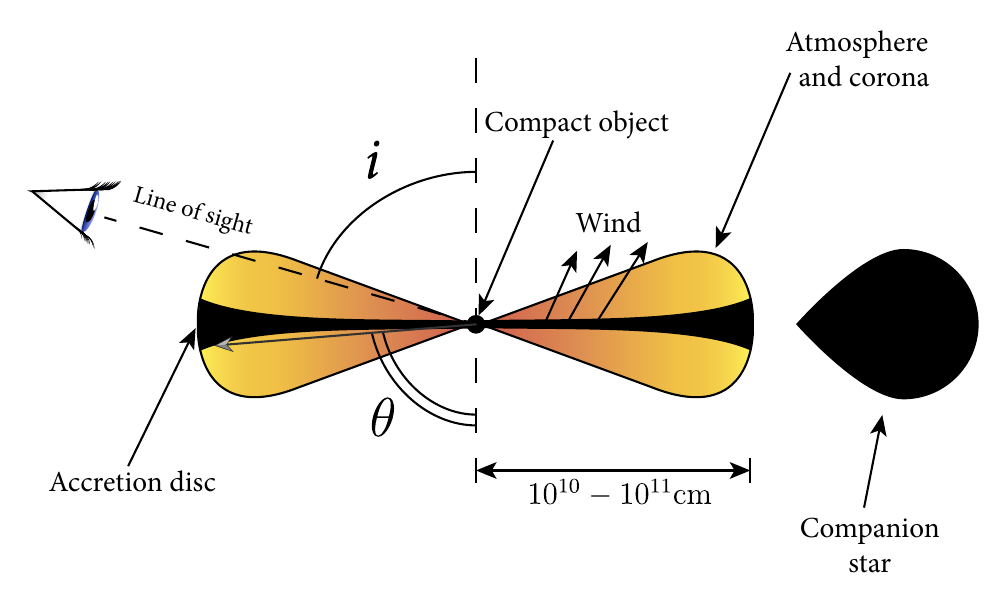}
    \caption{Schematic representation of a LMXB system with accretion disc wind. Angles $ i $ and $ \theta $ are the disc inclination and polar angle, respectively}
    \label{acc_disc}
    \end{center}
    \end{figure}

Theoretical models of accretion discs and observational data supply evidence that the radiation flux from the centre of the disc can irradiate its outer surface. In the standard accretion model by~\citet{SHAKURA_SUNAEV} the disc is concave, and becomes geometrically thicker with radius. As a result, the accretion disc's surface ( mostly it's outer parts) is exposed to the central radiation, which heats the disc material. Schematic picture of a disc in a binary system is presented in Fig.~\ref{acc_disc}.   
  
In the atmosphere of the outer disc, the energy of the incident photons transfers to the energy of particles through different processes: photoabsorption, photoionisation, scattering, bremsstrahlung.
The heating rate per particle is proportional to the radiation intensity. The cooling rate per particle depends on the two-particle processes and, therefore, it decreases with density, that is, with the distance from the equatorial plane of the disc. When the density drops to a critical value, the heating exceeds cooling, and the gas heats up to a high temperature determined by a balance between the principal processes at such conditions: the Compton and inverse Compton scatterings. In particular, for X-ray binaries and quasars hosting compact objects, the central radiation is sufficiently hard and the gas can be heated to temperatures exceeding $ 10^7 $ K.

\begin{figure}
    \begin{center}
    \includegraphics[width=\columnwidth]{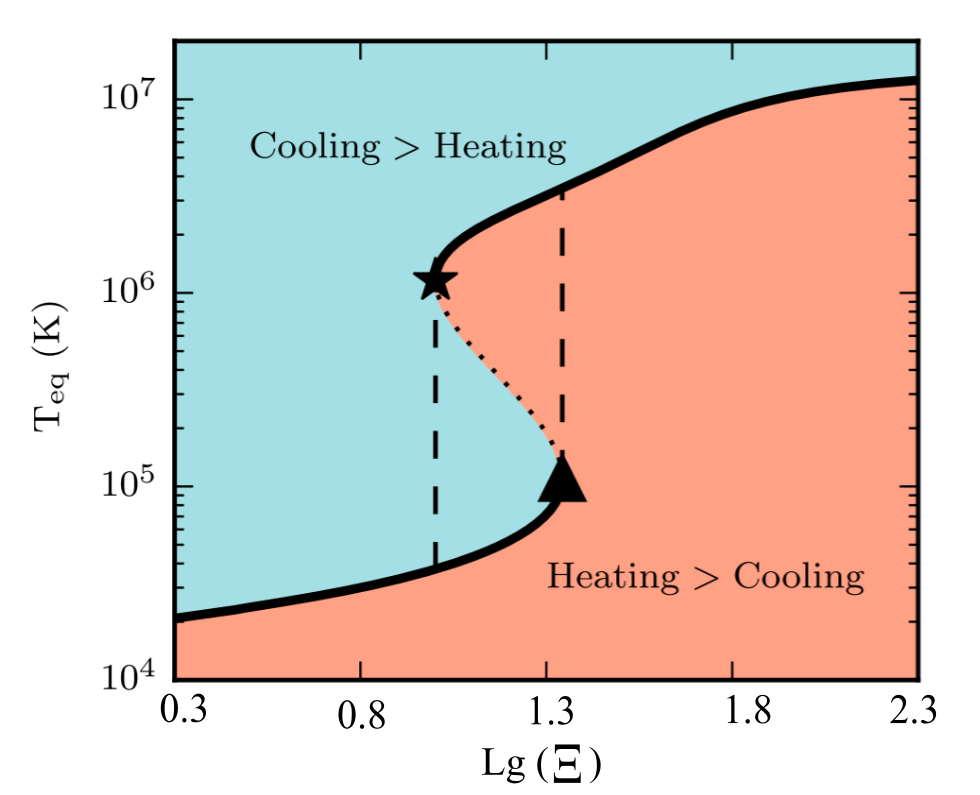}
    \caption{Schematic disc thermal equilibrium curve~\citep[basing on a figure from][]{Higgin17}. 
    The triangle denotes the position of the maximum value of the pressure ionisation parameter $\Xi_{\rm c, max}$, at which the gas is in a cold state. The asterisk corresponds to the minimum value $\Xi_{\rm h, min}$ in the hot state. The dotted part of the thermal equilibrium curve shows the instability zone.}
    \label{Teq}
    \end{center}
    \end{figure}

Let us briefly recall the model proposed by B83 and introduce basic parameters to describe winds from a disc irradiated by an X-ray or ultraviolet continuum. 
The thermal state of the irradiated  gas is determined by the shape of the X-ray spectrum along with the pressure ionisation parameter~(B83):
    \begin{equation}\label{XiForm}
    \Xi \equiv P_{\rm rad}\,/\,P_{\rm gas} = \mathcal{F}\,/\,(c\,P_{\rm gas})\, ,
    \end{equation}
where $P_{\rm rad}$ and $P_{\rm gas}$ are the radiation and gas pressure, respectively, $\mathcal{F}$ is the radiative flux and $c$ is the speed of light. Thermal equilibrium of a stellar atmosphere gas exposed to ionising X-ray radiation was calculated by \citet{Basko73, London}. Fig.~\ref{Teq} shows the schematic dependence of the equilibrium temperature on a value, which is proportional to the pressure ionisation parameter $ \Xi $ (the picture is adopted from \citet{Higgin17}). It follows that there are two stable phases of gas depending on the value of  $\Xi$: `cool' and 'hot', when $\Xi < \Xi_{\rm c, max}$ and $\Xi > \Xi_{\rm h, min}$. In the cool phase, close to the disc photosphere, the photoionisation is balanced by the recombination, so the gas is maintained at the temperature $T \approx  10^{4}$~K. Away from the disc surface, the pressure and density drop, consequently the pressure ionisation parameter grows, eventually exceeding the critical $\Xi_{\rm c, max}$, which leads to the heating of gas. This heating is quenched when the Compton heating by energetic photons is balanced by the inverse Compton cooling by softer photons. 

The equilibrium temperature in this state is generally close to the `inverse Compton temperature' (hereinafter the term 'Compton temperature' will be used), which is defined as:
    \begin{equation}\label{TIC}
     T_{\rm IC} \equiv \frac{1}{4\,k}\,\left<\epsilon\right> = \frac{1}{4\,k}\, L^{-1}\int_{0}^{+\infty} h\nu \,L_{\rm \nu} \,\mathrm{d} \nu\, ,
    \end{equation}
where $\left<\epsilon\right>$ is the mean photon energy, $h$ is the Planck constant, $k$ is the Boltzmann constant, $\nu$ is the radiation frequency, $L_{\rm \nu}$ is the luminosity per frequency interval and $L$ is the bolometric luminosity.

B83 also introduce the `escape temperature': $T_{\rm g} \equiv GM_{\rm x}\mu m_{\rm p} / (k r) $, which follows from the equality of the thermal and escape velocities of a particle. Here $M_{\rm x} $ is the mass of the compact object, $ r $ is the axial distance from the centre of the disc, $m_{\rm p}$ is the proton mass, and $\mu$ is the mean mass of ions and electrons per particle~(we use value of $0.61$ as for the solar abundances). If the medium temperature is higher than $T_{\rm g}$, it expands. Thus, the characteristic 'Compton radius' can be  defined as follows~(B83):
    \begin{equation}\label{R_IC}
     R_{\rm IC} \equiv \frac{GM_{\rm x}\,\mu\,m_{\rm p}}{kT_{\rm IC}} = \frac{1.0\times10^{10}}{T_{\rm IC8}}\,\left(\frac{M_{\rm x}}{M_{\odot}}\right) \, {\rm cm}\,,
    \end{equation}
where $T_{\rm IC8} = T_{\rm IC}/ 10^8$ K.
B83 suggest that the matter begins to flow out of the disc in the form of a wind at radii greater than $ 0.1 \times R_{\rm IC} $, where the height of the corona is already significant. The upper layer, which is heated up to the Compton temperature $T_{\rm IC}$, expands with the sound speed $c_{\rm IC} = \sqrt{k T_{\rm IC}/(\mu m_{\rm p})}$.

B83 also define some critical luminosity:
    \begin{equation}\label{L_cr}
     L_{\rm cr} \equiv \frac{1}{8}\,\left(\frac{m_{\rm e}}{\mu m_{\rm p}}\right)^{1/2}\,\left(\frac{m_{\rm e}c^2}{kT_{\rm IC}}\right)^{1/2}L_{\rm Edd} = 0.030\times{T_{\rm IC8}^{-1/2}}\,L_{\rm Edd}\, ,
    \end{equation}
where $m_{\rm e}$ is the mass of an electron, $L_{\rm Edd}$ is the Eddington luminosity limit, which characterises the efficiency of overcoming the gravity directly by the radiation pressure:
    \begin{equation}\label{L_edd}
    L_{\rm Edd} \equiv \frac{4\pi GM_{\rm x}\,\mu_e\, m_{\rm p}c}{\sigma_{\rm T}} \approx 1.5 \times 10^{38} \left( \frac{M}{M_{\rm \odot}}\right) \, {\rm erg \,/\, s}\,,
    \end{equation}
where $\sigma_{\rm T}$ is the Thomson cross section, and $\mu_e$ is the mean molecular weight per election (which we set to be equal $2\mu$).
Thus, if the ratio $L / L_{\rm Edd}$ specifies the effectiveness of the radiation pressure in overcoming gravity, the ratio $L / L_{\rm cr}$ characterises the efficiency of overcoming the gravity by electromagnetic radiation via the heating of the matter by the Compton processes.

On the other hand, at smaller radii and/or lower luminosity, most of the material of the disc remains gravitationally bound, forming a hydrostatic corona~\citep[][B83, S86, W96]{Garate}.

\subsection{Viscous disc evolution}\label{mod_irr}
\noindent

The evolution of the accretion disc with the wind is described by a modified equation of the diffusion type~(for example, S86):
    \begin{equation}\label{e1}
    \frac{\partial \Sigma}{\partial t} = \frac{1}{4\pi}\,\frac{(GM_{\rm x})^2}{h^3}\,\frac{\partial^2 F}{\partial h^2} - \Windrate(F,\,h) \, ,
    \end{equation}
where $ h = \sqrt{GM_{\rm x}\, r} $ is the specific angular momentum of the accretion disc's material, 
$ \Sigma $ is the surface density, $ F $ is the torque of the viscous forces acting on a given disc ring, which is expressed as $F = 2\pi \,W_{r \varphi}r^{2}$ using the height-integrated viscous stress tensor $W_{r \varphi}$~\citep[assumed to be positive, see, e.g.,][]{Lyubarskij_Shakura}. Function $ \Windrate $ is a local wind mass loss rate, the outflow of mass per unit disc surface area per unit time, and $ \Windrate > 0 $ if the wind is present. In the upcoming Section~\ref{windmod}, we will discuss the form of this function in more detail.

The surface density $\Sigma$ and viscous torque $F$ are related in the Keplerian disc as follows~\citep[e.g.,][chapter 1]{Shakura_book}:
\begin{equation}\label{F_S}
F = 3\pi \, h\,\nu_{\rm t} \,\Sigma \, ,
\end{equation}
where $\nu_{\rm t}$ is the kinematic coefficient of turbulent viscosity, whose dependence on the radius $ r $ and surface density $ \Sigma $ is determined by physical processes underlying the viscosity. It follows from conservation of angular momentum:
\begin{equation}\label{dfdh}
\dot{M} = \frac{\partial F}{\partial h} \,.
\end{equation}

To effectively describe a complex phenomenon of the turbulent viscosity, \citet{Shakura72} introduced the dimensionless turbulent parameter $\alpha$~\citep[see also][]{SHAKURA_SUNAEV}.
It relates the total pressure in the disc and the viscous stress tensor component:
    \begin{equation}\label{WaP}
    w_{r \varphi}^{\rm \,t} = \alpha \, P_{\rm tot} \, ,
    \end{equation}
where $P_{\rm tot}$ is the sum of the gas pressure $P_{\rm gas}$ and radiation pressure $P_{\rm rad}$.
The height-integrated viscous stress tensor $W_{r \varphi}$ is as follows:
    \begin{equation}\label{W_int}
    W_{r \varphi} (t,\, r) = \int_{-z_{0}}^{+z_0}w_{r \varphi}^{\rm \, t}(t, \, r, \, z)\,\mathrm{d}z \, ,
    \end{equation}
where $z_0$ is the disc half-thickness at the radius $r$.

The condition at the inner boundary of the disc (corresponding to the last stable orbit) is the zero viscous torque $ F $~\citep{SHAKURA_SUNAEV}:
\begin{equation}\label{inner}
F(h_{\rm in}, t) = 0 \, ,
 \end{equation}
where $h_{\rm in}$ is the specific angular momentum at the inner edge of the disc. We set outer boundary conditions in two different ways. Firstly, when we assume that the outer radius of a viscously evolving disc is fixed, we set the external boundary condition of the second type:
\begin{equation}\label{bound}
\frac{\partial F}{\partial h}\bigg|_{ 
h = h_{\rm out}} = \dot{M}_{\rm out}\, ,
 \end{equation}
where $ \dot{M}_{\rm out} $ is the mass rate of matter entering the disc, and $ h_{\rm out} $ is the specific angular momentum at its outer edge.
This approach is used for the oscillations test~(see Appendix~\ref{appendix_ver}), which we use to verify our code.

Alternatively, we assume that a disc consists of two zones: an ionised 'hot' inner part and a 'cold' one with recombined hydrogen.
Presumably, they have different values of the viscosity parameter $\alpha$, $\alpha_{\rm hot}$ and $\alpha_{\rm cold}$ for the 'hot' and 'cold' zones, respectively. The viscous evolution in the colder part has a much longer time scale, since $\alpha_{\rm cold} \ll \alpha_{\rm hot}$. This means that the viscous evolution in the colder part has almost no effect on the central accretion rate during an outburst. Near the outer boundary of the hot zone $ r_{\rm hot} $, the viscous torque $F$ reaches an extremum. For all simulated outbursts, we thus assume that the mass flow rate vanishes at the boundary: $\partial F/ \partial h\,(h = h_{\rm hot}) = 0$, where $ h_{\rm hot} $ corresponds to $ r_{\rm hot} $.
 
Finally, the initial distribution of the viscous torque must satisfy the boundary conditions:
\begin{equation}\label{init}
F(h, 0) = F_0(h)\, .
\end{equation}

The numerical technique of solving the system of equations \eqref{e1}, \eqref{inner}, \eqref{bound}, and \eqref{init} is described in detail in Appendix~\ref{appendix_num}.

\section{Self-irradiation and optical flux}\label{irrad_3}
\subsection{Self-irradiation of the disc and the radius of the hot zone}\label{Irrad_Rhot}
\noindent

We assume that during an outburst there are generally two zones in the disc, with ionised and neutral material, separated at $ R_{\rm hot} $~--~the `radius of the hot zone'. If the accretion rate is very high, the hot zone can reach its outer radius, close to the Roche lobe radius. In the binaries with large $P_{\rm orb}$, however, the hot zone is often smaller than the whole disc, and it shrinks during an outburst decay.

We find radius of the hot zone following \citet{Lipunova+2022}, considering irradiating and viscous heating rate:

\begin{equation}\label{Qirr}
Q_{\rm irr}  = \mathcal{C}_{\rm irr} \frac{L_{\rm x}}{4\, \pi \,r^2}\, ,
\end{equation}
\begin{equation}\label{Qvis}
Q_{\rm vis} = \frac 38\, \frac{\sqrt{GM_{\rm x}}}{\pi\,r^{7/2}}\,F, \qquad F \approx \dot{M}_{\rm acc}\,h\,,
\end{equation}
where $L_{\rm x} = \eta\,(a_{\rm Kerr})\,\dot{M}_{\rm acc} c^2$ is the X-ray luminosity, $\dot{M}_{\rm acc}$ is the central mass accretion rate, $\eta\,(a_{\rm Kerr})$ is the accretion efficiency, which depends on the dimensionless Kerr parameter $a_{\rm Kerr}$, and $\mathcal{C}_{\rm irr}$ is the irradiation parameter. Note that for the 'hot' and 'cold' parts of the accretion disc, the value of irradiation parameter can be different. Below, $\mathcal{C}_{\rm irr}$ denotes the value of the irradiation parameter for the 'hot' part of the disc, while $\mathcal{C}_{\rm irr}^{\rm\,cold}$ corresponds to the 'cold' one.
Let us define the 'viscous temperature' $T_{\rm vis}$ from $Q_{\rm vis} \equiv \sigma_{\rm SB} T_{\rm vis}^4$, where $\sigma_{\rm SB}$ is the Stephan–Boltzmann constant. By analogy, the `irradiation temperature' $T_{\rm irr}$ is defined from $Q_{\rm irr} = \sigma_{\rm SB} T_{\rm irr}^4$.

We assume that disc at radius $r$ is in the hot state if irradiation temperature is higher than $T_{\rm hot} = 10^{4}$~K~\citep[][]{tuchman_et1990, dubus_et1999, Tavleev23}. Thus, radius of the hot zone $R_{\rm hot}(t)$ can be calculated from the central accretion rate:
    \begin{equation}\label{hot_r}
    \sigma_{\rm SB}\, T_{\rm hot}^4 = \mathcal{C}_{\rm irr}\, \frac{\eta\,(a_{\rm Kerr})\,\dot{M}_{\rm acc} c^2}{4 \,\pi \, R_{\rm hot}^2}\,.
    \end{equation}
when $ Q_{\rm irr} > Q_{\rm vis}$.

Due to gradual cooling of the disc, shrinking of the hot zone, and higher radial dependence of the viscous flux, eventually irradiation ceases to be important.
If the irradiation parameter was constant, condition $ Q_{\rm irr} > Q_{\rm vis}$ would break down at some radius, independent of the accretion rate, since one can write the ratio $Q_{\rm irr}/Q_{\rm vis}$ as~\citep{suleimanov_et2007e}:
\begin{equation}\label{Qir_Qvis}
\frac{Q_{\rm irr}}{Q_{\rm vis}} = \frac 43\, \eta(a_{\rm Kerr}) \, \mathcal{C}_{\rm irr} \, \frac{r}{R_{\rm grav}}\,, 
\end{equation}
where the gravitational radius $R_{\rm grav} = 2GM_{\rm x}/c^{2}$.

Later, the irradiation-controlled evolution is superseded by an evolution determined by the cooling-front propagation, in accordance with the disc instability model~(DIM)~\citep[][LM17]{dubus_et2001, lasota2001}. To find a position of $R_{\rm hot}$ one can take into account results of a cooling-front parameter study in the framework of DIM~\citep{Ludwig+1994}.  
To determine if a radius belongs to the hot zone, we check if any of the following conditions is met:
(1)~$T_{\rm irr}> T_{\rm hot}$; 
(2)~ $\Sigma > \Sigma_{\rm crit}^-$~\citep{Lasota_etal2008}; 
(3)~$ r<R_{\rm hot}(t-\Delta t) - v_{\rm front} \Delta t$,
where the cooling front velocity is calculated following \citet{Ludwig+1994}.

Regarding the irradiation parameter $\mathcal{C}_{\rm irr}$ in \eqref{Qirr}, one can  estimate its value by the following analytic expression~\citep[e.g.,][]{suleimanov_et2007e}:
    \begin{equation}\label{irr_theor}
    \mathcal{C}_{\rm irr} = \Psi(\theta) \,(1-A) \, \frac{z_0}{r} \, q, \ \ \ {\rm ~~where~~} q \equiv \left (\frac{{\rm d}\ln z_0}{{\rm d}\ln r} -1 \right)\, ,
    \end{equation}
where $z_0$ is the semi-thickness of the disc, $q\, (z_0/r)$ is the angle of incident rays, $(1-A)$ is the portion of absorbed and thermally reprocessed incident flux, and $\Psi(\theta) = 2\cos(\theta) \approx 2 z_0/r$ is the angular distribution of the irradiating flux from the central plane disc, where $\theta$ is the polar angle~(see Fig.~\ref{acc_disc}).

To take into account the radial dependencies of the irradiation parameter (see also discussion in Appendix~\ref{appendix_cirr}) we use auxiliary constant parameters $\widetilde{\mathcal{C}_{\rm irr}}$ and $\widetilde{\mathcal{C}_{\rm irr}^{\rm\,cold}}$, for the hot and cold parts of the disc, respectively. Using them, the genuine parameters $\mathcal{C}_{\rm irr}$ and $\mathcal{C}_{\rm irr}^{\rm\,cold}$ are expressed as follows:

    \begin{equation}\label{irr_both}
    \begin{split}
    &\mathcal{C}_{\rm irr} = \widetilde{\mathcal{C}_{\rm irr}} \, \left(\frac{z_0/r}{0.05}\right)^k \Psi(\theta(r)), \qquad k=1\, ; \\
    &\mathcal{C}_{\rm irr}^{\rm\,cold} = \widetilde{\mathcal{C}_{\rm irr}^{\rm\,cold}}  \,  \Psi(\theta^{\rm\,cold})\, .
    \end{split}
    \end{equation}

\subsection{Observed flux from the disc}\label{flux_disc}

Both parts of the accretion disc~(hot and cold) contribute to the total optical flux of the system.
Observed flux from the outer disc, which falls in the optical range, is determined by the contributions from the viscous heat $Q_{\rm vis}$ and self-irradiation $Q_{\rm irr}$. Thus, the local effective temperature can be expressed as:
    \begin{equation}\label{Te_Ti_Tv}
    T_{\rm eff}^4 = T_{\rm irr}^4 + T_{\rm vis}^4\, .
    \end{equation}
    
Accordingly, the observed flux from the disc in a band is calculated by the formula:
      \begin{equation}\label{fl_d}
    \mathcal{F}_{\rm disc} = \frac{2\,\pi\,\cos{i}}{d^2} 
    \int_{\nu_{\rm min}}^{\nu_{\rm max}} \mathrm{d}\nu
    \int_{R_{\rm in}}^{R_{\rm hot}}  \,B_\nu({T_\mathrm{eff}(r)})\, r\,\mathrm{d}r\,,
    \end{equation}
where $d$ is the distance to the source, $i$ is the disc inclination, $\nu_\mathrm{min}$ and $\nu_\mathrm{max}$ are the limits of the observational spectral band, $B_\nu$ is the Planck intensity.
The viscous temperature is found from the viscous torque:
\begin{equation}\label{Teff_Tvis}
\sigma\, T_{\rm vis}^4 = \frac{3 \, (G\,M)^4\, F}{8\,\pi\, h^7} \, .
\end{equation} 
where $F$ is a solution to the viscous evolution equation \eqref{e1}.

Due to the fact that we do not find a solution for the vertical structure of the cold disc, its relative semi-thickness assumed as $z_0/r=0.05$ ($\theta^{\rm cold} = \mathrm{const}$) and its viscous temperature $T_{\rm visc}$ is set to zero. Value of $\widetilde{\mathcal{C}_{\rm irr}^{\rm\,cold}}$ parameterises all associated uncertainty.

\subsection{Flux from the companion star} \label{flux_Mopt}
\begin{figure}
    \centering{
    \includegraphics[width=\columnwidth]{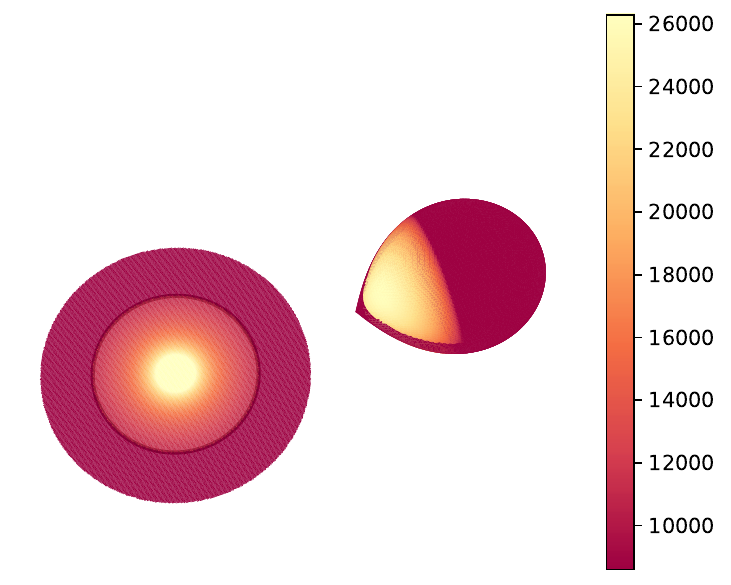}}
    \caption{Visualisation of the binary system~\lup{} at the peak of the 2002 outburst. Effective temperature is colour-coded. The circle on the disc shows the border between the hot and cold zones.}
    \label{3dsketch}
\end{figure}

The central X-ray source impacts not only the disc, but also the optical component of the binary system, if the latter is not completely shadowed. Code \textsc{freddi} is capable of calculating orbit-modulated light curves~\citep{Lipunova+2022} of a companion star irradiated with an X-ray flux when there is no self-occultation in the system. The effective temperature distribution in a close binary system~(namely, BH LMXB~\lup{}) is shown in Fig.~\ref{3dsketch}. Model parameters are listed in Table~\ref{tab:params} for the outburst in Fig.~\ref{mdotvstime}.

\section{Wind models and methods}\label{windmod}

In order to take into account the effect of the wind, it is required to specify a physically substantiated form of the inhomogeneity function $ \Windrate $ in~\eqref{e1}. The modified version of the~\textsc{freddi}~code can take into account any
user-specified wind model. In the subsection below we only describe the main thermal wind model adopted from S86 and W96 used to fit X-ray Novae light curves. All other available wind model options are listed and described in Appendix~\ref{appendix_val}. We also verify the work of our code with the 'toy' wind model adopted from S86. In this 'toy' model, wind's impact creates oscillations in luminosity if the constant inflow of matter into the disc occurs in the system ($\dot{M}_{\rm out} =$ const, see Eq.~\eqref{bound}). In Appendix~\ref{appendix_ver}, we describe this model in detail and compare {\sc freddi} and S86 results, reproducing $\dot{M}_{\rm acc}$ oscillations.

\subsection{Thermal wind model}\label{ther_mod}
\noindent

The dynamics of the Compton-heated wind and corona have been analytically studied in B83 (see basic definitions in Section~\ref{wind_desc}). The authors have estimated the mass loss rate as a function of distance along the disc's radius. They consider the flow carried away by the wind as a set of streamlines and solve equations of momentum and energy conservation along them. S86 have proposed analytical approximation for the results obtained in B83 (see Eqs.~\ref{3.7} --\ref{5/3e}) in Appendix~\ref{append_thermwind}).

In a later work, W96, two-dimensional magnetohydrodynamic calculations have been performed and the results of B83 and S86, generalised. W96 take into account the various processes contributing to X-ray heating and radiation cooling, while B83 consider heating and cooling only through the Compton processes. Secondly, while B83 suggest a disc is geometrically thin, W96 take into account the structure of the upper layers of the disc in an optically transparent approximation. The wind model elaborated by W96, applicable for winds in AGNs, is also added to the \textsc{freddi}'s list of choices.

\begin{figure} 
    \center{\includegraphics[width=\columnwidth]{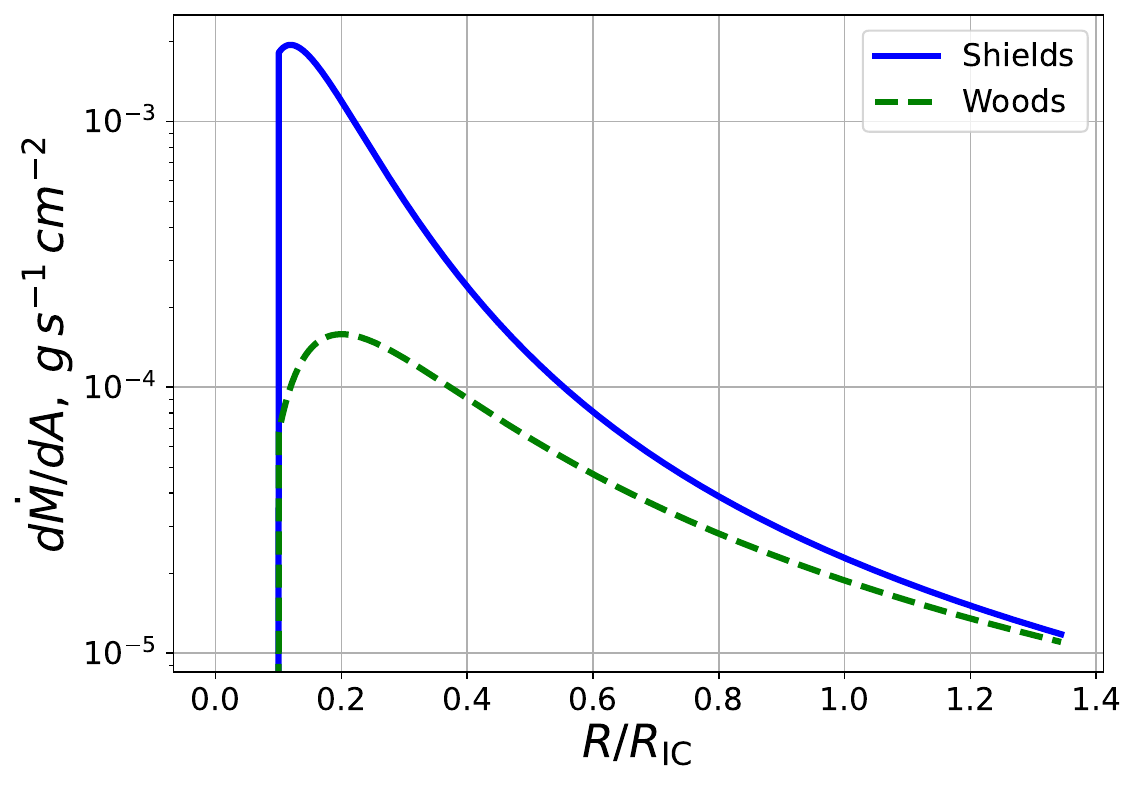}}
    \caption{Distribution of the surface flow rate due to wind along the radius for S86~\eqref{5/3e} and W96~\eqref{e3}. 
    The Compton temperature $T_{\rm IC}=1.5 \times 10^{7}$ K, $M_{\rm x} = 10 \, M_{\rm \odot}$, and bolometric luminosity $L = L_{\rm Edd}$.
    }
    \label{windvsrad}
    \end{figure}

To model a viscous evolution of a disc with a wind in the current study, we take the function $ \Windrate(h)$ as the approximation to the results of S86 in a form provided by W96:

    \begin{equation}\label{e3}
    \begin{gathered}
     \Windrate = 2\, \eta_{\rm wind}\,\dot{m}_{\rm ch} \times \left(\frac{1 + ((0.125l+0.00382)/\xi))^2}{1 +
    (l^{4}(1+262\xi^2))^{-2}}\right)^{1/6} \times
    \\
    \times \exp{[-(1-(1 + 0.25\xi^{-2})^{-1/2})^{2}/(2\xi)}]\,,
    \end{gathered}
    \end{equation}
where  $\xi = r/R_{\rm IC}$ and $ l = L / L_{\rm cr} $, see equations~\eqref{R_IC} and~\eqref{L_cr}, and the characteristic mass-loss rate per unit area $ \dot{m}_{\rm ch}(\Xi)$ can be found in Appendix~\ref{append_thermwind}. Value of the pressure ionisation parameter $\Xi$ at the wind-disc interface is assumed to be equal to $\Xi_{\rm c,max}$ on definition~(W96), see Fig.~\ref{Teq}. We also introduce parameter $\eta_{\rm wind}$ -- a dimensionless factor allowing easy wind-power adjustment. Corresponding distribution, as well as the function~\eqref{5/3e} from S86, is shown in Fig.~\ref{windvsrad}. 

Thus, parameters of the wind and its rate are controlled by parameters $T_{\rm IC}$, $L$, and $\Xi$ (see Appendix~\ref{append_thermwind}). The mass loss rate per unit area above is to be used in equation \eqref{e1}  whose method of  solution is described in the next subsection.

\subsection{Methods}\label{meth_merc}
\noindent

Using an implicit difference scheme, we reduce the solution of the differential equation~\eqref{e1} with boundary conditions to sequential solution of a system of algebraic equations at each time step, which is carried out using the tridiagonal matrix algorithm described in Appendix~\ref{appendix_num}. 
The new scheme supersedes the original one in the  code \textsc{freddi}~\citep[][LM17]{SOFT}, described also in \citet{Shakura_book}.

The thermal wind model described above is applicable for an outburst of the accretion disc in the case of X-ray nova, triggered by a critical accumulation of mass during a quiescent state or by a short-term mass transfer. As the initial condition, we take the distribution found by \citet{Lipunova-Shakura2002} for a disc without wind. It implies a zero inflow of matter ($\dot{M}_{\rm out} = 0$ ) at the external boundary and, thus, the zero first derivative of $ F(h) $ there~(see Eq.~\eqref{bound}). This quasi-stationary distribution describe  hot parts of $\alpha$-discs without winds after the peak of a fast-rise exponential-decay outburst,  as pointed out by \citet{lipunova2015}. Recall that the inner boundary condition is the same for all models, namely: the zero value of the viscous torque $ F $~\citep{SHAKURA_SUNAEV}.

\section{Results}\label{res_4u}
    
\subsection{The impact of wind on the course of the outburst}\label{res_im}
\noindent

To analyse the wind effect during an outburst in a LMXB, we use 
the wind model~\eqref{e3}. Outbursts are simulated for two values of the peak accretion rates: $\dot{M}_{\rm acc, 0} = \dot{M}_{\rm Edd}$  and $\dot{M}_{\rm acc, 0} = 0.1\times \dot{M}_{\rm Edd}$, where $\dot{M}_{\rm Edd} = 1.4 \times 10^{18}\,(M_{\rm x}/M_{\odot})$  g/s for a compact object of mass $ M_{\rm x} = 10 \, M_{\odot} $. We also vary the turbulent viscosity parameter $ \alpha $ and the self-irradiation coefficient of the disc $ C_{\rm irr} $.
For the calculations described in this subsection, the temperature $ T_{\rm IC} $ is fixed at $ 10 ^ 8 $ K. Parameter $ \Xi = 7.53 $ is taken from simulations of the heating and cooling processes~\citep{Higgin17}. 
 
\begin{figure}
    \center{\includegraphics[width=\columnwidth]{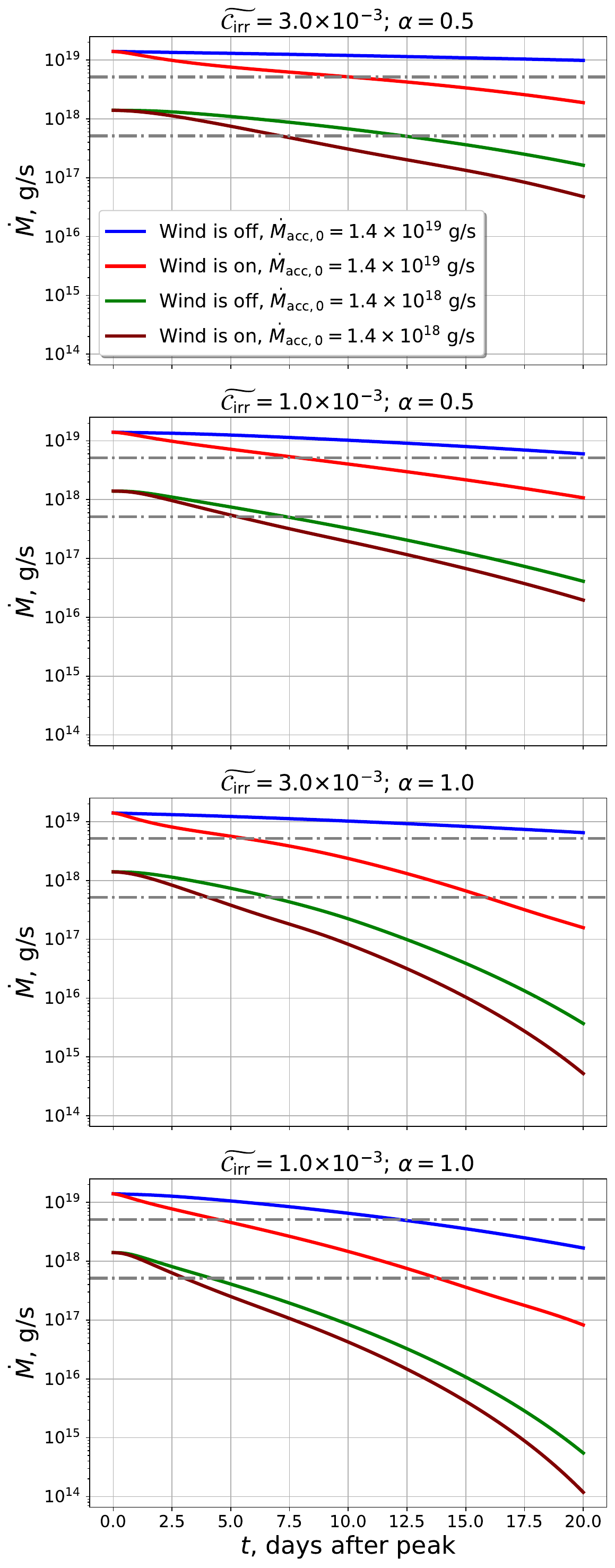}}
    \caption{Evolution of an accretion disc central accretion rate in the presence of the thermal wind and without it. Two sets of curves are shown for different initial peak accretion rates, namely: $\dot{M}_{\rm acc, 0} = \dot{M}_{\rm Edd}$ and $\dot{M}_{\rm acc, 0} = 0.1\,\dot{M}_{\rm Edd}$, where $\dot{M}_{\rm Edd} = 1.4 \times 10^{18}\,(M_{\rm x}/M_{\odot})$ g/s. Other parameters are $ M_{\rm x} = 10 \, M_{\odot} $, the external radius is the tidal one, $\Xi = 7.53$, and $T_{\rm IC} = 10^{8}$ K.  Two grey dash-dotted line indicate levels, where the accretion rate drops by $e$ times from the initial one.}
    \label{mdot_com}
    \end{figure}

\begin{figure} \center{\includegraphics[width=\columnwidth]{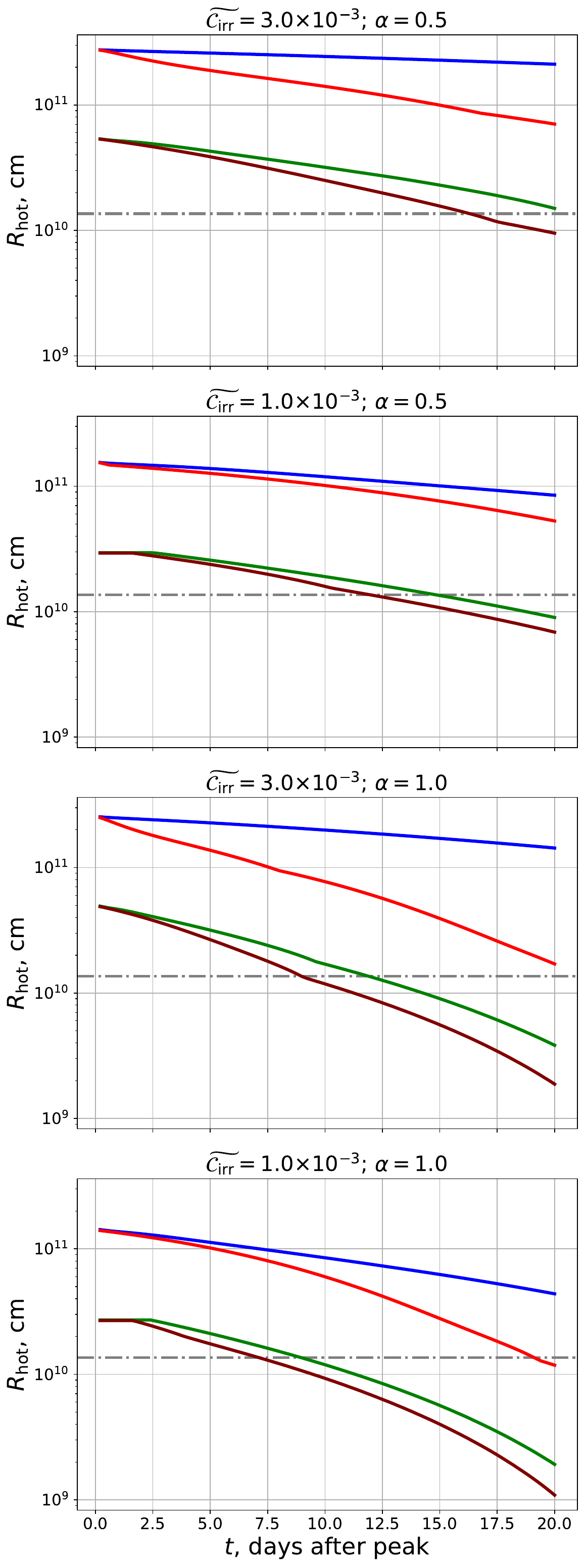}}
    \caption{Hot (ionised) zone radius $ R_{\rm hot} $ versus the time after the peak of a light curve. The parameters and curves styles (legends) are the same as in Fig.~\ref{mdot_com}. Grey dash-dotted line represents wind launching radius  $0.1\,R_{\rm IC}$.}
    \label{rhot_com}
    \end{figure}

\begin{figure} \center{\includegraphics[width=\columnwidth]{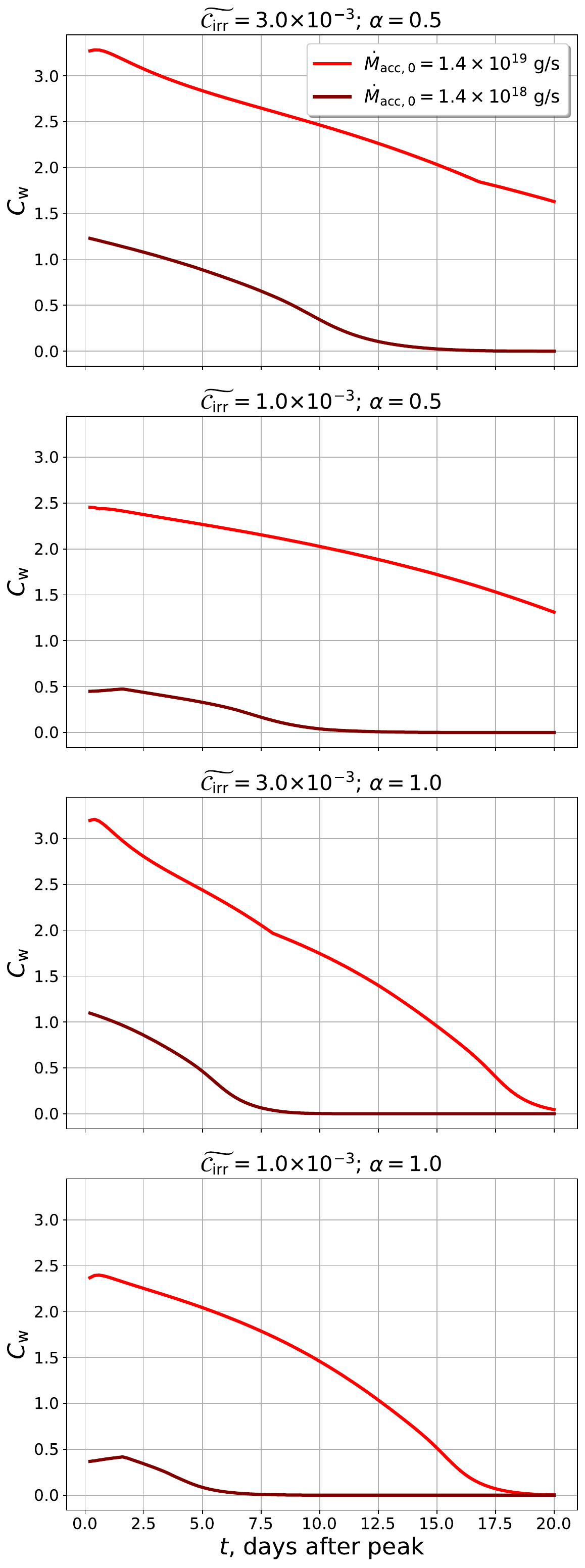}}
    \caption{The ratio between the mass loss rate due to the thermal wind and the central accretion rate ($ C_{\rm w} \equiv  \dot{M}_{\rm wind} / \dot{M}_{\rm acc}$) versus the time after the peak of a light curve. Parameters are as in Figs.~\ref{mdot_com} and ~\ref{rhot_com}.}
    \label{mwma_com}
    \end{figure}

Resulting accretion rate evolution is shown for models with and without wind in Fig.~\ref{mdot_com}. Figures~\ref{rhot_com} and~\ref{mwma_com} show the evolution of the radius of the hot zone $ R_{\rm hot} $ and the ratio of the mass loss rate due to wind to the accretion rate $ C_{\rm w} \equiv  \dot{M}_{\rm wind} / \dot{M}_{\rm acc}$ (where $\dot{M}_{\rm wind}>0$), respectively.

Figure~\ref{mdot_com} shows that the effect of the wind, accelerating the evolution of the central accretion rate, is more pronounced in models with higher values of the initial accretion rate $ \dot{M}_{\rm acc, 0} $ and self-irradiation parameter of the disc $ C_{\rm irr} $. This is explained as follows. 
Higher accretion rate means higher luminosity and a more powerful wind, according to its prescription \eqref{e3}. Furthermore, higher accretion rate and irradiation parameter, they both result in a bigger hot disc, enhancing the role of the wind and the mass lost with it (see Fig.~\ref{rhot_com}).

\begin{figure} 
    \center{\includegraphics[width=\columnwidth]{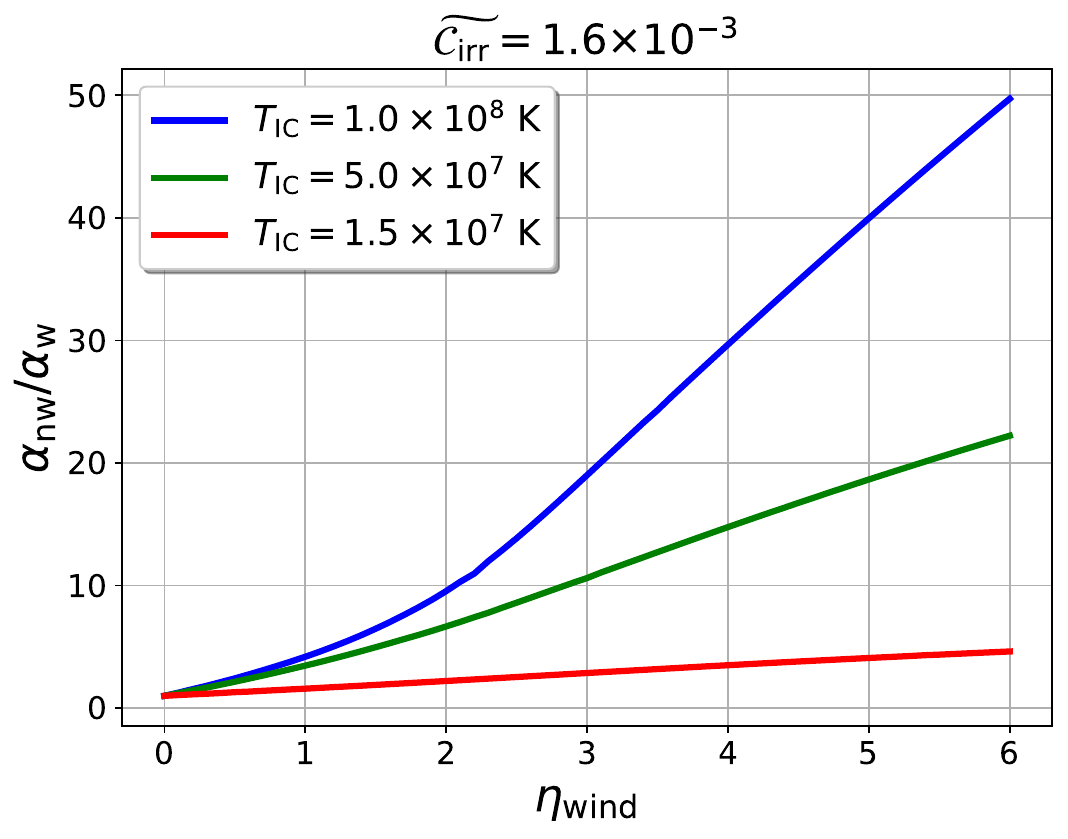}}
    \caption{Dependence of the thermal wind $\alpha$-correction on the wind's power $\eta_{\rm wind}$. This correction is the inverse ratio between calculated $\alpha$ parameter value in disc simulations with the wind presence~($\alpha_{\rm w}$) and without it~($\alpha_{\rm nw}$), while the same resulting decline rate of the central accretion rate is considered. All light curve simulations are made with $\alpha_{\rm nw}=1.0$, $M_{\rm x} = 10 M_{\odot}$ and $\dot{M}_{\rm acc, 0} = 1.4 \times 10^{19}$~g/s.} 
    \label{alpha3.10}
    \end{figure}

In the presence of a wind the outburst duration decreases: `wind models' demonstrate shorter characteristic times as if the turbulent parameter $\alpha$ was higher.
In Fig.~\ref{alpha3.10} we plot an `$\alpha$-correction': the value of the turbulent parameter $\alpha_{\rm nw}$ in a disc without wind, divided by the viscosity parameter $\alpha_{\rm w}$ in a disc with thermal wind and the same decline rate.
The wind force is parameterised by the dimensionless quantity $ \eta_{\rm wind} $ (see the formula \eqref{e3}). For each $ \eta_{\rm wind} $ and $\alpha_{\rm w}$ we calculate a model with wind. Afterwards, we fit the resulting $\dot{M}(t)$ curves with models without wind and find best-fit value of viscosity parameter $\alpha_{\rm nw}$. 
Since the wind takes away disc's mass and its angular momentum, the ratio $\alpha_{\rm nw}/\alpha_{\rm w}$ is always $>1$~(because with the same decline rate the viscosity processes contribute less when the wind acts). An apparent degeneracy between parameters $\alpha$ and $\eta_{\rm wind}$ is discussed in the beginning of Section~\ref{discus}.

In the framework of the thermal wind, the spectrum hardness plays a crucial role in changing the rate of an outburst decline (Fig.~\ref{alpha3.10}), because increasing the Compton temperature leads to the increase of the area where the wind is launched, see Eq.~\eqref{R_IC}. For parameters $ \eta_{\rm wind} = 1 $ and $ T_{\rm IC} = (0.15 - 1) \times 10^{8} $ K, the estimate of the turbulent viscosity parameter $ \alpha $ can be overestimated due to neglect of the influence of the thermal wind by approximately $ 1.4 - 4$ times.

\begin{figure} 
    \begin{center}
    \includegraphics[width=\columnwidth]{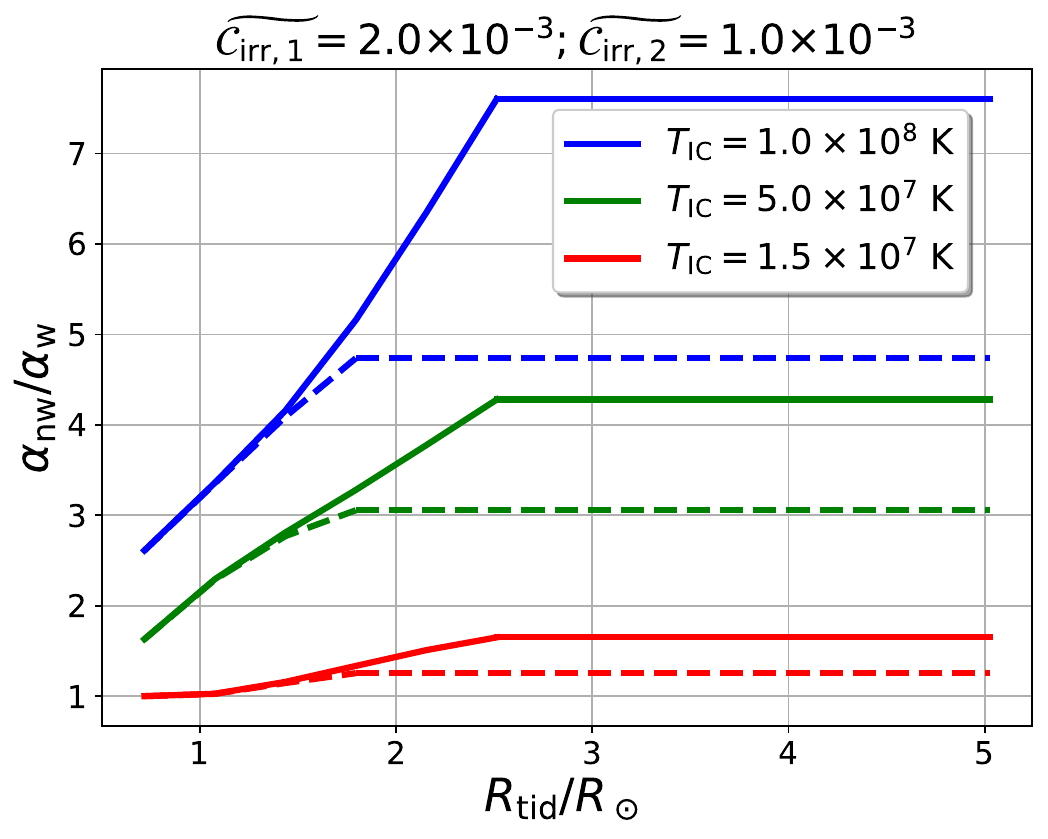}
    \caption{Thermal wind $\alpha$-correction 
    versus the outer radius of an accretion disc for the different Compton temperatures. This correction is the inverse ratio between calculated $\alpha$ parameter value in disc simulations with the wind presence~($\alpha_{\rm w}$) and without it~($\alpha_{\rm nw}$), while the same resulting decline rate of the central accretion rate is considered. 
    All models are calculated for $\alpha_{\rm nw}=1.0$, $M_{\rm x} = 10 M_{\odot}$ and $\dot{M}_{\rm acc, 0} = 1.4 \times 10^{19}$~g/s. All solid curves correspond to $\widetilde{C_{\rm irr}} = 2.0 \times 10^{-3}$, while dashed ones to $\widetilde{C_{\rm irr}} = 1.0 \times 10^{-3}$.}
    \label{alpha_rad}
    \end{center}
    \end{figure}

It is expected that the wind '$\alpha$-correction' depends on the disc size. Thus, we have calculated a number of models for different values of the outer radius.
Figure~\ref{alpha_rad} shows $\alpha$-correction for outbursts in discs of different sizes and Compton temperatures. It can bee seen that until some point $\alpha$-correction grows with the accretion disc size.
This is explained by the fact that the area, where the wind operates, grows as the square of the external (tidal) radius ($ R_{\rm tid}^2 $).
But it continues until the outer radius is smaller than $R_{\rm hot}$~\eqref{hot_r}. The $\alpha$-correction curve levels off for bigger discs because the hot disc area is limited and defined by $\dot{M}_{\rm acc, 0}$ and $C_{\rm irr}$. Thus, the $\alpha$-correction value never exceeds a few for a moderate Compton temperatures, for any binary periods. 

The wind power, which can be set by $ \eta_{\rm wind} $, affects simultaneously the shape of a light curve and the value of $ C_{\rm w} $. Thus, it is theoretically possible to impose restrictions on the wind strength relative to the accretion rate, namely, on the value of $ C_{\rm w} $ (see Fig.~\ref{shape} in Section~\ref{discus}) from a light curve profile.

\subsection{Modelling outburst of \texorpdfstring{~\lup{}}{4U 1543-47} in 2002}\label{4u_exact}
\noindent

To illustrate the effect of the wind for an outburst of a real transient, we fit the evolution of the central accretion rate and optical flux of the 2002 outburst of the X-ray nova~\lup{}. The observed central accretion rate $ \dot{M}_{\rm acc}(t) $ is obtained from spectral modelling of the archival data of \textit{RXTE}/PCA~(LM17).
The optical light curves in the $ V $ and $ J $ bands used in the current work are also described there. The system parameters are listed in Table~\ref{tab:params}.

\begin{figure} 
    \begin{center}
    \includegraphics[width=\columnwidth]{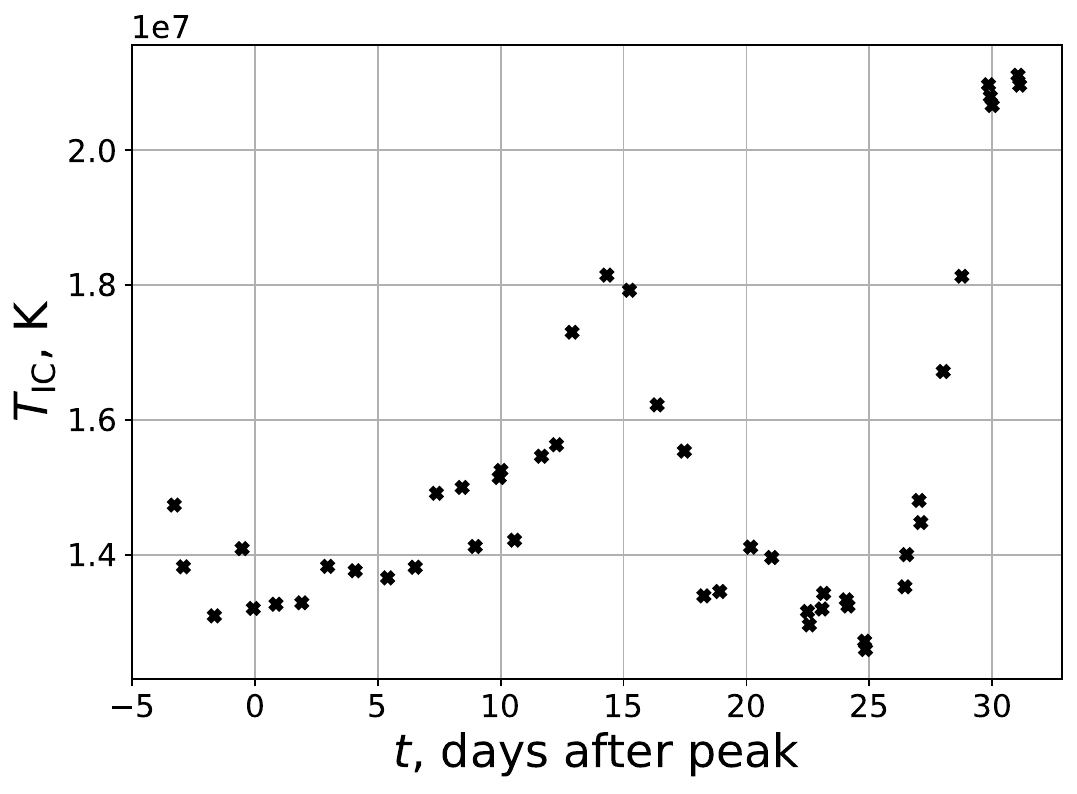}
    \caption{Inverse Compton temperature $T_{\rm IC}$ calculated for the 2002 outburst of \lup{}~(see Eq.~\eqref{TIC}).}
    \label{TICvst}
    \end{center}
    \end{figure}

Figure~\ref{TICvst} shows the evolution change of the $ T_{\rm IC} $ during the outburst~\lup{}\,(2002) obtained using results of the spectral modelling of the \textit{RXTE} data. To simulate the~\lup{} light curve in the present work, we take the average value of $ T_{\rm IC} $ over the shown time interval: $ 1.5 \times 10^{7} $~K or $1.3$~keV. The bump in the interval between the 10th and 20th day after the peak happened due to a turn-on of an additional spectral component, possibly, a jet~\citep[][LM17]{Bux_Bai2004, Russel_jet}. To assess impact of the variation in $ T_{\rm IC} $, we make auxiliary fits, adopting maximum and minimum values of $ T_{\rm IC} $ during this 30-day period for wind  model. The resulting $ \alpha $ parameter estimate change  within  15\% at the very worst.

\begin{table}
    \caption{\label{tab:params}  Parameters used in fitting of the 2002 outburst of~\lup{}.}
    \renewcommand{\arraystretch}{1.4} 
    \renewcommand{\tabcolsep}{2mm}   
    \begin{center}
    \begin{tabular}{c|c|c}
    \hline
    Parameter & Value & Ref.$^{\star}$
    \\
    \hline
    Mass of the BH, $M_{\rm x}$ &  $9.4 \, M_{\odot}$      & [1], [2]     
    \\
    Orbital period, $P$ &  $1.116$ days & [3]
    \\
    Inclination, $i$ &      $20.7^{\circ}$  &  [3]       
    \\
    Mass of the companion, $M_{\rm opt}$ &  $2.5\, M_{\odot}$ & [1], [2]
    \\
    Distance to the system, $d$ &  8.6 kpc & [4], [5]
    \\ 
    Kerr parameter, $a_{\rm Kerr}$      &    $0.4$     & [6]
    \\
    Pressure ionisation parameter, $\Xi$  &    $7.53$ & [7]
    \\
    Compton temperature, $T_{\rm IC}$      &    $1.5 \times 10^{7}$ K   &  Fig.~\ref{TICvst}
    \\
    Wind efficiency, $\eta_{\rm wind}$ &  $1.0$ & ---
    \\
    Effective temperature of the star surface &  $8600$ K & [8]
    \\
    Bolometric albedo of the star surface      &    $0.5$ & [9]
    \\
    Roche-lobe filling factor of the star     &    $1.0$
    & --- 
    \\
    \hline
    \end{tabular}
    \end{center}
    $^\star$ 
    [1] \citet{Orosz98}, 
    [2] \citet{Orosz2002}, 
    [3] \citet{Orosz2003}, 
    [4] \citet{Gandhi19}, 
    [5] LM17, 
    [6] \citet{Spin}, 
    [7] \citet{Higgin17}, 
    [8] \citet{A2V}, 
    [9] \cite{Basko+1974}
    \end{table} 

    \begin{figure} 
    \begin{center}
    \includegraphics[width=\columnwidth]{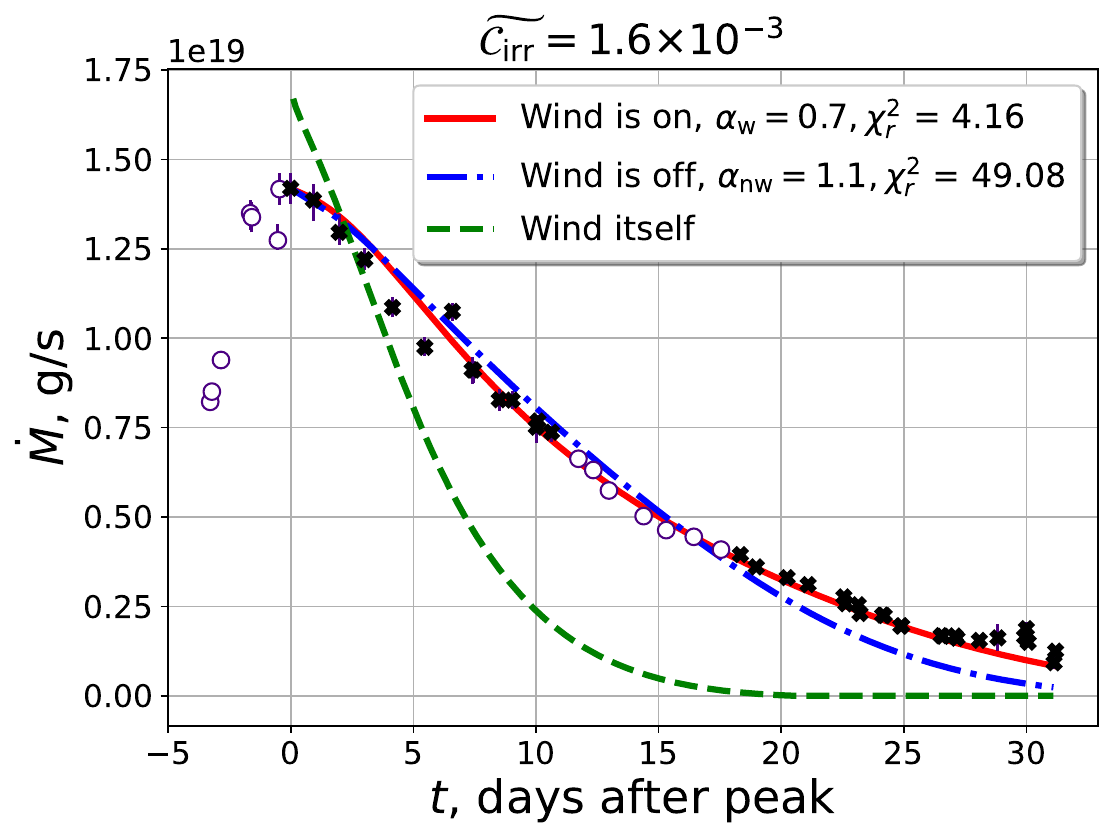}
    \caption{Central accretion rate evolution $\dot{M}_{\rm acc}(t)$  during the~\lup{} outburst (2002). The plot shows $\dot{M}_{\rm acc}(t)$ dependence based on the observations (black 'X'), which is fitted by two models, presented by the curves 'Wind is on' and 'Wind is off', meaning that the corresponding simulation fit takes or doesn't take thermal wind presence into account respectively. Empty circles are excluded from the fitting due to them not being a part of the decay. The 'Wind itself' curve shows the rate of the mass-loss in the thermal wind ($\dot{M}_{\rm wind}(t)$). All parameters used in simulations are listed in  Table~\ref{tab:params}.}
    \label{mdotvstime}
    \end{center}
    \end{figure}

Fits for models with and without thermal wind along with the observational data are presented in Fig.~\ref{mdotvstime}. To fit modelled accretion rate to  $ \dot{M}_{\rm acc}(t) $ obtained from observational data,  we use Levenberg-Marquardt non-linear least-squares minimisation curve fitting method from the python package LMFIT~\citep{lmfit}. We  exclude seven middle points from the fit because the spectral modelling for these data was carried out with frozen distance's value while all other points were modelled with thawed distance parameter (see LM17 for details). Nevertheless, it is remarkable how these seven data points lie neatly on the light curve. \citet{Tetarenko18} presented a bolometric light curve for this outburst, but its slope differs from that of our  $\dot M(t)$, apparently, because the bolometric corrections applied by them cannot correctly reflect the spectral evolution.

As expected, to explain the temporal evolution, lower value of the turbulent parameter is needed in the model with the thermal wind.
The fit results are: the initial accretion rate onto the BH is $ \dot{M}_{\rm acc, 0} = 1.4 \times 10^{19} $~g/s, the viscosity parameter $ \alpha $ for cases with and without wind is about $ 0.7 $ and $ 1.1 $, respectively. It should be also noted that in the absence of the wind the best-fit value of the turbulent parameter $\alpha$ is lower than the value found in LM17 for the same BH parameters. The explanation of this effect is given in Appendix~\ref{appendix_cirr}.

Figure~\ref{lastR} shows the hot radius evolution corresponding to Fig.~\ref{mdotvstime}. It can be seen that the radius $R_{\rm hot}(t)$ in the wind model (the red curve) becomes less than~$0.1 \times R_{\rm IC}$ (the grey horizontal line) on the 21st day. This leads to a complete shutdown of the thermal wind from the hot zone, see also Fig.~\ref{mdotvsmwind}. We note that first $\sim$30 days after the outburst's maximum, hot radius of the disc is controlled by irradiation. This is true for the both models (with and without wind). Thus, our resulting estimate of the alpha-correction due to wind is mostly free from effects of the later disc evolution when the cooling front advances.

    \begin{figure} 
    \begin{center}
    \includegraphics[width=\columnwidth]{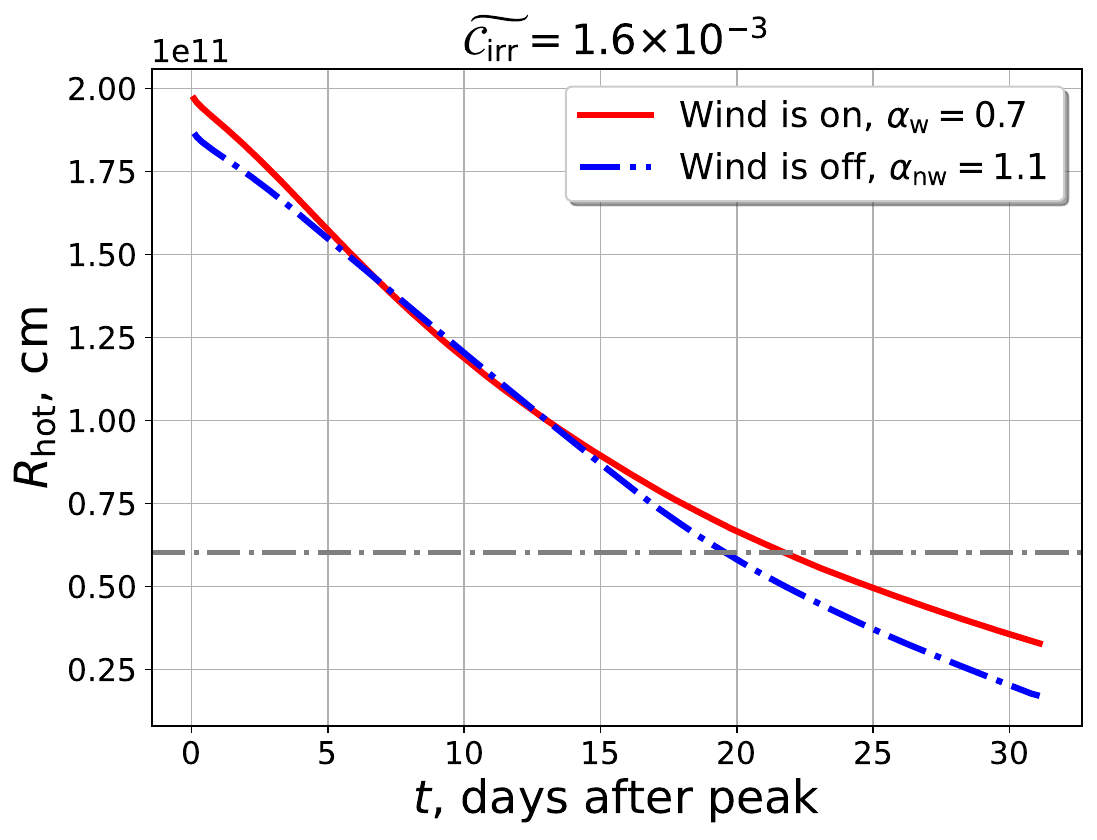}
    \caption{Modelled evolution of hot zone radius during the 2002 outburst of \lup{}~(see Fig.~\ref{mdotvstime}). Grey dash-dotted line represents wind launching radius  $0.1\,R_{\rm IC}$.}
    \label{lastR}
    \end{center}
    \end{figure}

    \begin{figure} 
    \begin{center}
    \includegraphics[width=\columnwidth]{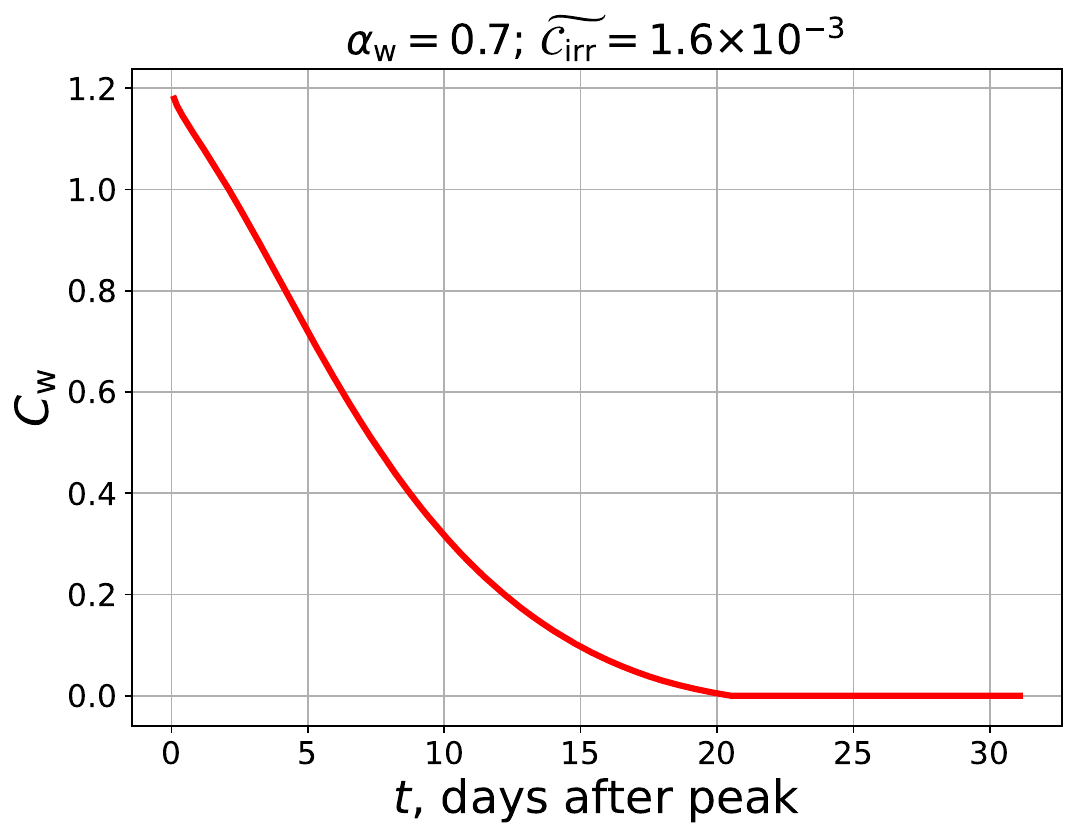}
    \caption{Evolution of the ratio of the mass loss rate due to wind to the accretion rate onto the BH ($ C_{\rm w} \equiv  \dot{M}_{\rm wind} / \dot{M}_{\rm acc}$), corresponding to the 'Wind is on' model in Fig.~\ref{mdotvstime}.}
    \label{mdotvsmwind}
    \end{center}
    \end{figure}

\begin{figure} 
    \begin{center}
    \includegraphics[width=\columnwidth]{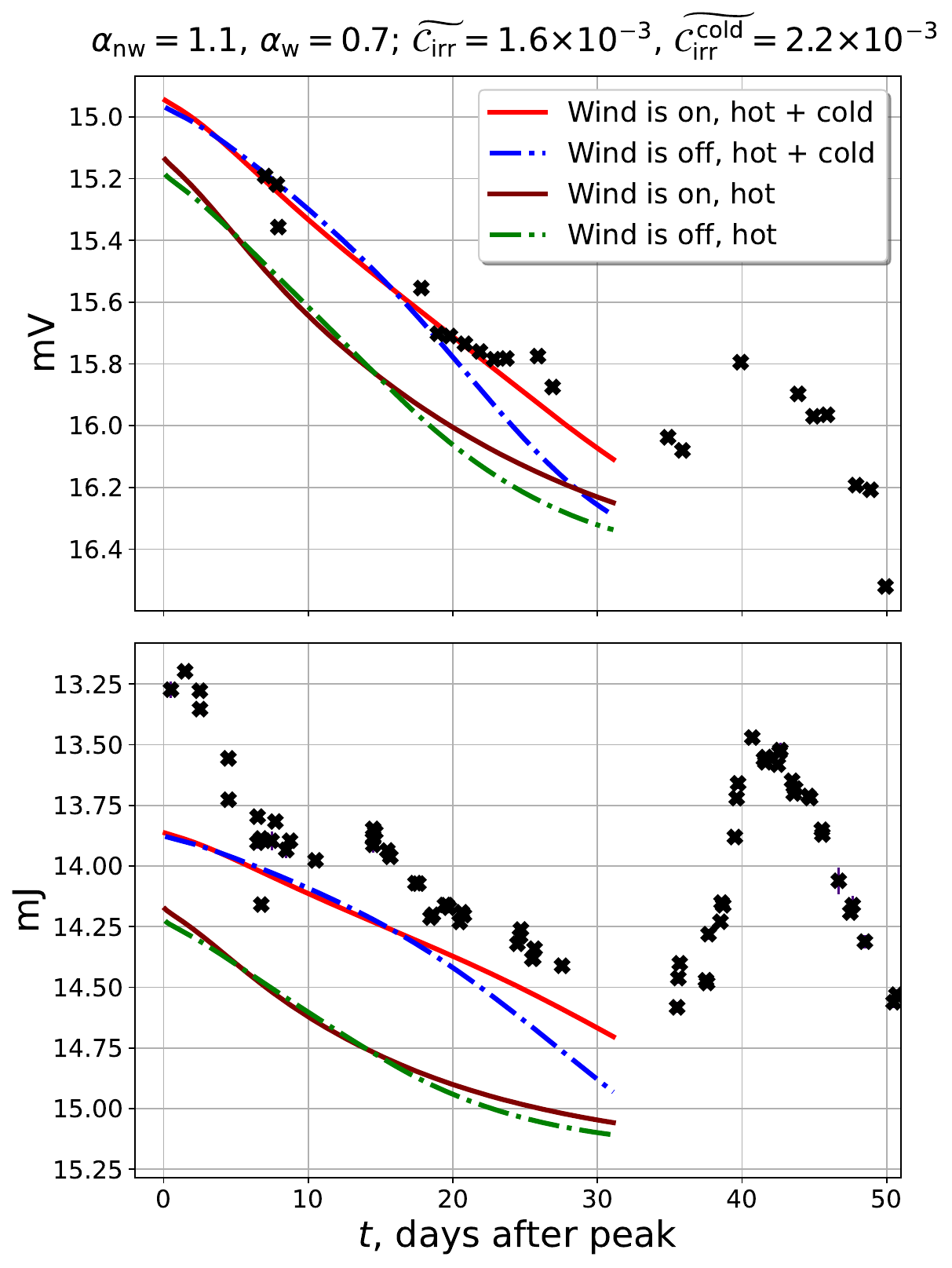}
    \caption{Observed and modelled optical/IR light curves in $V$ (top panel) and $J $ (bottom panel) bands during the 2002 outburst of~\lup{}. The plots show two types of modelled light curves: 'hot' and 'hot + cold disc',  which means that either the simulation considers that only the hot part of the disc emits optical/IR flux or its cold part does that as well, respectively.}
    \label{mVmJ_1}
    \end{center}
    \end{figure}

\begin{figure} 
    \begin{center} 

    \includegraphics[width=\columnwidth]{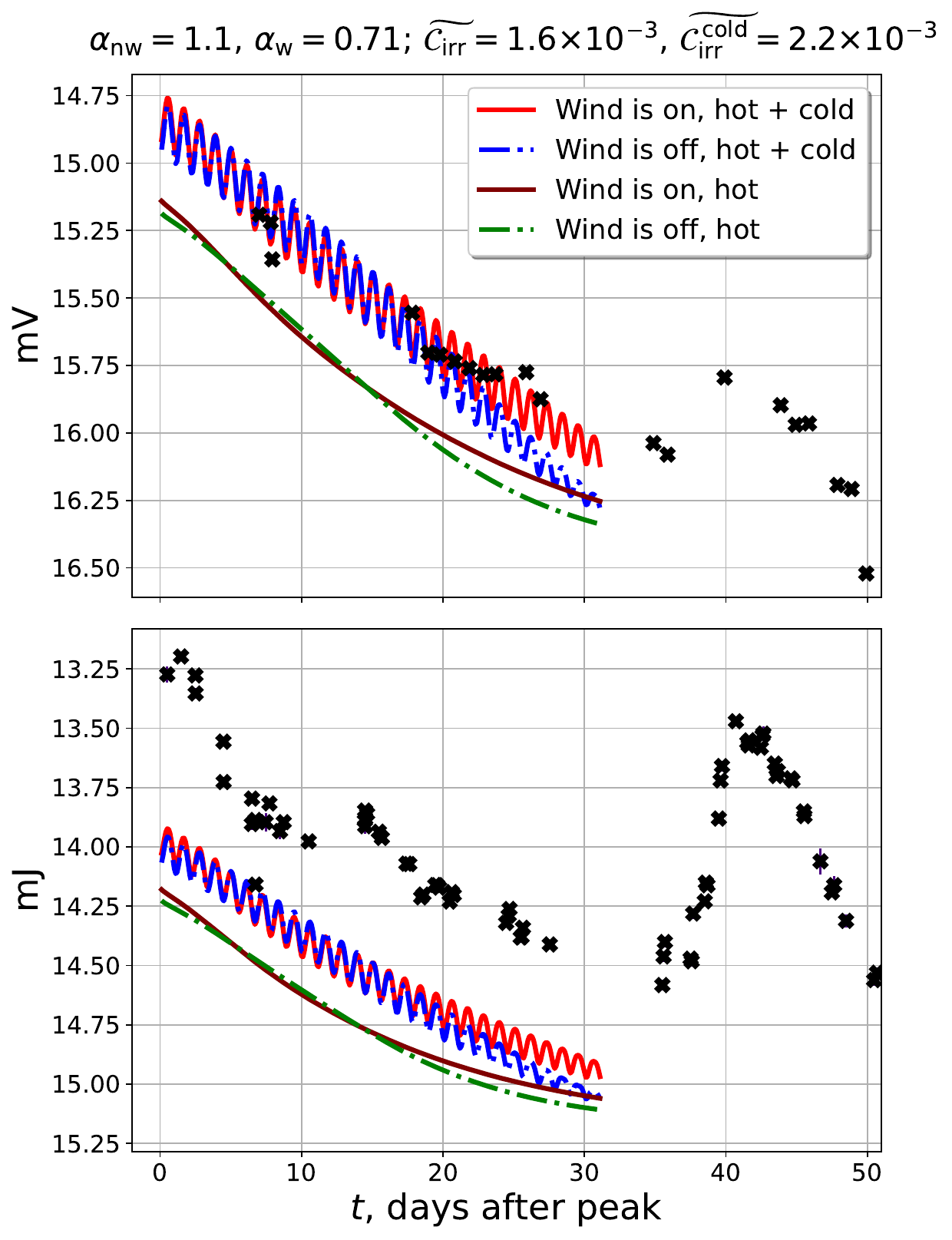}
    \caption{Observed and modelled optical/IR light curves in $V$ (top panel) and $J $ (bottom panel) bands during the 2002 outburst of~\lup{}. The plots show two types of modelled light curves: 'hot' and 'hot + star',  which means that either the simulation considers that only the hot part of the disc emits optical/IR flux or the irradiated donor star counterpart does it as well, respectively.}
    \label{mVmJ_2}
    \end{center}
    \end{figure}

The self-irradiation parameter $ C_{\rm irr} $ can be determined from the optical data. Within the framework of our model, three components contribute to the optical flux of the system: a hot disc, a cold disc, and an irradiated donor star, see~Sections~\hbox{\ref{Irrad_Rhot}--\ref{flux_Mopt}}. Figures~\ref{mVmJ_1} and~\ref{mVmJ_2} show the observed optical data in the $ V $ and $ J $ bands obtained by~\citet{Bux_Bai2004}, as well as the model light curves, corrected for the interstellar absorption with $A_V = 1.6$\, mag and $A_J = 0.282\,A_V$, see LM17. 

It can be seen that the emission from the hot disc alone cannot explain the observed data, in accordance with results of LM17.
To explain the optical flux in $ V $, it is necessary to take into account either re-emission from a cold disc or from irradiated optical companion (or maybe both). 
The self-irradiation parameter $ \widetilde{C_{\rm irr}} $ for the hot part of the disc is set to $ 1.6 \times 10^{-3} $. In Appendix~\ref{appendix_cirr} we discuss distinctions between irradiation parameter prescriptions in this work and LM17.
At the same time, in the $J$ band, full agreement cannot be reached. Perhaps this happens due to the fact that the contribution of the jet of the system is noticeable in this band~\citep[][ LM17]{Bux_Bai2004, Russel_jet}. 

\section{Discussion}\label{discus}
\noindent

Analysis of the LMXBs light curves observed during outbursts shows that $ \alpha $ lies in the range $ \sim 0.2-1$ \citep{Lipunova-Shakura2002, suleimanov_etal2008, malanch_shak15, lipunova_malanchev2017, VISCOS2019, Tetarenko18, Lipunova+2022}.
However, three-dimensional simulations of the magneto-rotational instability as a source of momentum transfer and matter in a disc yield values that are an order of magnitude smaller: $ \alpha \lesssim 0.1 $~\citep{Balbus,Hirose}. To a certain degree this discrepancy could be explained by an overestimate of the value of the parameter $\alpha$ from observations, which, in turn, can be caused by the presence of non-viscous mechanisms of the momentum transport in the accretion discs.

\citet{Tetarenko18_1} pointed to the importance of wind and the issue of overestimating the $\alpha$ parameter when analysing the light curves of the X-ray binaries. According to the authors, outburst's fast evolution in the considered systems  can be interpreted either as a strong intrinsic rate of angular momentum transfer in the disc, which can be achieved if a large-scale magnetic field threads the disc, or as a direct indication of mass outflow in the system.

    \begin{figure} 
    \begin{center}
    \includegraphics[width=\columnwidth]{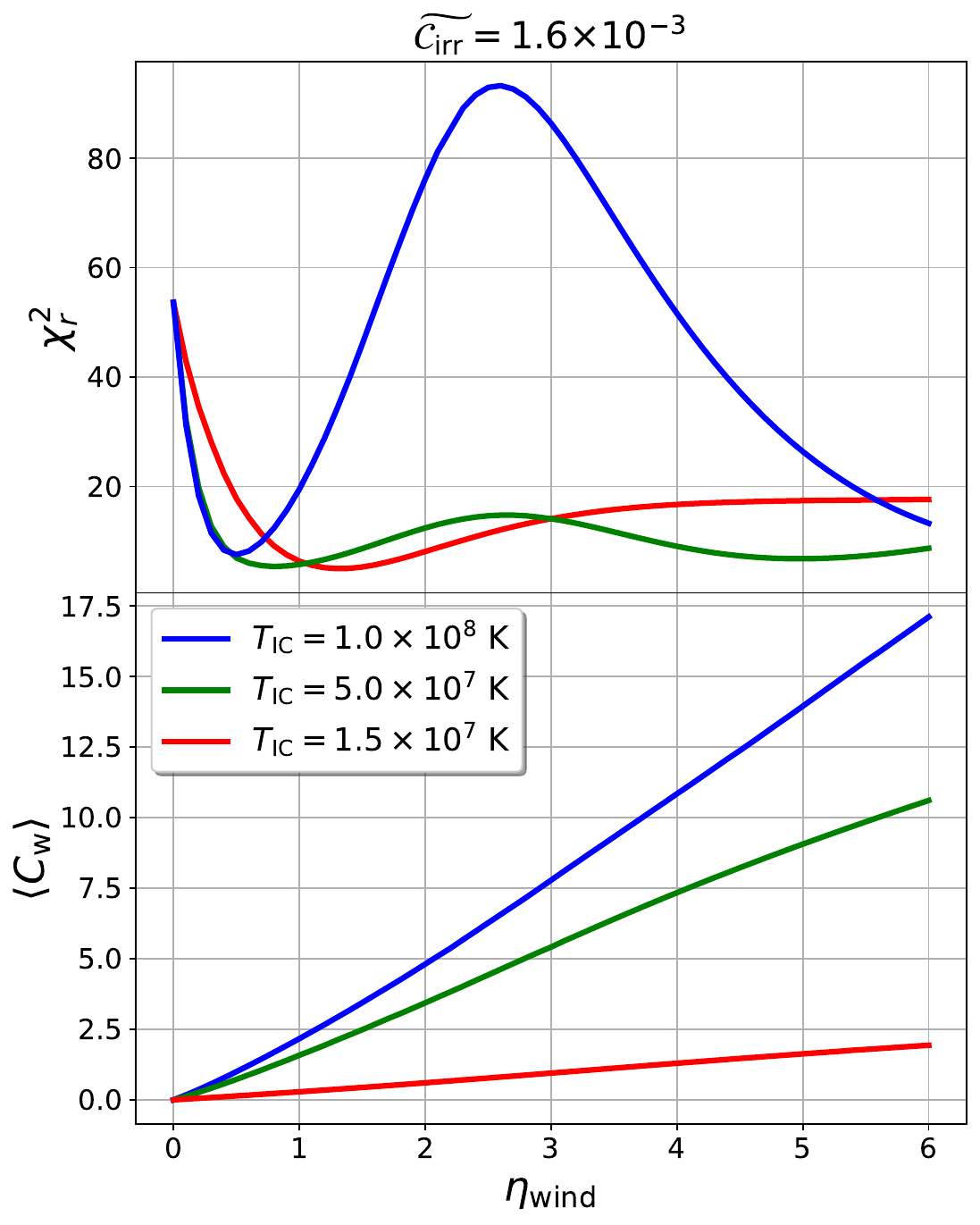}
    \caption{Reduced chi-square for the best-fit models of 2002 \lup{} outburst (top panel) and mean value of wind to accretion ratio $ \left<C_{\rm w}\right>=  \left< \dot{M}_{\rm wind} / \dot{M}_{\rm acc}\right>$ (bottom panel) depending on the wind's power parameter $\eta_{\rm wind}$.
    Model dependencies are taken with different values of $T_{\rm IC}$. 
    Value of $\alpha$ parameter without wind 
    $\alpha_{\rm nw} = 1.1$ and self-irradiation $C_{\rm irr} = 1.6 \times 10^{-3}$ is constant for all fits.}
    \label{shape}
    \end{center}
    \end{figure}

Both $\alpha$ and $\eta_{\rm wind}$ influence the decay rate of an outburst (see Fig.~\ref{alpha3.10}). To test the degree of degeneracy between them, we fit the same data for \lup{} changing $\eta_{\rm wind}$. Figure~\ref{shape} shows fits' results for the same 2002 outburst of \lup{}, obtained using different values of $\eta_{\rm wind}$, which parameterises the wind power, see Eq.~\eqref{e3}. On the top panel, reduced chi-square value is presented, while lower panel shows corresponding mean of mass loss ratio $ \left<C_{\rm w}\right> = \left<\dot{M}_{\rm wind} / \dot{M}_{\rm acc}\right>$ during the outburst. From the overall behaviour of fit statistics, one can conclude that there is a dependence of the shape of a modelled light curve on the wind power, which may help to resolve a degeneracy between parameters $\alpha$ and $\eta_{\rm wind}$. For particular outburst, Figure~\ref{shape} favours the model with the wind comparing to the model of disc evolution without a wind, when $\eta_{\rm wind} = 0$.
As one can see on the top panel, reduced chi-square curves all have a minimum near the value $\eta_{\rm wind} \sim 1$, i.e., the original wind model. This fact, while could be a coincidence, is quite remarkable.

Observational manifestation of an outflow depends on the inclination of the disc to our line of sight.
Let us estimate the number of hydrogen atoms in the wind along the line of sight. In a simplified picture, the wind starts at disc radius $R_{\rm w} \equiv 0.1 \times R_{\rm IC}$, but at a greater distance we can consider that it has spherical geometry with density depending solely on the angle $i$ with the disc normal. 
For the density distribution, we suppose that the number of particles decreases exponentially with the cosine of $i$.
With all that in mind, we get the following expression for the number of the hydrogen atoms $N_{\rm H}$:
\begin{equation}\label{Nh_eq}
    N_{\rm H}(t, i) = \int_{\frac{R_{\rm w}}{\sin{i}}}^{R_{\rm max}(t)} n_{\rm H}(t, R, i) \, {\rm d}R\,,
\end{equation}
where the hydrogen number density $n_{\rm H}$ is given by
\begin{equation}\label{nh_eq}
    \begin{gathered}
    n_{\rm H}(t, R, i) = \frac{k}{1 - \exp{(-k)}}\, \times \exp{(- k \cos{i})} \, \times 
    \\ 
    \frac1{4\pi R^2 v_{\rm out}(R)\mu m_{\rm p}} 
    \int_{R_{\rm w}}^{R_{\rm out}}
    {\!\Windrate\left(r, t - t_\mathrm{fly}\left(\frac{r}{\sin{i}}, R\right)\right)2 \pi r}\,{\rm d}r \,.
    \end{gathered}
\end{equation}

The upper limit $R_{\rm max}(t)$ is the external radius of the envelope, which expands with velocity $v_{\rm out} (R)$. The mass loss per unit area $\Windrate = \Windrate(r, t)$ is given by Eq.~\eqref{e3}. The time delay $t_{\rm fly}$ is calculated as the time required for the matter to move from `a start of expansion' $r/\sin{i}$ to $R$. In our case of constant velocity $t_{\rm fly} = (R - r/\sin{i})/v_{\rm out}$. The first two multipliers in Eq.~\eqref{nh_eq} are the damping factor for the density distribution over polar angles with a dimensionless parameter $k$ and its normalisation. 

    \begin{figure} 
    \begin{center}
    \includegraphics[width=\columnwidth]{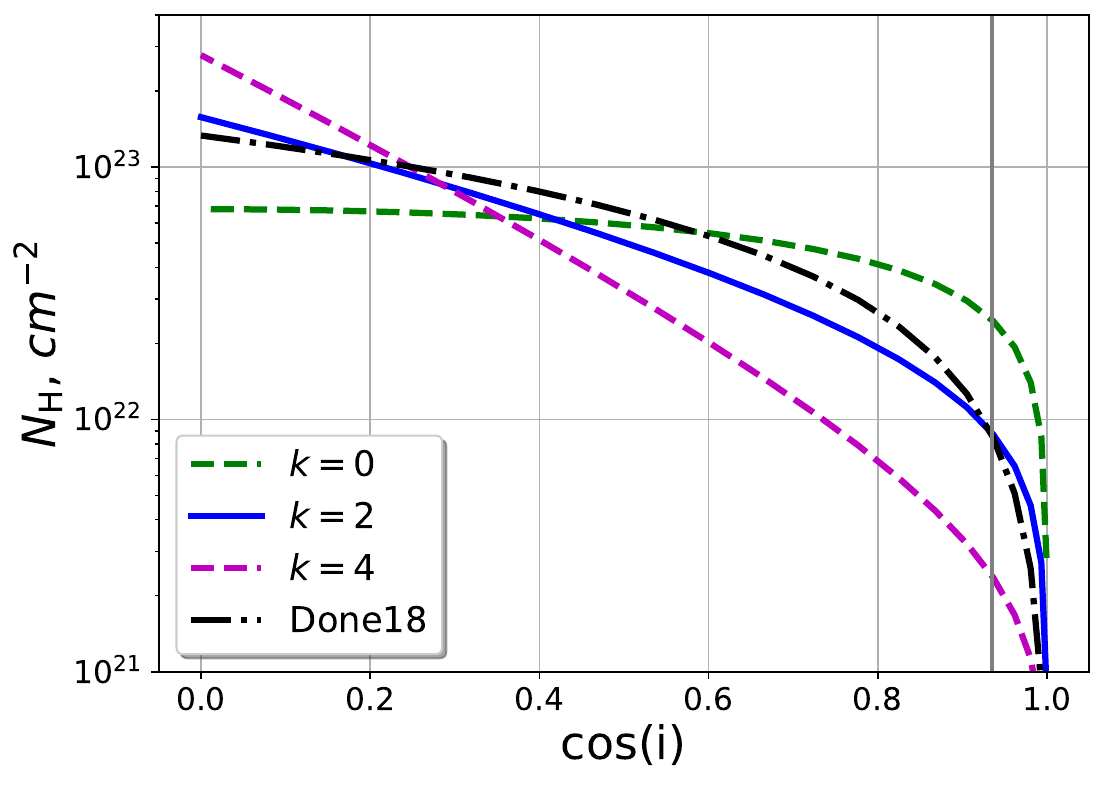}
    \caption{Column density at 10th day of an outburst with parameters as in Fig.~\ref{mdotvstime} versus the inclination of a binary for various values of the damping coefficient $k$ in the angular distribution of wind density~(Eq.~\eqref{Nh_eq},~\eqref{nh_eq}). Black dash-dotted line shows the dependence calculated following~\citet{Done18}, see Eq.~\eqref{Done18eq6}. Grey line indicates the adopted value of inclination for \lup{}~($20.7^{\circ}$).}
    \label{nh_1}
    \end{center}
    \end{figure}

The above estimate leaves out the complexity of density and streamlines' configuration in the proximity of the disc. Time delay between the moment when matter leaves the disc and the moment that we consider as the `start of expansion' is ignored, so the accuracy of the estimate should deteriorate for very small angles $i$. 

For a typical wind velocity $\sim 500$ km/s \citep{Higgin19}, a spherical layer expands to $\sim 10^{13}$~cm in three days, which shows that for not very small inclination angle $i$ the simple spherical geometry should be satisfactory. During first days after the peak of an outburst, the estimate is rough, since we ignore the matter lunched before the peak.

Figure~\ref{nh_1} shows the column density versus the system inclination at the 10th day after the 2002 outburst of \lup{} for different damping parameters $k$. It can be seen that the number of particles corresponding to the uniform angular distribution ($k = 0$) hardly depends on $i$. Its mild variation is explained by the dependence of the starting radius of the envelope $R_{\rm w}/\sin{i}$ on the inclination $i$. In Fig.~\ref{nh_1}, the black dash-dotted line shows the dependence calculated following equation(6) from~\citet{Done18}, with a proper normalisation applied:

\begin{equation}\label{Done18eq6}
N_{\rm H}^{\rm Done18} = \frac{\dot{M}_{\rm wind}\,(1- \cos i)}{2\pi R_{\rm w} v_{\rm out}\, \mu\, m_{\rm p}}\,,
\end{equation} 
with the same $v_{\rm out}=500~$km/s and $R_{\rm w}=0.1 \times R_{\rm IC}$\footnote{\cite{Done18} adopt the angular density distribution $\rho(R, \cos i) = \rho_0(R) (1 - \cos i)$. From the mass conservation law, $\dot{M}_{\rm wind} = R^2 v_{\rm out} \int_\Omega \rho(R, \cos i) {\rm d}\Omega$, one obtains $\rho_0(R) = \dot{M}_{\rm wind} / (2\pi R^2 v_{\rm out})$.}.

    \begin{figure} 
    \begin{center}
    \includegraphics[width=\columnwidth]{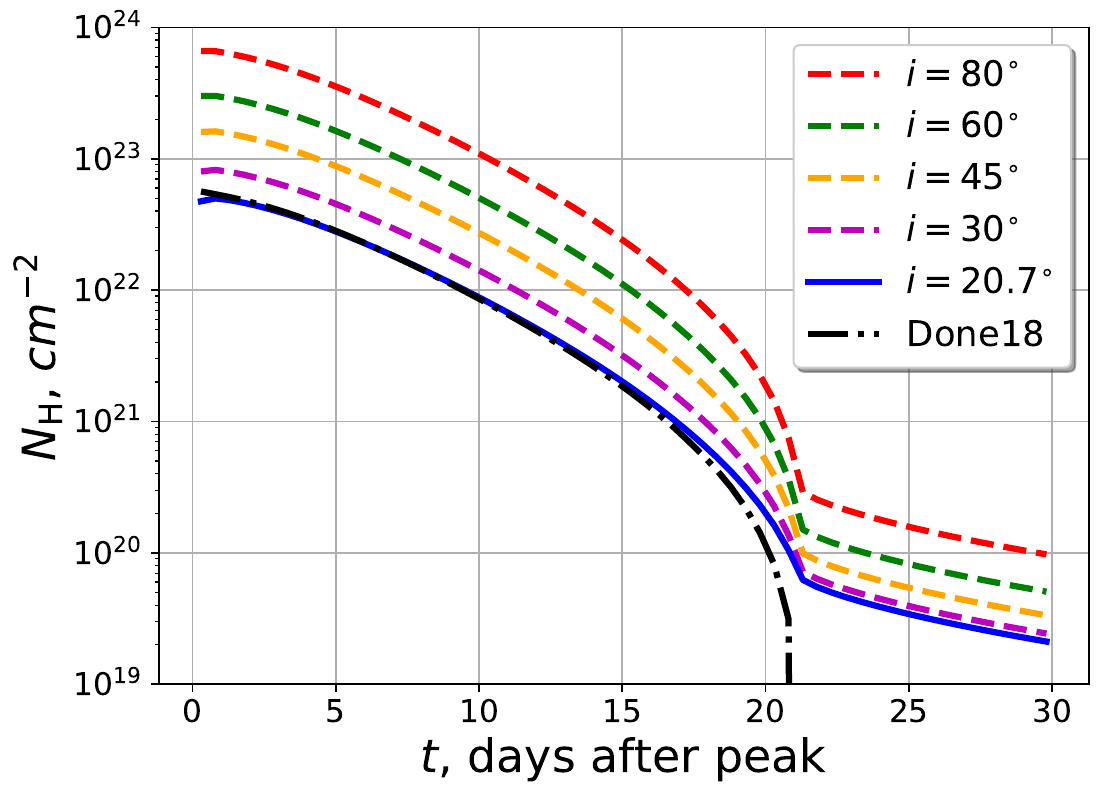}
    \caption{Column density versus the time after the peak of the 2002 outburst of~\lup{} for damping coefficient $k=2$, see Eq.~\eqref{Nh_eq},~\eqref{nh_eq}. Black dash-dotted line shows the dependence calculated following~\citet{Done18} with $i = 20.7^{\circ}$, see Eq.~\eqref{Done18eq6}.} 
    \label{nh_2}
    \end{center}
    \end{figure}

As the envelope expands, the number of atoms along the line of sight should decrease due to spherical divergence principle. Fig.~\ref{nh_2} shows the column density $N_{\rm H}(t)$ versus time for the same set of the wind parameters as in Fig.~\ref{nh_1} and fixed $k=2$. Different curves correspond to different values of the binary inclination. Approximately on the 21st after the peak, all curves show a break due to the thermal wind shutdown, which happens after the hot zone radius becomes less than $R_{\rm w}$, see Section~\ref{4u_exact} and Fig.~\ref{lastR}. This characteristic trend is explained by the fact that the main contribution to the number of $N_{\rm H}$ comes from the wind matter close to the disc.

\citet{Done18} have analysed effects of the spectral evolution  (with spectrum parameterised using bolometric luminosity) on the wind using the analytical model for the thermal wind by W96. Our present work neglects the spectral evolution because we assume the constant Compton temperature (based on observational data; see \S\ref{4u_exact}). Thus we focus on the effects of the thermal wind on the viscous disc evolution. Naturally, to apply our scheme to a source with strong spectral evolution, one should take into account evolution of $T_{\rm IC}$.

Figure~\ref{nh_2} demonstrates that the column density is not proportional to the mass accretion rate, or the luminosity, as in \citet{Done18}, but drops faster following the mass loss rate. The difference between our results apparently results from different treatment of the wind-launching zone. We assume that the wind operates only in the hot zone (because it affects disc evolution only there), while \cite{Done18} assume constant outer radius of the launching zone. 
Actually, the outer recombined disc can be shadowed by the thicker inner parts of the disc and this could explain no thermal wind there. On the contrary, if scattered but effective irradiation works, the thermal wind is expected from the outer disc part as well (see \citealt{Dubus_19}).

    \begin{figure} 
    \begin{center}
    \includegraphics[width=\columnwidth]{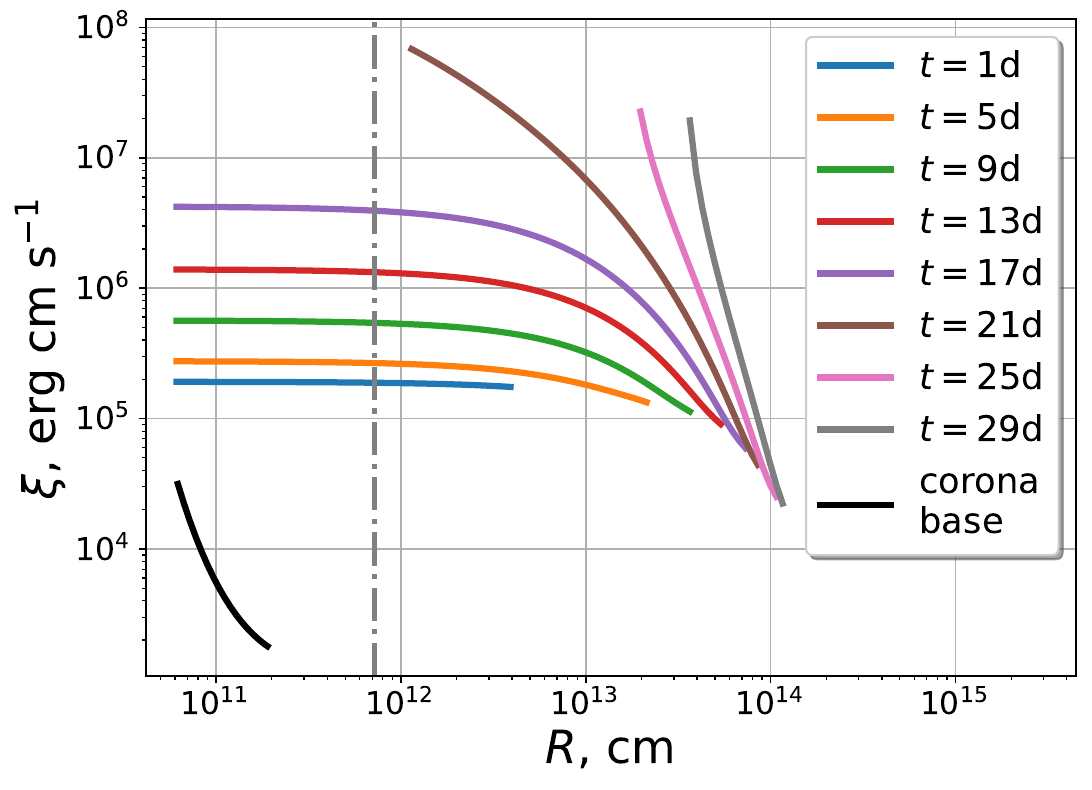}
    \caption{Radial distribution of $\xi= L_{\rm x}^{(i)} / (n_{\rm H} R^2)$, where $n_{\rm H}$ is given by \eqref{nh_eq}. Colour lines show profile of $\xi$ along the line of sight of the observer ($i = 20.7^\circ$) for an indicated number of days after the peak. Line ranges show expansion of the wind. Dash-dotted grey line marks the semi-major axis.}
    \label{xi_R}
    \end{center}
    \end{figure}

The dash-dotted line in Fig.~\ref{nh_2} shows $N_{\rm H}(t)$ calculated using \eqref{Done18eq6} from \citet{Done18} for $i = 20.7^{\circ}$. It demonstrates that slightly different angular distributions of mass in the wind (our $k=2$ and formulation of \citealt{Done18}) does not affect the evolution of $N_{\rm H}$. More crucial is how the wind-launching zone changes.

When the cooling front reaches the innermost wind-launch radius ($6 \times 10^{10}$~cm), $N_{\rm H}(t)$ changes remarkably. At late times this evolution has the asymptotics $t^{-2}$ expected for an envelope expanding with the constant speed and mass. As a result, if such deceleration of $N_{\rm H}(t)$ is observed, that will confirm the concept of the evolving viscous disc and shadowing of the outer disc at the same time. However, if $N_{\rm H}(t)$ is proportional to luminosity, this will indicate the whole surface of the disc persistently launches the wind. A hardening of X-ray spectrum will result in the increase of $N_{\rm H}$ \citep{Done18}.

In spectral modelling of the X-ray data~(LM17), the absorption of soft X-rays by ISM was accounted for with Xspec model {\tt TBabs} assuming column density $N_{\rm H} = 4 \times 10^{21} {\rm cm}^{-2}$ (absorption by photoionisation in gas, molecules and grains of ISM). This number is less than expected in the wind during the first $\sim 15$ days after the peak, see Fig.~\ref{nh_2}. This fact bears no conflict, since, if the wind material is highly ionised, the absorption is mainly due to electron scattering. The optical depth of the wind by the Thomson scattering is 0.04 near the peak of the outburst. 

Approximate evolution of the ionisation parameter $\xi$ (cf. pressure ionisation parameter $\Xi$, $\xi=4\,\pi\,\Xi\, k\,T\,c = L_{{\rm x}}^{(i)} / (n_{\rm H} R^2)$, $n_{\rm H} = n_{\rm H}(t, R, i)$) in the distant spherical wind along the assumed line of sight of \lup{} is shown in Fig.~\ref{xi_R}. Here
$L_{\rm x}^{(i)} = 4\pi d^2\, F_{\rm x}^{\rm obs} $ is the isotropic luminosity derived from the flux in $0.5-50$~keV. A constancy of $\xi$ over radii close to the origin is explained by an assumed constant speed of expansion \citep[see also equation 7 in][]{Done18}. At first, $\xi$ rises at each radius, since the mass loss rate drops faster than accretion rate (and $L_x$). Along the particular line of sight, no lines can be produced in the distant wind for such high values of the ionisation parameter. After the 21th day the wind stops: it is seen as an outward shift of the left edge of the curves since the wind material moves away. The black line shows a distribution of $\xi$ parameter just near the disc surface at the base of the corona where $n_{H} = \dot{\Sigma}_{\rm w} / (2 v_{\rm out}\, \mu\,m_p)$. The corona base distribution of $\xi$ is plotted for constant $i$ corresponding to $z_0/r = 0.05$ at time $t = 1$d and covers the radial range from $R_{\rm W}$ to $R_{\rm hot}$. This value has a very weak time dependency and evolves as $\sim 1 / \log L_{\rm X}$. Evidently, sources at high inclination would be better candidates to observe spectral lines produced in the wind matter.

It would be reasonable to assume that all the material ejected in the thermal wind, not only increases $N_{\rm H}$, but can also influence the process of self-irradiation of the disc through scattering of X-rays. \citet{Dubus_19} have actually shown that including a dependence of $C_{\rm irr}$ on the mass loss in the wind can lead to a more complicated form of light curves of X-ray novae. For instance, it is theoretically predicted that $C_{\rm irr}$ is proportional to the column density in the wind material. This could affect the $\alpha$-correction that we obtain and requires a dedicated study. As of now, we can evaluate the irradiation parameter according to equation (10) of \citet{Dubus_19}. Resulting $C_{\rm irr}$ is about $ 6 \times 10^{-2} $ at the maximum of the outburst which is about 2 orders higher than the irradiation parameter obtained from comparison of observed and model optical flux in $V$, $C_{\rm irr} \lesssim 3\times 10^{-4}$~(see Fig.~\ref{Cirr_comp} in Appendix~\ref{appendix_cirr}). A similar result was obtained in LM17, where it was shown (for the same BH parameters), that $C_{\rm irr}$ cannot be significantly higher than $(3-6)\times 10^{-4}$. This is also consistent with the results of \citet{Tetarenko20} for BH LMXB GX\,339$-$4 on analytic estimates of self-irradiation from scattering in thermal wind. Namely, in the beginning of an outburst of GX\,339$-$4, the value of $C_{\rm irr}$ is highly overestimated even from the scattering in thermal wind alone~\citep[Fig.~6 in][folowing \citealt{Dubus_19}]{Tetarenko20}. And this is despite the fact that after transitioning to the intermediate state the situation actually reverses, which is pointed out by the authors.

\section{Conclusions}\label{concl}
\noindent

In this work, we studied how the course of the viscous evolution of an accretion disc changes in the presence of the thermal wind. A numerical method for solving this problem was developed and implemented as a part of the open code {\sc{freddi}}.

We have incorporated stationary thermal wind models of B83, S86 and W96 into {\sc{freddi}}. 

We have simulated typical light curves during LMXB outbursts with the thermal wind by W96 and compared them with light curves obtained for accretion discs without wind. All simulations are conducted through solving the evolution equation for a viscous $\alpha$-disc. It was found that in the presence of the thermal wind, the evolution of accretion disc noticeably accelerates, which is especially pronounced for brighter outbursts. This is due to the fact that the power of the wind increases with the accretion rate as well as with the disc size.

Ignorance of the thermal wind can lead to an overestimation of the viscosity parameter $\alpha$ determined from observations. A factor of several is suggested by Fig.~\ref{alpha3.10}; it depends on the spectral hardness of central X-rays.
This change may not be enough to fully explain the discrepancy between MHD calculations and observations concerning the magnitude of the turbulent parameter; however, the change is significant and must be taken into account when modelling disc bursts evolution.

We have tested our method by comparing its results with those of~S86 in the context of the accretion disc luminosity oscillations. 
Oscillations of luminosity arise and do not decay if the rate of mass loss in the wind is proportional to the rate of accretion onto a compact object and the ratio $ C_{\rm w} = \dot{M}_{\rm wind} / \dot{M}_{\rm acc} $ is large enough, namely, $ C_{\rm w} \gtrsim 12.8 $, which is reasonably consistent with the results of S86.

Using implemented thermal wind model W96, we have fitted $\dot{M}_{\rm acc}(t) $ of the  BH LMXB \lup{} outburst of 2002. 
We find that, when the thermal wind is ignored, the observational value of the turbulent viscosity parameter $ \alpha $ for the particular outburst is overestimated by about a factor of 1.6. 
Moreover, the fits to the $\dot{M}_{\rm acc}(t)$ dependence are in favour of a model-predicted power of the thermal wind. 

With an updated version of {\sc freddi} now also capable of reproducing more complex optical/IR light curves, we simulated these as well. We have compared observed optical/IR ($V$ and $J$) light curves of the 2002 \lup{} outburst with our model ones. Ultimately, we conclude that, while $V$ band flux could be explained within our models, the emission from the hot disc, cold disc and irradiated by X-rays optical companion still cannot fully account for $J$ band flux. This may be a sign of the presence of some unaccounted by us source of additional optical/IR radiation, possibly a jet.

We have calculated the wind column density $N_{\rm H}$ evolution along the line of sight. The model predicts that 
after the wind stops (when disc hot-zone radius becomes smaller than the wind launch radius), the number of hydrogen atoms along the line of sight undergoes a break of a temporal evolution. At the same time, the residual envelope is hardly observable due a very high ionisation parameter.

It should be said that if during an outburst the accretion rate becomes higher than the super-Eddington accretion rate, an outflow starts from from the central, radiation-dominated region of the disc~\citep[][]{SHAKURA_SUNAEV}, that can partly screen the outer parts of the disc from central X-rays. Such outbursts should be modelled with another wind function $\Windrate$.

\section*{Acknowledgements}

Authors are grateful to Nikolai Shakura and Valery Suleimanov for the helpful discussions.

\section*{Data Availability}

Spectral parameters evolution obtained by modelling the archival data from \textit{RXTE} observations of \lup{} in 2002 is available at \url{https://vizier.u-strasbg.fr/viz-bin/VizieR-3?-source=J/MNRAS/468/4735}. Optical data from \citet{Bux_Bai2004} is available from us upon request. The
{\sc{freddi}} code is available at 
\url{https://github.com/hombit/freddi/tree/dev}.



\bibliographystyle{mnras}
\bibliography{main.bib} 

\begin{thebibliography}{}
\makeatletter
\relax
\def\mn@urlcharsother{\let\do\@makeother \do\$\do\&\do\#\do\^\do\_\do\%\do\~}
\def\mn@doi{\begingroup\mn@urlcharsother \@ifnextchar [ {\mn@doi@}
  {\mn@doi@[]}}
\def\mn@doi@[#1]#2{\def\@tempa{#1}\ifx\@tempa\@empty \href
  {http://dx.doi.org/#2} {doi:#2}\else \href {http://dx.doi.org/#2} {#1}\fi
  \endgroup}
\def\mn@eprint#1#2{\mn@eprint@#1:#2::\@nil}
\def\mn@eprint@arXiv#1{\href {http://arxiv.org/abs/#1} {{\tt arXiv:#1}}}
\def\mn@eprint@dblp#1{\href {http://dblp.uni-trier.de/rec/bibtex/#1.xml}
  {dblp:#1}}
\def\mn@eprint@#1:#2:#3:#4\@nil{\def\@tempa {#1}\def\@tempb {#2}\def\@tempc
  {#3}\ifx \@tempc \@empty \let \@tempc \@tempb \let \@tempb \@tempa \fi \ifx
  \@tempb \@empty \def\@tempb {arXiv}\fi \@ifundefined
  {mn@eprint@\@tempb}{\@tempb:\@tempc}{\expandafter \expandafter \csname
  mn@eprint@\@tempb\endcsname \expandafter{\@tempc}}}

\bibitem[\protect\citeauthoryear{{Avakyan}, {Malanchev}  \&
  {Lipunova}}{{Avakyan} et~al.}{2019}]{Avakyan19}
{Avakyan} A.~L.,  {Malanchev} K.~L.,   {Lipunova} G.~V.,  2019, in The
  Multi-Messenger Astronomy: Gamma-Ray Bursts, Search for Electromagnetic
  Counterparts to Neutrino Events and Gravitational Waves. pp 25--31,
  \mn@doi{10.26119/SAO.2019.1.35452}

\bibitem[\protect\citeauthoryear{{Avakyan}, {Lipunova}, {Malanchev}  \&
  {Shakura}}{{Avakyan} et~al.}{2021}]{Avakyan21}
{Avakyan} A.~L.,  {Lipunova} G.~V.,  {Malanchev} K.~L.,   {Shakura} N.~I.,
  2021, \mn@doi [Astronomy Letters] {10.1134/S1063773721050017}, \href
  {https://ui.adsabs.harvard.edu/abs/2021AstL...47..377A} {47, 377}

\bibitem[\protect\citeauthoryear{{Avakyan}, {Neumann}, {Zainab}, {Doroshenko},
  {Wilms}  \& {Santangelo}}{{Avakyan} et~al.}{2023}]{AvakyanLMXBcat}
{Avakyan} A.,  {Neumann} M.,  {Zainab} A.,  {Doroshenko} V.,  {Wilms} J.,
  {Santangelo} A.,  2023, \mn@doi [\aap] {10.1051/0004-6361/202346522}, \href
  {https://ui.adsabs.harvard.edu/abs/2023A&A...675A.199A} {675, A199}

\bibitem[\protect\citeauthoryear{{Balbus} \& {Hawley}}{{Balbus} \&
  {Hawley}}{1991}]{Balbus}
{Balbus} S.~A.,  {Hawley} J.~F.,  1991, \mn@doi [\apj] {10.1086/170270}, \href
  {http://adsabs.harvard.edu/abs/1991ApJ...376..214B} {376, 214}

\bibitem[\protect\citeauthoryear{{Basko} \& {Sunyaev}}{{Basko} \&
  {Sunyaev}}{1973}]{Basko73}
{Basko} M.~M.,  {Sunyaev} R.~A.,  1973, \mn@doi [\apss] {10.1007/BF00647654},
  \href {https://ui.adsabs.harvard.edu/abs/1973Ap&SS..23..117B} {23, 117}

\bibitem[\protect\citeauthoryear{{Basko}, {Sunyaev}  \& {Titarchuk}}{{Basko}
  et~al.}{1974}]{Basko+1974}
{Basko} M.~M.,  {Sunyaev} R.~A.,   {Titarchuk} L.~G.,  1974, \aap, \href
  {https://ui.adsabs.harvard.edu/abs/1974A&A....31..249B} {31, 249}

\bibitem[\protect\citeauthoryear{{Begelman}, {McKee}  \& {Shields}}{{Begelman}
  et~al.}{1983}]{Begelman1983}
{Begelman} M.~C.,  {McKee} C.~F.,   {Shields} G.~A.,  1983, \mn@doi [\apj]
  {10.1086/161178}, \href {http://adsabs.harvard.edu/abs/1983ApJ...271...70B}
  {271, 70}

\bibitem[\protect\citeauthoryear{{Blandford} \& {Payne}}{{Blandford} \&
  {Payne}}{1982}]{MagnetOrigin}
{Blandford} R.~D.,  {Payne} D.~G.,  1982, \mn@doi [\mnras]
  {10.1093/mnras/199.4.883}, \href
  {https://ui.adsabs.harvard.edu/abs/1982MNRAS.199..883B} {199, 883}

\bibitem[\protect\citeauthoryear{{Buxton} \& {Bailyn}}{{Buxton} \&
  {Bailyn}}{2004}]{Bux_Bai2004}
{Buxton} M.~M.,  {Bailyn} C.~D.,  2004, \mn@doi [\apj] {10.1086/424503}, \href
  {https://ui.adsabs.harvard.edu/abs/2004ApJ...615..880B} {615, 880}

\bibitem[\protect\citeauthoryear{{Casares}, {Mu{\~n}oz-Darias}, {Mata
  S{\'a}nchez}, {Charles}, {Torres}, {Armas Padilla}, {Fender}  \&
  {Garc{\'\i}a-Rojas}}{{Casares} et~al.}{2019}]{Casares}
{Casares} J.,  {Mu{\~n}oz-Darias} T.,  {Mata S{\'a}nchez} D.,  {Charles} P.~A.,
   {Torres} M.~A.~P.,  {Armas Padilla} M.,  {Fender} R.~P.,
  {Garc{\'\i}a-Rojas} J.,  2019, \mn@doi [\mnras] {10.1093/mnras/stz1793},
  \href {https://ui.adsabs.harvard.edu/abs/2019MNRAS.488.1356C} {488, 1356}

\bibitem[\protect\citeauthoryear{{Charles}, {Matthews}, {Buckley}, {Gandhi},
  {Kotze}  \& {Paice}}{{Charles} et~al.}{2019}]{Charles}
{Charles} P.,  {Matthews} J.~H.,  {Buckley} D. A.~H.,  {Gandhi} P.,  {Kotze}
  E.,   {Paice} J.,  2019, \mn@doi [\mnras] {10.1093/mnrasl/slz120}, \href
  {https://ui.adsabs.harvard.edu/abs/2019MNRAS.489L..47C} {489, L47}

\bibitem[\protect\citeauthoryear{Cherepaschuk}{Cherepaschuk}{2013}]{cherep}
Cherepaschuk A.~M.,  2013, {Close Binary Stars, II}.
MOSCWA FIZMATLIT

\bibitem[\protect\citeauthoryear{{Chevalier} \& {Ilovaisky}}{{Chevalier} \&
  {Ilovaisky}}{1992}]{A2V}
{Chevalier} C.,  {Ilovaisky} S.~A.,  1992, \iaucirc, \href
  {https://ui.adsabs.harvard.edu/abs/1992IAUC.5520....1C} {5520, 1}

\bibitem[\protect\citeauthoryear{{D{\'\i}az Trigo} \& {Boirin}}{{D{\'\i}az
  Trigo} \& {Boirin}}{2016}]{Trigo}
{D{\'\i}az Trigo} M.,  {Boirin} L.,  2016, \mn@doi [Astronomische Nachrichten]
  {10.1002/asna.201612315}, \href
  {https://ui.adsabs.harvard.edu/abs/2016AN....337..368D} {337, 368}

\bibitem[\protect\citeauthoryear{{Done}, {Tomaru}  \& {Takahashi}}{{Done}
  et~al.}{2018}]{Done18}
{Done} C.,  {Tomaru} R.,   {Takahashi} T.,  2018, \mn@doi [\mnras]
  {10.1093/mnras/stx2400}, \href
  {https://ui.adsabs.harvard.edu/abs/2018MNRAS.473..838D} {473, 838}

\bibitem[\protect\citeauthoryear{{Dubus}, {Lasota}, {Hameury}  \&
  {Charles}}{{Dubus} et~al.}{1999}]{dubus_et1999}
{Dubus} G.,  {Lasota} J.-P.,  {Hameury} J.-M.,   {Charles} P.,  1999, \mnras,
  \href
  {http://adsabs.harvard.edu/cgi-bin/nph-bib_query?bibcode=1999MNRAS.303..139D&db_k
  ey=AST} {303, 139}

\bibitem[\protect\citeauthoryear{{Dubus}, {Hameury}  \& {Lasota}}{{Dubus}
  et~al.}{2001}]{dubus_et2001}
{Dubus} G.,  {Hameury} J.-M.,   {Lasota} J.-P.,  2001, \mn@doi [\aap]
  {10.1051/0004-6361:20010632}, \href
  {http://adsabs.harvard.edu/cgi-bin/nph-bib_query?bibcode=2001A\%26A...373..251D&db_key=AST}
  {373, 251}

\bibitem[\protect\citeauthoryear{{Dubus}, {Done}, {Tetarenko}  \&
  {Hameury}}{{Dubus} et~al.}{2019}]{Dubus_19}
{Dubus} G.,  {Done} C.,  {Tetarenko} B.~E.,   {Hameury} J.-M.,  2019, \mn@doi
  [\aap] {10.1051/0004-6361/201936333}, \href
  {https://ui.adsabs.harvard.edu/abs/2019A&A...632A..40D} {632, A40}

\bibitem[\protect\citeauthoryear{{Feldmeier} \& {Shlosman}}{{Feldmeier} \&
  {Shlosman}}{1999}]{Feld99v2}
{Feldmeier} A.,  {Shlosman} I.,  1999, \mn@doi [\apj] {10.1086/307976}, \href
  {https://ui.adsabs.harvard.edu/abs/1999ApJ...526..344F} {526, 344}

\bibitem[\protect\citeauthoryear{{Fijma}, {Castro Segura}, {Degenaar},
  {Knigge}, {Higginbottom}, {Hern{\'a}ndez Santisteban}  \&
  {Maccarone}}{{Fijma} et~al.}{2023}]{2023MNRAS.526L.149F}
{Fijma} S.,  {Castro Segura} N.,  {Degenaar} N.,  {Knigge} C.,  {Higginbottom}
  N.,  {Hern{\'a}ndez Santisteban} J.~V.,   {Maccarone} T.~J.,  2023, \mn@doi
  [\mnras] {10.1093/mnrasl/slad125}, \href
  {https://ui.adsabs.harvard.edu/abs/2023MNRAS.526L.149F} {526, L149}

\bibitem[\protect\citeauthoryear{{Fortin}, {Garc{\'\i}a}, {Simaz Bunzel}  \&
  {Chaty}}{{Fortin} et~al.}{2023}]{FotinHMXBcat}
{Fortin} F.,  {Garc{\'\i}a} F.,  {Simaz Bunzel} A.,   {Chaty} S.,  2023,
  \mn@doi [\aap] {10.1051/0004-6361/202245236}, \href
  {https://ui.adsabs.harvard.edu/abs/2023A&A...671A.149F} {671, A149}

\bibitem[\protect\citeauthoryear{{Fukumura}, {Kazanas}, {Contopoulos}  \&
  {Behar}}{{Fukumura} et~al.}{2010}]{2010ApJ...715..636F}
{Fukumura} K.,  {Kazanas} D.,  {Contopoulos} I.,   {Behar} E.,  2010, \mn@doi
  [\apj] {10.1088/0004-637X/715/1/636}, \href
  {https://ui.adsabs.harvard.edu/abs/2010ApJ...715..636F} {715, 636}

\bibitem[\protect\citeauthoryear{{Fukumura}, {Kazanas}, {Shrader}, {Behar},
  {Tombesi}  \& {Contopoulos}}{{Fukumura} et~al.}{2017}]{2017NatAs...1E..62F}
{Fukumura} K.,  {Kazanas} D.,  {Shrader} C.,  {Behar} E.,  {Tombesi} F.,
  {Contopoulos} I.,  2017, \mn@doi [Nature Astronomy]
  {10.1038/s41550-017-0062}, \href
  {https://ui.adsabs.harvard.edu/abs/2017NatAs...1E..62F} {1, 0062}

\bibitem[\protect\citeauthoryear{{Gandhi}, {Rao}, {Johnson}, {Paice}  \&
  {Maccarone}}{{Gandhi} et~al.}{2019}]{Gandhi19}
{Gandhi} P.,  {Rao} A.,  {Johnson} M. A.~C.,  {Paice} J.~A.,   {Maccarone}
  T.~J.,  2019, \mn@doi [\mnras] {10.1093/mnras/stz438}, \href
  {https://ui.adsabs.harvard.edu/abs/2019MNRAS.485.2642G} {485, 2642}

\bibitem[\protect\citeauthoryear{{Harmon}, {Wilson}, {Finger}, {Paciesas},
  {Rubin}  \& {Fishman}}{{Harmon} et~al.}{1992}]{Harmon92}
{Harmon} B.~A.,  {Wilson} R.~B.,  {Finger} M.~H.,  {Paciesas} W.~S.,  {Rubin}
  B.~C.,   {Fishman} G.~J.,  1992, \iaucirc, \href
  {https://ui.adsabs.harvard.edu/abs/1992IAUC.5504....1H} {5504, 1}

\bibitem[\protect\citeauthoryear{{Higginbottom} \& {Proga}}{{Higginbottom} \&
  {Proga}}{2015}]{HigProg}
{Higginbottom} N.,  {Proga} D.,  2015, \mn@doi [\apj]
  {10.1088/0004-637X/807/1/107}, \href
  {https://ui.adsabs.harvard.edu/abs/2015ApJ...807..107H} {807, 107}

\bibitem[\protect\citeauthoryear{{Higginbottom}, {Proga}, {Knigge}  \&
  {Long}}{{Higginbottom} et~al.}{2017}]{Higgin17}
{Higginbottom} N.,  {Proga} D.,  {Knigge} C.,   {Long} K.~S.,  2017, \mn@doi
  [\apj] {10.3847/1538-4357/836/1/42}, \href
  {https://ui.adsabs.harvard.edu/abs/2017ApJ...836...42H} {836, 42}

\bibitem[\protect\citeauthoryear{{Higginbottom}, {Knigge}, {Long}, {Matthews}
  \& {Parkinson}}{{Higginbottom} et~al.}{2019}]{Higgin19}
{Higginbottom} N.,  {Knigge} C.,  {Long} K.~S.,  {Matthews} J.~H.,
  {Parkinson} E.~J.,  2019, \mn@doi [\mnras] {10.1093/mnras/stz310}, \href
  {https://ui.adsabs.harvard.edu/abs/2019MNRAS.484.4635H} {484, 4635}

\bibitem[\protect\citeauthoryear{{Hirose}, {Blaes}, {Krolik}, {Coleman}  \&
  {Sano}}{{Hirose} et~al.}{2014}]{Hirose}
{Hirose} S.,  {Blaes} O.,  {Krolik} J.~H.,  {Coleman} M.~S.~B.,   {Sano} T.,
  2014, \mn@doi [\apj] {10.1088/0004-637X/787/1/1}, \href
  {http://adsabs.harvard.edu/abs/2014ApJ...787....1H} {787, 1}

\bibitem[\protect\citeauthoryear{{Iglesias} \& {Rogers}}{{Iglesias} \&
  {Rogers}}{1996}]{Iglesias-Rogers1996}
{Iglesias} C.~A.,  {Rogers} F.~J.,  1996, \mn@doi [\apj] {10.1086/177381},
  \href {https://ui.adsabs.harvard.edu/abs/1996ApJ...464..943I} {464, 943}

\bibitem[\protect\citeauthoryear{{Jahoda}, {Swank}, {Giles}, {Stark},
  {Strohmayer}, {Zhang}  \& {Morgan}}{{Jahoda} et~al.}{1996}]{RXTE}
{Jahoda} K.,  {Swank} J.~H.,  {Giles} A.~B.,  {Stark} M.~J.,  {Strohmayer} T.,
  {Zhang} W.,   {Morgan} E.~H.,  1996, in {Siegmund} O.~H.,  {Gummin} M.~A.,
  eds,  Society of Photo-Optical Instrumentation Engineers (SPIE) Conference
  Series Vol. 2808, EUV, X-Ray, and Gamma-Ray Instrumentation for Astronomy
  VII. pp 59--70, \mn@doi{10.1117/12.256034}

\bibitem[\protect\citeauthoryear{{Jimenez-Garate}, {Raymond}  \&
  {Liedahl}}{{Jimenez-Garate} et~al.}{2002}]{Garate}
{Jimenez-Garate} M.~A.,  {Raymond} J.~C.,   {Liedahl} D.~A.,  2002, \mn@doi
  [\apj] {10.1086/344364}, \href
  {http://adsabs.harvard.edu/abs/2002ApJ...581.1297J} {581, 1297}

\bibitem[\protect\citeauthoryear{{Kalitkin}}{{Kalitkin}}{1978}]{Kalitkin}
{Kalitkin} N.~N.,  1978, {Numerical methods}.
Nauka, Moskwa

\bibitem[\protect\citeauthoryear{{King} \& {Ritter}}{{King} \&
  {Ritter}}{1998}]{KingRit}
{King} A.~R.,  {Ritter} H.,  1998, \mn@doi [\mnras]
  {10.1046/j.1365-8711.1998.01295.x}, \href
  {http://adsabs.harvard.edu/abs/1998MNRAS.293L..42K} {293, L42}

\bibitem[\protect\citeauthoryear{{Kitamoto}, {Miyamoto}, {Tsunemi}, {Makishima}
   \& {Nakagawa}}{{Kitamoto} et~al.}{1984}]{Kitam84}
{Kitamoto} S.,  {Miyamoto} S.,  {Tsunemi} H.,  {Makishima} K.,   {Nakagawa} M.,
   1984, \pasj, \href {https://ui.adsabs.harvard.edu/abs/1984PASJ...36..799K}
  {36, 799}

\bibitem[\protect\citeauthoryear{{Kosec}, {Fabian}, {Pinto}, {Walton}, {Dyda}
  \& {Reynolds}}{{Kosec} et~al.}{2020}]{Kosec}
{Kosec} P.,  {Fabian} A.~C.,  {Pinto} C.,  {Walton} D.~J.,  {Dyda} S.,
  {Reynolds} C.~S.,  2020, \mn@doi [\mnras] {10.1093/mnras/stz3200}, \href
  {https://ui.adsabs.harvard.edu/abs/2020MNRAS.491.3730K} {491, 3730}

\bibitem[\protect\citeauthoryear{{Kotko} \& {Lasota}}{{Kotko} \&
  {Lasota}}{2012}]{Kotko-Lasota2012}
{Kotko} I.,  {Lasota} J.~P.,  2012, \mn@doi [\aap]
  {10.1051/0004-6361/201219618}, \href
  {https://ui.adsabs.harvard.edu/abs/2012A&A...545A.115K} {545, A115}

\bibitem[\protect\citeauthoryear{{Lasota}}{{Lasota}}{2001}]{lasota2001}
{Lasota} J.-P.,  2001, New Astronomy Review, \href
  {http://adsabs.harvard.edu/cgi-bin/nph-bib_query?bibcode=2001NewAR..45..449L&db_k
  ey=AST} {45, 449}

\bibitem[\protect\citeauthoryear{{Lasota}}{{Lasota}}{2016}]{DIM}
{Lasota} J.-P.,  2016, {Black Hole Accretion Discs}.
Springer, p.~1, \mn@doi{10.1007/978-3-662-52859-4_1}

\bibitem[\protect\citeauthoryear{{Lasota}, {Dubus}  \& {Kruk}}{{Lasota}
  et~al.}{2008}]{Lasota_etal2008}
{Lasota} J.~P.,  {Dubus} G.,   {Kruk} K.,  2008, \mn@doi [\aap]
  {10.1051/0004-6361:200809658}, \href
  {https://ui.adsabs.harvard.edu/abs/2008A&A...486..523L} {486, 523}

\bibitem[\protect\citeauthoryear{{Lipunova}}{{Lipunova}}{2015}]{lipunova2015}
{Lipunova} G.~V.,  2015, \mn@doi [\apj] {10.1088/0004-637X/804/2/87}, \href
  {http://adsabs.harvard.edu/abs/2015ApJ...804...87L} {804, 87}

\bibitem[\protect\citeauthoryear{{Lipunova} \& {Malanchev}}{{Lipunova} \&
  {Malanchev}}{2017}]{lipunova_malanchev2017}
{Lipunova} G.~V.,  {Malanchev} K.~L.,  2017, \mn@doi [\mnras]
  {10.1093/mnras/stx768}, \href
  {http://adsabs.harvard.edu/abs/2017MNRAS.468.4735L} {468, 4735}

\bibitem[\protect\citeauthoryear{{Lipunova} \& {Shakura}}{{Lipunova} \&
  {Shakura}}{2000}]{Lipunova-Shakura2000}
{Lipunova} G.~V.,  {Shakura} N.~I.,  2000, \mn@doi [\aap]
  {10.48550/arXiv.astro-ph/0103274}, \href
  {https://ui.adsabs.harvard.edu/abs/2000A&A...356..363L} {356, 363}

\bibitem[\protect\citeauthoryear{{Lipunova} \& {Shakura}}{{Lipunova} \&
  {Shakura}}{2002}]{Lipunova-Shakura2002}
{Lipunova} G.~V.,  {Shakura} N.~I.,  2002, \mn@doi [Astronomy Reports]
  {10.1134/1.1479424}, \href
  {https://ui.adsabs.harvard.edu/abs/2002ARep...46..366L} {46, 366}

\bibitem[\protect\citeauthoryear{{Lipunova}, {Malanchev}, {Tsygankov},
  {Shakura}, {Tavleev}  \& {Kolesnikov}}{{Lipunova}
  et~al.}{2022}]{Lipunova+2022}
{Lipunova} G.,  {Malanchev} K.,  {Tsygankov} S.,  {Shakura} N.,  {Tavleev} A.,
   {Kolesnikov} D.,  2022, \mn@doi [\mnras] {10.1093/mnras/stab3343}, \href
  {https://ui.adsabs.harvard.edu/abs/2022MNRAS.510.1837L} {510, 1837}

\bibitem[\protect\citeauthoryear{{London}, {McCray}  \& {Auer}}{{London}
  et~al.}{1981}]{London}
{London} R.,  {McCray} R.,   {Auer} L.~H.,  1981, \mn@doi [\apj]
  {10.1086/158661}, \href
  {https://ui.adsabs.harvard.edu/abs/1981ApJ...243..970L} {243, 970}

\bibitem[\protect\citeauthoryear{{Ludwig}, {Meyer-Hofmeister}  \&
  {Ritter}}{{Ludwig} et~al.}{1994}]{Ludwig+1994}
{Ludwig} K.,  {Meyer-Hofmeister} E.,   {Ritter} H.,  1994, \aap, \href
  {https://ui.adsabs.harvard.edu/abs/1994A&A...290..473L} {290, 473}

\bibitem[\protect\citeauthoryear{{Luketic}, {Proga}, {Kallman}, {Raymond}  \&
  {Miller}}{{Luketic} et~al.}{2010}]{Luketic}
{Luketic} S.,  {Proga} D.,  {Kallman} T.~R.,  {Raymond} J.~C.,   {Miller}
  J.~M.,  2010, \mn@doi [\apj] {10.1088/0004-637X/719/1/515}, \href
  {https://ui.adsabs.harvard.edu/abs/2010ApJ...719..515L} {719, 515}

\bibitem[\protect\citeauthoryear{{Lyubarskij} \& {Shakura}}{{Lyubarskij} \&
  {Shakura}}{1987}]{Lyubarskij_Shakura}
{Lyubarskij} Y.~E.,  {Shakura} N.~I.,  1987, Soviet Astronomy Letters, \href
  {https://ui.adsabs.harvard.edu/abs/1987SvAL...13..386L} {13, 386}

\bibitem[\protect\citeauthoryear{{Malanchev} \& {Lipunova}}{{Malanchev} \&
  {Lipunova}}{2016}]{SOFT}
{Malanchev} K.~L.,  {Lipunova} G.~V.,  2016, {Freddi: Fast Rise Exponential
  Decay accretion Disk model Implementation}, Astrophysics Source Code Library
  (\mn@eprint {ascl} {1610.014})

\bibitem[\protect\citeauthoryear{{Malanchev} \& {Shakura}}{{Malanchev} \&
  {Shakura}}{2015}]{malanch_shak15}
{Malanchev} K.~L.,  {Shakura} N.~I.,  2015, \mn@doi [Astronomy Letters]
  {10.1134/S1063773715120087}, \href
  {https://ui.adsabs.harvard.edu/abs/2015AstL...41..797M} {41, 797}

\bibitem[\protect\citeauthoryear{{Martin}, {Nixon}, {Pringle}  \&
  {Livio}}{{Martin} et~al.}{2019}]{VISCOS2019}
{Martin} R.~G.,  {Nixon} C.~J.,  {Pringle} J.~E.,   {Livio} M.,  2019, \mn@doi
  [\na] {10.1016/j.newast.2019.01.001}, \href
  {http://adsabs.harvard.edu/abs/2019NewA...70....7M} {70, 7}

\bibitem[\protect\citeauthoryear{{McClintock} \& {Remillard}}{{McClintock} \&
  {Remillard}}{2006}]{2006csxs.book..157M}
{McClintock} J.~E.,  {Remillard} R.~A.,  2006, in , Vol.~39, Compact stellar
  X-ray sources.
pp 157--213, \mn@doi{10.48550/arXiv.astro-ph/0306213}

\bibitem[\protect\citeauthoryear{{Meyer} \& {Meyer-Hofmeister}}{{Meyer} \&
  {Meyer-Hofmeister}}{1984}]{MeyerHofm}
{Meyer} F.,  {Meyer-Hofmeister} E.,  1984, \aap, \href
  {http://adsabs.harvard.edu/abs/1984A\%26A...132..143M} {132, 143}

\bibitem[\protect\citeauthoryear{{Middleton}, {Higginbottom}, {Knigge}, {Khan}
  \& {Wiktorowicz}}{{Middleton} et~al.}{2022}]{Middleton+2022}
{Middleton} M.~J.,  {Higginbottom} N.,  {Knigge} C.,  {Khan} N.,
  {Wiktorowicz} G.,  2022, \mn@doi [\mnras] {10.1093/mnras/stab2991}, \href
  {https://ui.adsabs.harvard.edu/abs/2022MNRAS.509.1119M} {509, 1119}

\bibitem[\protect\citeauthoryear{{Miller} \& {Remillard}}{{Miller} \&
  {Remillard}}{2002}]{Miller02}
{Miller} J.~M.,  {Remillard} R.~A.,  2002, The Astronomer's Telegram, \href
  {https://ui.adsabs.harvard.edu/abs/2002ATel...98....1M} {98, 1}

\bibitem[\protect\citeauthoryear{{Miller} et~al.,}{{Miller}
  et~al.}{2016}]{2016ApJ...821L...9M}
{Miller} J.~M.,  et~al., 2016, \mn@doi [\apjl] {10.3847/2041-8205/821/1/L9},
  \href {https://ui.adsabs.harvard.edu/abs/2016ApJ...821L...9M} {821, L9}

\bibitem[\protect\citeauthoryear{{Mi{\v{s}}kovi{\v{c}}ov{\'a}}
  et~al.,}{{Mi{\v{s}}kovi{\v{c}}ov{\'a}} et~al.}{2016}]{2016A&A...590A.114M}
{Mi{\v{s}}kovi{\v{c}}ov{\'a}} I.,  et~al., 2016, \mn@doi [\aap]
  {10.1051/0004-6361/201322490}, \href
  {https://ui.adsabs.harvard.edu/abs/2016A&A...590A.114M} {590, A114}

\bibitem[\protect\citeauthoryear{{Morningstar} \& {Miller}}{{Morningstar} \&
  {Miller}}{2014}]{Spin}
{Morningstar} W.~R.,  {Miller} J.~M.,  2014, \mn@doi [\apjl]
  {10.1088/2041-8205/793/2/L33}, \href
  {https://ui.adsabs.harvard.edu/abs/2014ApJ...793L..33M} {793, L33}

\bibitem[\protect\citeauthoryear{{Mu{\~n}oz-Darias} et~al.,}{{Mu{\~n}oz-Darias}
  et~al.}{2016}]{2016Natur.534...75M}
{Mu{\~n}oz-Darias} T.,  et~al., 2016, \mn@doi [\nat] {10.1038/nature17446},
  \href {https://ui.adsabs.harvard.edu/abs/2016Natur.534...75M} {534, 75}

\bibitem[\protect\citeauthoryear{{Mu{\~n}oz-Darias} et~al.,}{{Mu{\~n}oz-Darias}
  et~al.}{2019}]{Munoz}
{Mu{\~n}oz-Darias} T.,  et~al., 2019, \mn@doi [\apjl]
  {10.3847/2041-8213/ab2768}, \href
  {https://ui.adsabs.harvard.edu/abs/2019ApJ...879L...4M} {879, L4}

\bibitem[\protect\citeauthoryear{{Murray}, {Chiang}, {Grossman}  \&
  {Voit}}{{Murray} et~al.}{1995}]{Murray95}
{Murray} N.,  {Chiang} J.,  {Grossman} S.~A.,   {Voit} G.~M.,  1995, \mn@doi
  [\apj] {10.1086/176238}, \href
  {https://ui.adsabs.harvard.edu/abs/1995ApJ...451..498M} {451, 498}

\bibitem[\protect\citeauthoryear{{Negoro} et~al.,}{{Negoro}
  et~al.}{2021}]{Negoro+2021}
{Negoro} H.,  et~al., 2021, The Astronomer's Telegram, \href
  {https://ui.adsabs.harvard.edu/abs/2021ATel14701....1N} {14701, 1}

\bibitem[\protect\citeauthoryear{{Neumann}, {Avakyan}, {Doroshenko}  \&
  {Santangelo}}{{Neumann} et~al.}{2023}]{NeumannHMXBcat}
{Neumann} M.,  {Avakyan} A.,  {Doroshenko} V.,   {Santangelo} A.,  2023,
  \mn@doi [\aap] {10.1051/0004-6361/202245728}, \href
  {https://ui.adsabs.harvard.edu/abs/2023A&A...677A.134N} {677, A134}

\bibitem[\protect\citeauthoryear{Newville et~al.,}{Newville
  et~al.}{2023}]{lmfit}
Newville M.,  et~al., 2023, lmfit/lmfit-py: 1.2.2,
  \mn@doi{10.5281/zenodo.8145703}, \url
  {https://doi.org/10.5281/zenodo.8145703}

\bibitem[\protect\citeauthoryear{{Orosz}}{{Orosz}}{2003}]{Orosz2003}
{Orosz} J.~A.,  2003, in {van der Hucht} K.,  {Herrero} A.,   {Esteban} C.,
  eds,  Proc. IAU Symp Vol. 212, A Massive Star Odyssey: From Main Sequence to
  Supernova. p.~365 (\mn@eprint {arXiv} {astro-ph/0209041})

\bibitem[\protect\citeauthoryear{{Orosz}, {Jain}, {Bailyn}, {McClintock}  \&
  {Remillard}}{{Orosz} et~al.}{1998}]{Orosz98}
{Orosz} J.~A.,  {Jain} R.~K.,  {Bailyn} C.~D.,  {McClintock} J.~E.,
  {Remillard} R.~A.,  1998, \mn@doi [\apj] {10.1086/305620}, \href
  {https://ui.adsabs.harvard.edu/abs/1998ApJ...499..375O} {499, 375}

\bibitem[\protect\citeauthoryear{{Orosz}, {Polisensky}, {Bailyn},
  {Tourtellotte}, {McClintock}  \& {Remillard}}{{Orosz}
  et~al.}{2002}]{Orosz2002}
{Orosz} J.~A.,  {Polisensky} E.~J.,  {Bailyn} C.~D.,  {Tourtellotte} S.~W.,
  {McClintock} J.~E.,   {Remillard} R.~A.,  2002, in American Astronomical
  Society Meeting Abstracts. p. 15.11

\bibitem[\protect\citeauthoryear{{Park} et~al.,}{{Park}
  et~al.}{2004}]{2004ApJ...610..378P}
{Park} S.~Q.,  et~al., 2004, \mn@doi [\apj] {10.1086/421511}, \href
  {https://ui.adsabs.harvard.edu/abs/2004ApJ...610..378P} {610, 378}

\bibitem[\protect\citeauthoryear{{Pelletier} \& {Pudritz}}{{Pelletier} \&
  {Pudritz}}{1992}]{Pelletier92}
{Pelletier} G.,  {Pudritz} R.~E.,  1992, \mn@doi [\apj] {10.1086/171565}, \href
  {https://ui.adsabs.harvard.edu/abs/1992ApJ...394..117P} {394, 117}

\bibitem[\protect\citeauthoryear{{Ponti}, {Fender}, {Begelman}, {Dunn},
  {Neilsen}  \& {Coriat}}{{Ponti} et~al.}{2012}]{Ponti12}
{Ponti} G.,  {Fender} R.~P.,  {Begelman} M.~C.,  {Dunn} R.~J.~H.,  {Neilsen}
  J.,   {Coriat} M.,  2012, \mn@doi [\mnras]
  {10.1111/j.1745-3933.2012.01224.x}, \href
  {https://ui.adsabs.harvard.edu/abs/2012MNRAS.422L..11P} {422, L11}

\bibitem[\protect\citeauthoryear{{Proga} \& {Kallman}}{{Proga} \&
  {Kallman}}{2002}]{Proga2002}
{Proga} D.,  {Kallman} T.~R.,  2002, \mn@doi [\apj] {10.1086/324534}, \href
  {https://ui.adsabs.harvard.edu/abs/2002ApJ...565..455P} {565, 455}

\bibitem[\protect\citeauthoryear{{Proga} \& {Kallman}}{{Proga} \&
  {Kallman}}{2004}]{ProgaKall04}
{Proga} D.,  {Kallman} T.~R.,  2004, \mn@doi [\apj] {10.1086/425117}, \href
  {https://ui.adsabs.harvard.edu/abs/2004ApJ...616..688P} {616, 688}

\bibitem[\protect\citeauthoryear{{Russell}, {Casella}, {Kalemci}, {Vahdat
  Motlagh}, {Saikia}, {Pirbhoy}  \& {Maitra}}{{Russell}
  et~al.}{2020}]{Russel_jet}
{Russell} D.~M.,  {Casella} P.,  {Kalemci} E.,  {Vahdat Motlagh} A.,  {Saikia}
  P.,  {Pirbhoy} S.~F.,   {Maitra} D.,  2020, \mn@doi [\mnras]
  {10.1093/mnras/staa1182}, \href
  {https://ui.adsabs.harvard.edu/abs/2020MNRAS.495..182R} {495, 182}

\bibitem[\protect\citeauthoryear{{Scepi}, {Begelman}  \& {Dexter}}{{Scepi}
  et~al.}{2023}]{2023arXiv230210226S}
{Scepi} N.,  {Begelman} M.~C.,   {Dexter} J.,  2023, \mn@doi [arXiv e-prints]
  {10.48550/arXiv.2302.10226}, \href
  {https://ui.adsabs.harvard.edu/abs/2023arXiv230210226S} {p. arXiv:2302.10226}

\bibitem[\protect\citeauthoryear{{Shakura}}{{Shakura}}{1972}]{Shakura72}
{Shakura} N.~I.,  1972, \azh, \href
  {https://ui.adsabs.harvard.edu/abs/1972AZh....49..921S} {49, 921}

\bibitem[\protect\citeauthoryear{{Shakura} \& {Sunyaev}}{{Shakura} \&
  {Sunyaev}}{1973}]{SHAKURA_SUNAEV}
{Shakura} N.~I.,  {Sunyaev} R.~A.,  1973, \aap, \href
  {https://ui.adsabs.harvard.edu/abs/1973A&A....24..337S} {500, 33}

\bibitem[\protect\citeauthoryear{{Shakura} et~al.,}{{Shakura}
  et~al.}{2018}]{Shakura_book}
{Shakura} N.,  et~al., 2018, {Accretion Flows in Astrophysics}.
 Astrophysics and Space Science Library Vol. 454, {Springer},
  \mn@doi{10.1007/978-3-319-93009-1}

\bibitem[\protect\citeauthoryear{{Shields}, {McKee}, {Lin}  \&
  {Begelman}}{{Shields} et~al.}{1986}]{Shields}
{Shields} G.~A.,  {McKee} C.~F.,  {Lin} D.~N.~C.,   {Begelman} M.~C.,  1986,
  \mn@doi [\apj] {10.1086/164322}, \href
  {http://adsabs.harvard.edu/abs/1986ApJ...306...90S} {306, 90}

\bibitem[\protect\citeauthoryear{{Shlosman}, {Vitello}  \& {Shaviv}}{{Shlosman}
  et~al.}{1985}]{Shlosman85}
{Shlosman} I.,  {Vitello} P.~A.,   {Shaviv} G.,  1985, \mn@doi [\apj]
  {10.1086/163278}, \href
  {https://ui.adsabs.harvard.edu/abs/1985ApJ...294...96S} {294, 96}

\bibitem[\protect\citeauthoryear{{Smak}}{{Smak}}{1984}]{Smak1984_IV}
{Smak} J.,  1984, \actaa, \href
  {https://ui.adsabs.harvard.edu/abs/1984AcA....34..161S} {34, 161}

\bibitem[\protect\citeauthoryear{{Suleimanov}}{{Suleimanov}}{1995}]{Suleimanov95}
{Suleimanov} V.~F.,  1995, Astronomy Letters, \href
  {https://ui.adsabs.harvard.edu/abs/1995AstL...21..126S} {21, 126}

\bibitem[\protect\citeauthoryear{{Suleimanov}, {Lipunova}  \&
  {Shakura}}{{Suleimanov} et~al.}{2007}]{suleimanov_et2007e}
{Suleimanov} V.~F.,  {Lipunova} G.~V.,   {Shakura} N.~I.,  2007, \mn@doi
  [Astronomy Reports] {10.1134/S1063772907070049}, \href
  {http://adsabs.harvard.edu/abs/2007ARep...51..549S} {51, 549}

\bibitem[\protect\citeauthoryear{{Suleimanov}, {Lipunova}  \&
  {Shakura}}{{Suleimanov} et~al.}{2008a}]{suleimanov_etal2008}
{Suleimanov} V.~F.,  {Lipunova} G.~V.,   {Shakura} N.~I.,  2008a, \mn@doi
  [\aap] {10.1051/0004-6361:200810155}, \href
  {http://adsabs.harvard.edu/abs/2008A%26A...491..267S} {491, 267}

\bibitem[\protect\citeauthoryear{{Suleimanov}, {Lipunova}  \&
  {Shakura}}{{Suleimanov} et~al.}{2008b}]{Suleimanov+2008}
{Suleimanov} V.~F.,  {Lipunova} G.~V.,   {Shakura} N.~I.,  2008b, \mn@doi
  [\aap] {10.1051/0004-6361:200810155}, \href
  {https://ui.adsabs.harvard.edu/abs/2008A&A...491..267S} {491, 267}

\bibitem[\protect\citeauthoryear{{Tavleev}, {Lipunova}  \&
  {Malanchev}}{{Tavleev} et~al.}{2023}]{Tavleev23}
{Tavleev} A.~S.,  {Lipunova} G.~V.,   {Malanchev} K.~L.,  2023, \mn@doi
  [\mnras] {10.1093/mnras/stad1881}, \href
  {https://ui.adsabs.harvard.edu/abs/2023MNRAS.524.3647T} {524, 3647}

\bibitem[\protect\citeauthoryear{{Tetarenko}, {Sivakoff}, {Heinke}  \&
  {Gladstone}}{{Tetarenko} et~al.}{2016}]{WATCH}
{Tetarenko} B.~E.,  {Sivakoff} G.~R.,  {Heinke} C.~O.,   {Gladstone} J.~C.,
  2016, VizieR Online Data Catalog, \href
  {http://adsabs.harvard.edu/abs/2016yCat..22220015T} {222}

\bibitem[\protect\citeauthoryear{{Tetarenko}, {Dubus}, {Lasota}, {Heinke}  \&
  {Sivakoff}}{{Tetarenko} et~al.}{2018a}]{Tetarenko18}
{Tetarenko} B.~E.,  {Dubus} G.,  {Lasota} J.~P.,  {Heinke} C.~O.,   {Sivakoff}
  G.~R.,  2018a, \mn@doi [\mnras] {10.1093/mnras/sty1798}, \href
  {https://ui.adsabs.harvard.edu/abs/2018MNRAS.480....2T} {480, 2}

\bibitem[\protect\citeauthoryear{{Tetarenko}, {Lasota}, {Heinke}, {Dubus}  \&
  {Sivakoff}}{{Tetarenko} et~al.}{2018b}]{Tetarenko18_1}
{Tetarenko} B.~E.,  {Lasota} J.~P.,  {Heinke} C.~O.,  {Dubus} G.,   {Sivakoff}
  G.~R.,  2018b, \mn@doi [\nat] {10.1038/nature25159}, \href
  {https://ui.adsabs.harvard.edu/abs/2018Natur.554...69T} {554, 69}

\bibitem[\protect\citeauthoryear{{Tetarenko}, {Dubus}, {Marcel}, {Done}  \&
  {Clavel}}{{Tetarenko} et~al.}{2020}]{Tetarenko20}
{Tetarenko} B.~E.,  {Dubus} G.,  {Marcel} G.,  {Done} C.,   {Clavel} M.,  2020,
  \mn@doi [\mnras] {10.1093/mnras/staa1367}, \href
  {https://ui.adsabs.harvard.edu/abs/2020MNRAS.495.3666T} {495, 3666}

\bibitem[\protect\citeauthoryear{{Tuchman}, {Mineshige}  \&
  {Wheeler}}{{Tuchman} et~al.}{1990}]{tuchman_et1990}
{Tuchman} Y.,  {Mineshige} S.,   {Wheeler} J.~C.,  1990, \mn@doi [\apj]
  {10.1086/169045}, \href {http://adsabs.harvard.edu/abs/1990ApJ...359..164T}
  {359, 164}

\bibitem[\protect\citeauthoryear{{Ueda}, {Yamaoka}  \& {Remillard}}{{Ueda}
  et~al.}{2009}]{Ueda}
{Ueda} Y.,  {Yamaoka} K.,   {Remillard} R.,  2009, \mn@doi [\apj]
  {10.1088/0004-637X/695/2/888}, \href
  {https://ui.adsabs.harvard.edu/abs/2009ApJ...695..888U} {695, 888}

\bibitem[\protect\citeauthoryear{{Woods}, {Klein}, {Castor}, {McKee}  \&
  {Bell}}{{Woods} et~al.}{1996}]{Woods}
{Woods} D.~T.,  {Klein} R.~I.,  {Castor} J.~I.,  {McKee} C.~F.,   {Bell} J.~B.,
   1996, \mn@doi [\apj] {10.1086/177101}, \href
  {http://adsabs.harvard.edu/abs/1996ApJ...461..767W} {461, 767}

\makeatother
\end{thebibliography}



\appendix

\section{Numerical solution of the nonlinear diffusion equation}\label{appendix_num}

\subsection{Formulation of the problem}

It is necessary to construct a method for the numerical solution of equation~\eqref{e1} and study it for stability, taking into account the influence of the wind. Let us consider the problem in a more general form which is equivalent to the task of solving the equation~\eqref{e1} with boundary~\eqref{inner},~\eqref{bound} and initial~\eqref{init} conditions. This problem can be presented as the following system of equations:
\begin{equation}\label{b1}
 \begin{cases}
 \dfrac{\partial Z(U (x, t), x)}{\partial t} = \dfrac{\partial^2 U(x,t)}{\partial x^2} \, + \, V\left(\dfrac{\partial U}{\partial x}, U, x\right) \,, \, x \in [a, b]\,, t \in [0, T]\,;
  \\
  U (a, t)\, = \,0\,;
  \\
  \dfrac{\partial U(x,t)}{\partial x}\Big|_{\rm x = b} \,=\, 0\,;
  \\
  U (x, 0) \,= \,\Phi(x)\,;
 \end{cases}
\end{equation}
where $ T $ is the considered evolution time of the disc, $ V $ and $ Z $ are known functions of their variables, $ U $ is the desired function, $ \Phi (x) $ is the initial condition. The functions $ Z $, $ U $ and $ \Phi $ are strictly positive in the interval $ x \in [a, b] $.

We represent the function $ V $ in the form:
\begin{equation}\label{b2}
V\left(\frac{\partial U}{\partial x}, U, x\right)\, =\, A(x)\,\frac{\partial U(x,t)}{\partial x} \,+\, B(x)\,U(x,t) \,+\, C(x)\,,
\end{equation}
where $ A $, $ B $ and $ C $ are the expansion coefficients, which in a known manner depend on $ x $. Then, taking into account~\eqref{b2}, the first equation in the system~\eqref{b1} will look as follows:
\begin{equation}\label{b3}
\begin{gathered}
\frac{\partial Z(U (x, t), x)}{\partial t} \,=\, \frac{\partial^2 U(x,t)}{\partial x^2}\, + \, A(x\,)\frac{\partial U(x,t)}{\partial x} \, + \\ + \, B(x)\,U(x,t) \, + \,C(x)\,.
\end{gathered}
\end{equation}

Note that the functions $ Z $, $ U $ and $ V $ from the first equation of system~\eqref{b1} correspond to the functions $ \Sigma $, $ F $ and $ - \Windrate $ from equation~\eqref{e1}, respectively.

\subsection{Constructing a solution to equation~\texorpdfstring{\eqref{b3}}{(A3)}}
We introduce a difference scheme by constructing a grid:
\begin{equation}\label{b4}
\begin{gathered}
x_0 < x_1 < ... < x_{\rm n} < ... < x_{\rm N-1} < x_{\rm N} \,, 
\\
h_{\rm n} = x_{\rm n} - x_{\rm n-1},\,n = 1..N\,;
\\
t_0 < t_1 < ... < t_{\rm m} < ... < t_{\rm M-1} < t_{\rm M}\,; 
\\ 
\tau_{\rm m} = t_{\rm m+1} - t_{\rm m},\,m = 0..M\mathrm{-1}\,.
\end{gathered}
\end{equation}

Thus, the initial functions $ Z $, $ U $ and $ \Phi $ are replaced with the corresponding grid ones:
\begin{equation}\label{b5}
\begin{gathered}
Z (U(x_{\rm n}, t_{\rm m}), x_{\rm n}) \rightarrow z_{\rm n}^{\rm m} \,,
\\
U (x_{\rm n}, t_{\rm m}) \rightarrow u_{\rm n}^{\rm m} \,,
\\
\Phi (x_{\rm n}) \rightarrow \phi_{\rm n}\,.
\end{gathered}
\end{equation}

First, we write the difference equations for the boundary and initial conditions. In our case of a BH, the internal boundary condition of the first type is written as:
\begin{equation}\label{b6}
u_{\rm 0}^{\rm m } = 0\,, \ \  \forall m \,.
\end{equation}
In turn, the initial condition is as follows:
\begin{equation}\label{b6_2}
u_{\rm n}^{0} = \phi_{n}\,, \ \  \forall n \,.
\end{equation}

In order to write down an external boundary condition of the second type, we expand 
$ u_{\rm N-1}^{\rm m+1} $ 
in a Taylor series at the point $ x_{\rm N}$:

\begin{equation}\label{b7}
u_{\rm N - 1}^{\rm m+1} = u_{\rm N}^{\rm m+1} \,-\, h_{\rm N}\frac{\partial U}{\partial x}\bigg|_{\rm x_{\rm N}} \,+\, \frac{h_{\rm N}^2}{2}\frac{\partial^2 U}{\partial x^2}\bigg|_{\rm x_{\rm N}} \,+\, o(h_{\rm N}^2)\,.
\end{equation}

If we substitute the value of the second derivative $ u_{\rm n}^{\rm m+1} $ with respect to $ x $, expressed from the initial equation~\eqref{b3}, into the equation~\eqref{b7}, then, in terms of differences up to the second order $ h_{\rm N} $ inclusively, we can write the outer boundary condition (third equation in \ref{b1}) as follows:

\begin{equation}\label{b8}
\begin{gathered}
\frac{u_{\rm N}^{\rm m+1} - u_{\rm N - 1}^{\rm m+1}}{h_{\rm N}}  \,+\, \frac{h_{\rm N}}{2}\frac{z_{\rm N}^{\rm m+1} - z_{\rm N}^{\rm m}}{\tau_{\rm m}} \,-\, \frac{A_{\rm N}}{2}(u_{\rm N}^{\rm m+1} - u_{\rm N - 1}^{\rm m+1}) \,- 
\\
-\,\frac{B_{\rm N}}{2}u_{\rm N}^{\rm m+1}h_{\rm N} \,-\, \frac{C_{\rm N}}{2}h_{\rm N} \,+\,  o(h_{\rm N}^2) \,+\, o(\tau_{\rm m}) \,=\, 0\,.
\end{gathered}
\end{equation}

Here, the time derivative of the function 
$ Z(x_{\rm N}, t) $ is written as:
\begin{equation}\label{b9}
\frac{\partial Z(x_{\rm N}, t)}{\partial t} = \frac{z_{\rm N}^{\rm m+1} - z_{\rm N}^{\rm m}}{\tau_{\rm m}} \,+\, o(\tau_{\rm m})\,.
\end{equation}
Henceforth, in the expression for the external boundary condition, the terms of order of smallness $o(h_{\rm N}^2) + o(\tau_{\rm m})$, $\forall m$ and higher will be omitted.

Now we obtain the difference form of the differential equation~\eqref{b3} itself. We write the Taylor series expansion of the function $ u_{\rm n}^{\rm m+1} $ for $ x_{\rm n + 1} $ and $ x_{\rm n - 1} $ at the point $ x_{\rm n } $.
\begin{equation}\label{b10}
\begin{gathered}
u_{\rm n + 1}^{\rm m + 1 } = u_{\rm n}^{\rm m+1} \,+\, h_{\rm n+1}\frac{\partial u^{\rm m+1}}{\partial x}\bigg|_{\rm x_{\rm n}} \,+\, \frac{h_{\rm n+1}^2}{2}\frac{\partial^2 u^{\rm m+1}}{\partial x^2}\bigg|_{\rm x_{\rm n}} \,+\, o(h_{\rm n+1}^2)\,,
\\
u_{\rm n - 1}^{\rm m + 1 } = u_{\rm n}^{\rm m+1} \,-\, h_{\rm n}\frac{\partial u^{\rm m+1}}{\partial x}\bigg|_{\rm x_{\rm n}} \,+\, \frac{h_{\rm n}^2}{2}\frac{\partial^2 u^{\rm m+1}}{\partial x^2}\bigg|_{\rm x_{\rm n}} \,+\, o(h_{\rm n}^2)\,,
\end{gathered}
\end{equation}
where $n = 1..N$--1 and $m = 0..M$--1.

Hence, omitting terms of order $ o(h_{\rm n+1}^{2}) $,  $ o(h_{\rm n}^{2}) $
and higher, we obtain that the first derivative $ \partial u^{\rm m + 1} / \partial x $ at the point $ x_{\rm n} $:
\begin{equation}\label{b11}
\frac{\partial u^{\rm m + 1}}{\partial x}\bigg|_{\rm x_{\rm n}} = \frac{u_{\rm n + 1}^{\rm m+1}\dfrac{h_{\rm n}^2}{h_{\rm n} + h_{\rm n+1}} \,+\, u_{\rm n}^{\rm m+1}(h_{\rm n+1} - h_{\rm n}) \,-\, u_{\rm n - 1}^{\rm m+1}\dfrac{h_{\rm n+1}^2}{h_{\rm n} + h_{\rm n+1}} }{h_{\rm n}h_{\rm n + 1}}\,
\end{equation} 
and the second derivative:
\begin{equation}\label{b12}
\frac{\partial^2 u^{\rm m + 1}}{\partial x^2}\bigg|_{\rm x_{\rm n}} = 2\,\frac{ u_{\rm n + 1 }^{\rm m + 1}\dfrac{h_{\rm n}}{h_{\rm n} + h_{\rm n + 1}} \,-\,  u_{\rm n }^{\rm m + 1}  \,+\, u_{\rm n - 1
}^{\rm m + 1}\dfrac{h_{\rm n+1}}{h_{\rm n} + h_{\rm n + 1}}}{h_{\rm n}h_{\rm n + 1}}\,.
\end{equation}

Finally, substituting the values of the derivatives~\eqref{b9},~\eqref{b11} and~\eqref{b12} into the differential equation~\eqref{b3} and replacing the outer boundary condition from~\eqref{b1} with its difference analogue~\eqref{b8}, we get a difference scheme for our problem:
\begin{equation}\label{b13}
 \begin{cases}
   \dfrac{z_{\rm n}^{\rm m+1} - z_{\rm n}^{\rm m}}{\tau_{\rm m}} \,=\, 2\,\dfrac{ u_{\rm n + 1 }^{\rm m + 1}\dfrac{h_{\rm n}}{h_{\rm n} + h_{\rm n + 1}} -  u_{\rm n }^{\rm m + 1}  + u_{\rm n - 1
   }^{\rm m + 1}\dfrac{h_{\rm n+1}}{h_{\rm n} + h_{\rm n + 1}}}{h_{\rm n}h_{\rm n + 1}} \,+\,
   \\ \\
   \,+\, A_{\rm n}\dfrac{u_{\rm n + 1}^{\rm m+1}\dfrac{h_{\rm n}^2}{h_{\rm n} + h_{\rm n+1}} + u_{\rm n}^{\rm m+1}(h_{\rm n+1} - h_{\rm n}) - u_{\rm n - 1}^{\rm m+1}\dfrac{h_{\rm n+1}^2}{h_{\rm n} + h_{\rm n+1}} }{h_{\rm n}h_{\rm n + 1}} \,+
   \\ \\ 
   +\,B_{\rm n}u_{\rm n - 1}^{\rm m+1} \,+\, C_{\rm n}\,,
   \\ \\
   \dfrac{u_{\rm N}^{\rm m+1} - u_{\rm N - 1}^{\rm m + 1 }}{h_{\rm N}}  \,+\, \dfrac{h_{\rm N}}{2}\,\dfrac{z_{\rm N}^{\rm m+1} - z_{\rm N}^{\rm m}}{\tau_{\rm m}} \,-\, \dfrac{A_{\rm N}}{2}\,(u_{\rm N}^{\rm m+1} - u_{\rm N - 1}^{\rm m + 1}) \,- 
   \\ \\
   - \, \dfrac{B_{\rm N}}{2\,}u_{\rm N}^{\rm m + 1}h_{\rm N} \,-\, \dfrac{C_{\rm N}}{2}\,
   h_{\rm N} \,=\, 0\,, 
 \end{cases}
\end{equation}
where $n = 1..N$--1 and $m = 0..M$--1, with internal boundary condition~\eqref{b6} and initial condition~\eqref{b6_2}.

Note that the accuracy of the resulting system~\eqref{b13} is of the order of $\,o(h_{\rm max}^2) \,+\, o(\tau_{\rm max}) $, where $ h_{\rm max} = \max(h_{\rm n}) $, and $ \tau_{\rm max} = \max(\tau_{\rm m}) $.

To find the values of function $ u $ at the next time step $m+$1, it is necessary to solve system~\eqref{b13}. Consider the transition from the known values of $ u $ at $ t = t_{\rm m} $ to the unknown values of $ u $ at $ t = t_{\rm m + 1} $.

For convenience, we denote:
\begin{equation}\label{b14}
\begin{gathered}
u_{\rm n} \,  \equiv \,  u_{\rm n}^{\rm m}\,, \qquad \widetilde{u}_{\rm n} \, \equiv \, u_{\rm n}^{\rm m+1}\,;
\\
z_{\rm n} \, \equiv \, z_{\rm n}^{\rm m}\,,\qquad \widetilde{z}_{\rm n} \, \equiv \, z_{\rm n}^{\rm m+1}\,;
\\
\tau \, \equiv \, \tau_{\rm m}\,.
\end{gathered}
\end{equation}

We also introduce the new grid function $ \widetilde{\lambda} $:
\begin{equation}\label{b15}
\widetilde{\lambda}_{\rm n} \,\equiv\, \frac{\widetilde{z}_{\rm n}}{\widetilde{u}_{\rm n}}\,.
\end{equation}

Note that due to the strict positiveness of the functions $ Z $ and $ U $ from the system~\eqref{b1}, $ \widetilde{\lambda} $ is also a positive definite function.

We transform the first equation of system~\eqref{b13} using the new notation:
\begin{equation}\label{b16}
\begin{gathered}
\left( \frac{2h_{\rm n+1}}{h_{\rm n} + h_{\rm n+1}} - A_{\rm n}\frac{h_{\rm n+1}^2}{h_{\rm n} + h_{\rm n+1}}\right)\widetilde{u}_{\rm n-1} \,-\, \left( 2 + \frac{h_{\rm n}h_{\rm n+1}}{\tau}\widetilde{\lambda}_{\rm n} \right)\widetilde{u}_{\rm n} \,-
\\
- \, \left(\, - A_{\rm n}(h_{\rm n+1} - h_{\rm n}) - B_{n}h_{\rm n}h_{\rm n+1} \,\right) \widetilde{u}_{\rm n} \,+\, \frac{2h_{\rm n+1}}{h_{\rm n} + h_{\rm n+1}}\widetilde{u}_{\rm n+1} \,+ 
\\
+ \, A_{\rm n}\,\frac{h_{\rm n+1}^2}{h_{\rm n} + h_{\rm n+1}}\widetilde{u}_{\rm n+1} \,= \, - \frac{h_{\rm n}h_{\rm n+1}}{\tau}\,z_{\rm n} \,-\, C_{\rm n}\,h_{\rm n}h_{\rm n+1}\,,
\end{gathered}
\end{equation}
where $n = 1..N$--1.
We also rewrite the initial and boundary conditions for the grid function $ u $:
\begin{equation}\label{b17}
 \begin{cases}
   u_{\rm 0} = 0\,,
   \\
   \\
\left(1- \dfrac{A_{\rm N}h_{\rm N}}{2}\right)\,\widetilde{u}_{\rm N-1} \,-\, \left(1 + \dfrac{h_{\rm N}^2}{2\tau}\widetilde{\lambda}_{\rm n} - \dfrac{A_{\rm N}h_{\rm N}}{2} - \dfrac{B_{\rm N}h_{\rm N}^2}{2}\right)\,\widetilde{u}_{\rm N} \,=\, \\
  = -\, \dfrac{h_{\rm N}^2}{2}\,\left(\dfrac{z_{\rm N}}{\tau} + C_{\rm N}\right)\,,
   \\
   \\
   u_{\rm n}^{0} = \phi_{n}\,.
 \end{cases}
\end{equation}

\subsection{Solution of a system of nonlinear equations}

Equations~\eqref{b16} and~\eqref{b17} form a system of nonlinear equations. 
To find the values $ \widetilde{u} $ and $ \widetilde{z} $ at the next step of time, we will use the iteration method. Thus, it is necessary to solve the system of $N$--1 linear equations:
\begin{equation}\label{b18}
\begin{gathered}
\left( \frac{2h_{\rm n+1}}{h_{\rm n} + h_{\rm n+1}} \,-\, A_{\rm n}\,\frac{h_{\rm n+1}^2}{h_{\rm n} + h_{\rm n+1}}\right)\,\widetilde{u}_{\rm n-1}^{\rm (s+1)} \,-\, 
\\
-\, \left( 2 \,+\, \frac{h_{\rm n}h_{\rm n+1}}{\tau}\,\widetilde{\lambda}_{\rm n}^{\rm (s)}  \,-\, A_{\rm n}\,(h_{\rm n+1} - h_{\rm n})  \,-\, B_{n}\,h_{\rm n}h_{\rm n+1}\right)\,\widetilde{u}_{\rm n}^{\rm (s+1)} \,+\, 
\\
+\,  \left( \frac{2h_{\rm n+1}}{h_{\rm n} + h_{\rm n+1}} \,+\,  A_{\rm n}\,\frac{h_{\rm n+1}^2}{h_{\rm n} + h_{\rm n+1}}\right)\,\widetilde{u}_{\rm n+1}^{\rm (s+1)}\, =\, - \frac{h_{\rm n}h_{\rm n+1}}{\tau}z_{\rm n}^{\rm (s)}\,-\, C_{\rm n}h_{\rm n}h_{\rm n+1}\,,
\end{gathered}
\end{equation}
where $n = 1..N$--1 and $s$ is the number of iteration step. The initial and boundary conditions for this system of linear equations are defined in~\eqref{b17}.

This system is solved using the tridiagonal matrix algorithm~(see Section~\ref{progon}).

Iterations stop when the criterion is met:
\begin{equation}\label{b19}
\max_{n = 1..N\mathrm{-1}} \frac{|\widetilde{\lambda}^{\rm (s+1)} - \widetilde{\lambda}^{\rm (s)}|}{\widetilde{\lambda}^{\rm (s+1)}} < \epsilon\,,
\end{equation}
where $\epsilon$ is the requested relative accuracy.

\subsection{Tridiagonal matrix algorithm for solving a system of linear equations}\label{progon}

Let us consider a system of equations, where the desired variables are $y_{\rm 
n}$:
\begin{equation}\label{b20}
 \begin{cases}
 D_{\rm n}y_{\rm n-1} \,-\, E_{\rm n}y_{\rm n} \,+\, F_{\rm n}y_{\rm n+1} \,=\, - G_{\rm n}\,, & n = 1..(N-1)
   \\
   y_{0} = \mu_{1} \,+\, k_{1}\,y_{1} \,,
   \\
   y_{\rm N} = \mu_{2} \,+\, k_{2}\,y_{\rm N-1}\,. 
   \\
 \end{cases}
\end{equation}

To resolve this system, we use the tridiagonal matrix algorithm~\citep{Kalitkin}.

The forward sweep allows finding auxiliary coefficients $ \alpha_{\rm n} $ and $ \beta_{\rm n} $:
\begin{equation}\label{b21}
\begin{gathered}
\alpha_{\rm 1} \, = \, k_{1}\,,
\\
\beta_{\rm 1} \, = \, \mu_{1}\,;
\\
\alpha_{\rm n +1} \, = \, \frac{F_{\rm n}}{E_{\rm n} - \alpha_{\rm n}D_{\rm n}}, \qquad n= 1..N\mathrm{-1}\,;
\\
\beta_{\rm n +1} \, = \, \frac{\beta_{\rm n}D_{\rm n} + G_{\rm n} }{E_{\rm n} - \alpha_{\rm n}D_{\rm n}}\,, \qquad n = 1..N\mathrm{-1}\,.
\end{gathered}
\end{equation}

Back substitution gives the desired values $y_{\rm n}$:
\begin{equation}\label{b22}
\begin{gathered}
y_{\rm n} \, = \, \alpha_{\rm n+1}y_{\rm n+1} \,+\, \beta_{\rm n+1}\,,
\\
y_{\rm N} \, = \, \frac{\mu_{2} \,+\, \beta_{\rm N}k_{2}}{1 \,-\, \alpha_{\rm N}k_2}\,.
\end{gathered}
\end{equation}

Sufficient stability conditions for the system~\eqref{b20} are as follows~\citep{Kalitkin}:
\begin{equation}\label{b23}
\begin{gathered}
|E_{\rm n}| \, \geq \, |D_{\rm n}| \, + \, |F_{\rm n}|\,, \qquad n = 1..N\mathrm{-1}\,;
\\
|k_1| \, \leq \, 1, |k_2| \, \leq \, 1\,;
\\
|k_1| \, +  \, |k_2| \,< \, 2 \,.
\end{gathered}
\end{equation}

Using this method to solve the system of linear equations~\eqref{b18}, with initial and boundary conditions~\eqref{b17}, we find that the criterion for stability of the system~\eqref{b18} is the condition:
\begin{equation}\label{b24}
B_{\rm n} \, \leq  \, \widetilde{\lambda}_{\rm n}\,/\,\tau\,, \qquad n = 1..N\mathrm{-1}\,.
\end{equation}

This condition is satisfied automatically for the case $ V \left(\dfrac{\partial U}{\partial x}, U, x \right) = V(U) = B(x)U (x, t) $, because $ \widetilde{\lambda} $, $ \tau $ and $ U $ are strictly positive, and the coefficient $ B $ of the expansion of the function $ V $~\eqref{b2} in this case should be negative~(the function $ V $ corresponds to the function $ - \Windrate,  \Windrate > 0$ from~\eqref{e1}, thus $ V < 0$).

In the general case, the condition~\eqref{b24} may not be fulfilled~(the function $ V $ may be negative for positive $ B $ and negative $ A $ and/or $ C $). But, for our set of wind patterns~(see Section~\ref{windmod} and Appendices~\ref{append_thermwind},\ref{appendix_ver},\ref{appendix_val}), the condition is always satisfied, because the functions $ \Windrate $ from those sections do not simultaneously contain more than one term in decomposition~\eqref{b2}. This means that the solution of
system~\eqref{b18} is stable for all models presented in the manuscript.

\section{Equations of the Thermal Compton wind}\label{append_thermwind}
B83 have proposed the following model. 
The rate of mass loss per unit area $ \dot{m} = \rho Vy^{ \left<\beta\right> } $ is constant along a streamline, and its value depends on the sonic point position, pressure and temperature there, as well as on a divergence parameter $\beta$. Here $ V $ and $ \rho $ are the velocity and density of the flow, respectively. An increase in the cross-sectional area of streamlines is described by function $ y^{ \left<\beta\right> } $, $ y = r' / r $, and $ (r'- r) $ is the distance along the streamline from the base of the corona/wind, where $y=1$. Spherical flow corresponds to $\beta=2$.

B83 have established that the mass loss rates are insensitive to the trajectories of streamlines. At the same time, the complexity of the model follows from the fact that the escape, characteristic, and Compton temperatures relate to each other differently, depending on the disc radius and luminosity. S86 have proposed a formula that approximates the results obtained in B83. It can be written in terms of dimensionless rate as follows:
\begin{equation}\label{3.7}
\dot{m}^{*} \equiv \frac{\dot{m}}{\dot{m}_{\rm ch}} = \mathsf{M}\, \frac{c_{\rm ch}}{c_{\rm s}(r')}\frac{p(r')}{p_{0}}\,y^{\left<\beta\right>},
\end{equation}
where the characteristic mass-loss rate per unit area $ \dot{m}_{\rm ch} \equiv p_{0} / c_{\rm ch} $ is expressed from the gas pressure at the base of the corona/wind $ p_0 \equiv L / (4 \pi r^{2} \, \Xi \, c) $ and a characteristic speed $ c_{\rm ch}$ such that a flow, which has been moving at this speed along a streamline to a height equal to the radius of the streamline's base $r$, is heated to temperature $T_{\rm ch}\equiv\mu\,c_{\rm ch}^2 /k$. Since the optically thin heating rate at radius $r$:
    \begin{equation}\label{gamma}
      \Gamma = \frac{kT_{\rm IC}}{m_{\rm e}c^{2}}\frac{\sigma_{\rm T}\,L}{\pi r^{2}}\,
    \end{equation}
determines the characteristic temperature as follows:
\begin{equation}\label{T_ch}
    kT_{\rm ch} = \Gamma \frac{r}{c_{\rm ch}}\,,
\end{equation}
it is possible to obtain the exact expression for the  characteristic mass loss rate per unit area~\citep[see also][]{Middleton+2022}:
  \begin{equation}\label{m_ch}
    \dot{m}_{\rm ch} = \frac{1}{2^{8/3}\,\Xi}\,\frac{m_{\rm e} c}{\sigma_{\rm T}\, R_{\rm IC}}\,\left(\frac{\mu_{\rm e}}{\mu}\right)^{2/3}\,l^{2/3}\,\xi^{-5/3}\,,
    \end{equation}
    where $\xi = r/R_{\rm IC}$ and $ l = L / L_{\rm cr} $, see the definitions~\eqref{R_IC} and~\eqref{L_cr}.
In \eqref{3.7}, the isothermal Mach number $ \mathsf{M} \equiv V / c_{\rm s}(r')  = \sqrt{\rho V^{2}
/p}$  and  $ c_{\rm s}(r') $ is the speed of sound at location $r'$ along the streamline.

Ultimately, S86 propose the following relations that approximate the result of B83 in different zones of the wind: 
\begin{equation}\label{3.8}
p/p_{0} \approx 0.5 \exp{[-( 1- y^{-1} )^2/(2\xi)]},
\end{equation}
\begin{equation}\label{3.9}
\frac{\mathsf{M}}{c_{\rm s}/c_{\rm ch}} = \left[ \frac{ 1 + ( l + 1 )\xi^{-1}}{1 + l^{-4}(1 + \xi^{2})^{-1}}\right]^{1/3},
\end{equation}
\begin{equation}\label{3.10}
y^{<\beta>} = y^{2}= 1 + \frac{1}{4\xi^{2}} + \frac{\xi^{2}}{1 + \xi^{2}}\left[ \frac{1.2\xi}{\xi + l} + \frac{2.2}{(1 + l^2\xi)} \right]^{2}.
\end{equation}

Substituting the above expressions into \eqref{3.7}, one obtains the mass loss rate from two sides of the disc:

    \begin{equation}
    \Windrate = 2\,\dot{m} = 2\,\dot{m}_{\rm ch}(\xi, l)\,\dot{m}^{*}(\xi, l)\,.
   \label{5/3e}
    \end{equation}

\section{Verifying the code with the S86 'toy' wind model}\label{appendix_ver}

S86 have shown that the instability arising in accretion discs under the influence of a wind can lead to oscillations in the brightness of an object. This phenomenon occurs only if the wind from a disc is quite powerful compared to the accretion rate.

Therefore, in order to test our code and verify the presence of oscillations, we use the following wind term from S86:
\begin{equation}\label{e4}
\Windrate = \frac{1}{2\pi}\left( \frac{C_{\rm w}(t)\dot{M}_{\rm acc}}{\ln{(R_{\rm out}/R_{\rm w})}\,r^2}\right) , 
\end{equation}
where $ R_{\rm out} $ is the outer radius of the accretion disc, $ R_{\rm w} $ is the inner radius of the wind launching zone, and $ C_{\rm w} \equiv \dot{M}_{\rm wind} / \dot{M}_{\rm acc} $. The wind occurs only in the area outside the radius $R_{\rm w}$~(i.e. $ r \geq R_{\rm w}$), in other parts of the disc the function $ \Windrate $ is assumed to be zero.

The physical reasoning of the parameter $C_{\rm w}$ here is the following. Thermal wind is caused by the irradiation of the disc photosphere by a central source of X-ray (or UV) flux. Presumably, the mass loss rate due to the wind rises with the accretion rate onto a compact object. Hence, parameter $C_{\rm w}$ formalises an assumption
that the rate of wind mass loss is proportional to the accretion rate onto a compact object.

Note that this 'toy' wind model does not include the notion of the hot and cold zone in the disc (Section~\ref{Irrad_Rhot}). We rather fix the outer radius of an evolving disc zone, i.e. $ R_{\rm hot} = R_{\rm out} $. Furthermore, a constant inflow of matter into the disc is required. Thus we set a non-zero external boundary condition~($\dot{M}_{\rm out} = 10^{17}$ g/s). An initial condition should comply with boundary conditions. Thus, a stationary, linear dependence of $F$ on $ h $ is chosen, $F_0(h) = \dot{M}_{\rm acc, 0} \times h_{\rm out}(h - h_{\rm in})/(h_{\rm out} - h_{\rm in})$.

    \begin{figure} 
    \begin{center}
    \includegraphics[width=\columnwidth]{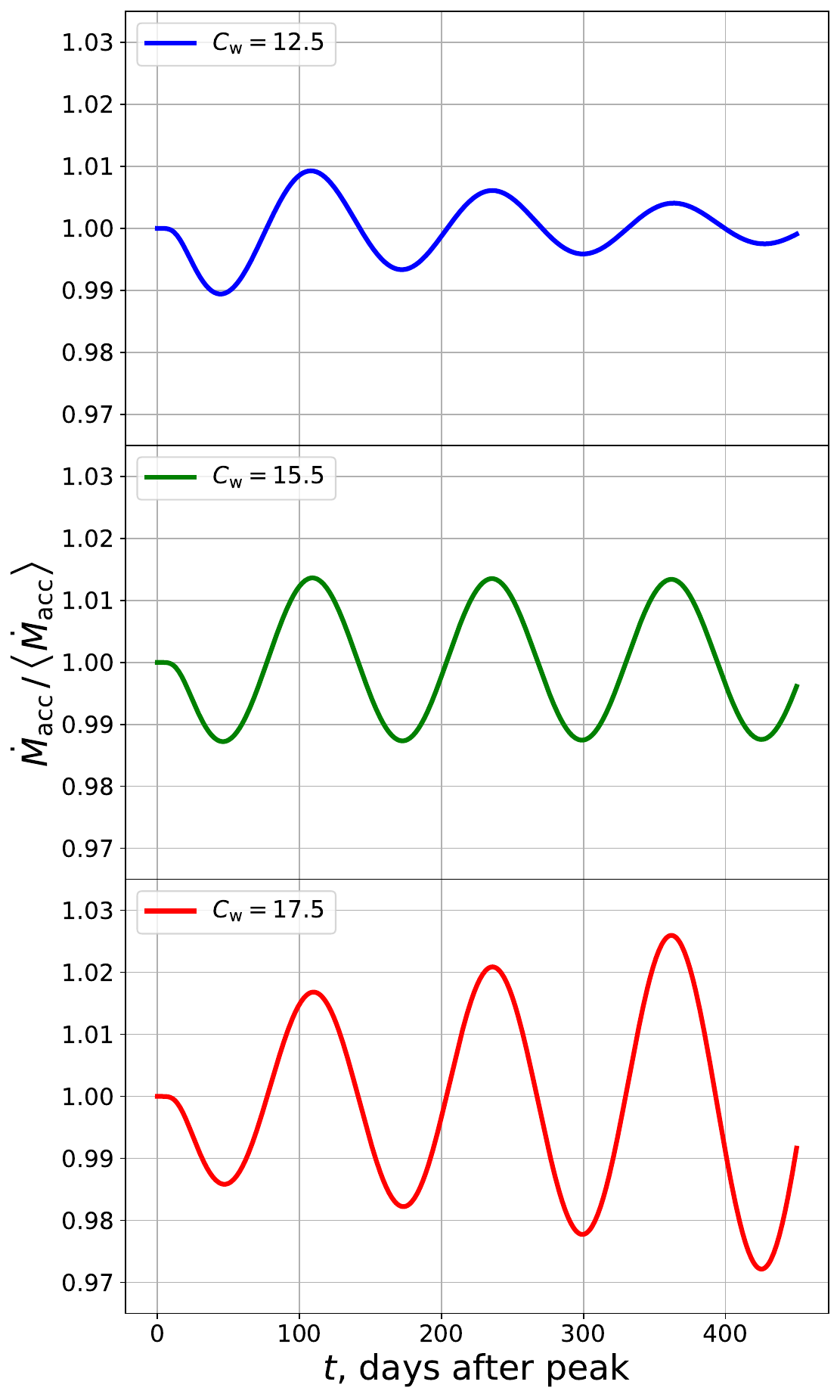}
    \caption{The time dependence of the central accretion rate for a disc with wind \eqref{e4}. The innermost wind launch radius is
    $ 0.9 \times R_{\rm out} $. 
    The top, middle, and bottom panels show the cases for $ C_{\rm w} \equiv \dot{M}_{\rm wind} / \dot{M}_{\rm acc} $ ratio: $C_{\rm w} = 12.8$, $C_{\rm w} = C_{\rm w}^{\, \rm crit} = 15.5$, and $C_{\rm w} = 17.8$,  respectively. 
    } 
    \label{fig:oscil}
    \end{center}
    \end{figure}
    
In the end, to test our numerical method, we reproduce oscillations of the accretion rate on a compact object arising in the disc in the presence of a sufficiently strong Compton thermal wind~(S86).
For this, we use the wind function~\eqref{e4}, which contains a dependence on the luminosity (or the rate of accretion on the central object), the parameter {\tt windtype} in {\sc freddi} is `ShieldsOscil1986'. The accretion rate evolution is calculated for various values of the parameter $ C_{\rm w} $~(Fig.~\ref{fig:oscil}). The launch radius of the wind $ R_{\rm w} $ is set equal to $0.9 \times R_{\rm out} $.

When the central luminosity $ L $ increases, the rate of the wind mass loss also increases. If the wind dominates, the disc mass decreases, and, after a short delay, $ L $ and $ \dot{M}_{\rm acc} $ decline, as well as $\dot{M}_{\rm wind} $, which allows the disc mass to increase again, starting a new cycle of oscillations. The authors give the critical value $ C_{\rm w}^{\, \rm crit} $, above which the oscillations of the central accretion rate grow with time (see their figure~1). The critical value $ C_{\rm w}^{\, \rm crit}$ depends on the functional form of the viscosity. For example, $C_{\rm w}^{\, \rm crit}=\cosh{\pi} \approx 11.6 $ for the viscosity law  $\nu_{\rm t} \propto r$.

The oscillations that we obtain confirm the results of S86 for the OPAL opacity~\citep{Iglesias-Rogers1996}, approximated by the power law $\propto \rho\, T^{-5/2}$ (see the last row in table 1 of S86), when $\nu_{\rm t} \propto r \Sigma^{1/2}$. For such viscosity law, S86 give the critical value \hbox{$ C_{\rm w}^{\, \rm crit} = 15.2 $}. To find $C_{\rm w}^{\, \rm crit}$, we have calculated the relative difference in amplitudes of the central accretion rate deviation at the first and seventh oscillation cycle for different values of the parameter $C_{\rm w}$ spaced by 0.1. Examples of resulting evolution are shown in Fig.~\ref{fig:oscil}. Resulting $C_{\rm w}^{\, \rm crit}\approx 15.5$, which is in a very good agreement with that found by S86. We note, that this also agrees with possibility of appearance of accretion rate oscillation suggested by \citet{SHAKURA_SUNAEV}.

Additional wind models to verify \textsc{freddi} work are described in Appendix~\ref{appendix_val}. These tests reproduce the radial structure of stationary discs with different types of wind. All solutions for the corresponding $ \Windrate(h) $ functions were found analytically.

\section{Validation of the numerical method by analytical solutions for stationary disc}\label{appendix_val}

To validate the numerical method, we have checked that a numerical solution converges successfully to a analytical one in the case of the constant $\dot{M}_\mathrm{out}$.

A solution for a stationary accretion disc can be found from equation \eqref{e1} with $\partial \Sigma / \partial t = 0$:
\begin{equation}\label{stationary_equation}
    \frac1{4 \piup} \frac{(G M_{\rm x})^2}{h^3} \frac{\partial^2 F}{\partial h^2} = \Windrate(F,h)\,.
\end{equation}
In every test, we set an initial condition, which did not satisfy \eqref{stationary_equation}, and run a simulation long enough to reach a stationary solution.
For all the tests it takes up to a hundred time iterations to reach relative accuracy of $F(h)$ of 1\%.
Below, we describe different wind types and find corresponding analytical solutions.

\subsection{S86 'toy' wind model}
This type of wind is described by Sec.~\ref{appendix_ver} and Eq.~\eqref{e4}. The solution of Eq.~\eqref{stationary_equation} for such a wind is
\begin{equation}
    F(h) = 
    \begin{cases}
      \dfrac{\dot{M}_\mathrm{out}\, h}{1 + C_\mathrm{W}} \left(1 - \dfrac{h_\mathrm{in}}{h}\right)\,, & h < h_\mathrm{W} ;
        \\
        \\
        \dfrac{\dot{M}_\mathrm{out}\, h}{1 + C_\mathrm{W}} \left(1 - \dfrac{h_\mathrm{in}}{h} + C_\mathrm{W} \dfrac{\ln{(h / h_\mathrm{W})} + h_\mathrm{W}/h - 1}{\ln{(h_\mathrm{out} / h_\mathrm{W})}}\right)\,, & h \geq h_{\rm W}\,,
    \end{cases}
\end{equation}
where $C_\mathrm{W} < C^\mathrm{crit}_\mathrm{W}$.
Here we use the fact that for the asymptotic solution $\dot{M}_\mathrm{acc} = \dot{M}_\mathrm{out} / (1 + C_\mathrm{W})$.

\subsection{Wind term is proportional to local accretion rate}
In view of Eq.\eqref{dfdh} we define:
\begin{equation}\label{windA}
    \Windrate(F, h) = k \, \frac{(G M_{\rm x})^2}{4 \piup \, h^3}\, \frac{h - h_\mathrm{in}}{(h_\mathrm{out} - h_\mathrm{in})^2} \frac{\partial F}{\partial h}\,,
\end{equation}
where $k$ is a dimensionless coefficient.

This case corresponds to a non-zero value of $A$ coefficient and zero values of $B$ and $C$ coefficients in Eq.~\eqref{b3}. In other words, $\Windrate(h, l) \propto \partial F / \partial h \propto \dot{M} $.
The corresponding solution of Eq.~\eqref{stationary_equation} is
\begin{equation}
    F(h) = \dot{M}_\mathrm{out} (h_\mathrm{out} - h_\mathrm{in}) \sqrt{\frac{\piup}{2 k}} \mathrm{e}^{-k/2} \mathrm{erfi}{\left(\sqrt{k/2} \frac{h - h_\mathrm{in}}{h_\mathrm{out} - h_\mathrm{in}}\right)}\,,
\end{equation}
where $\mathrm{erfi}$ is the imaginary error function.

\subsection{Wind term is proportional to viscous torque}
We define:
\begin{equation}\label{windB}
    \Windrate(F, h) = \frac{(G M_{\rm x})^2}{4 \piup \, h^3}\, \frac{k}{(h_\mathrm{out} - h_\mathrm{in})^2} F(h)\,,
\end{equation}
where $k$ is a dimensionless coefficient. This case corresponds to a non-zero $B$ coefficient and zero $A$ and $C$ coefficients in Eq.~\eqref{b3}, that is, $\Windrate(F, h) \propto F(h)$.
The corresponding solution of Eq.~\eqref{stationary_equation} is
\begin{equation}
    F(h) = \dot{M}_\mathrm{out} (h_\mathrm{out} - h_\mathrm{in}) \frac1{\sqrt{k} \sinh{\sqrt{k}}} \sinh{\left( \sqrt{k} \frac{h - h_\mathrm{in}}{h_\mathrm{out} - h_\mathrm{in}}\right)} \,.
\end{equation}

\section{Relation between the self-irradiation prescription and estimate of~\texorpdfstring{$\alpha$}{alpha}}\label{appendix_cirr}

The estimate of $\alpha$, obtained from comparison with observations, depends on the assumed self-irradiation model.
In this work, we parameterise the irradiation parameter taking into account its dependence on the thickness of the disc which comes from the changing incident angle and the angular distribution of the central radiation, see Eq.~\eqref{irr_both}. The relative thickness determines the angle of incident photons and also enters in the angular distribution of the central emission. In LM17 the irradiation parameter was taken constant throughout the modelled time interval.

\begin{figure}
    \begin{center}
    \includegraphics[width=\columnwidth]{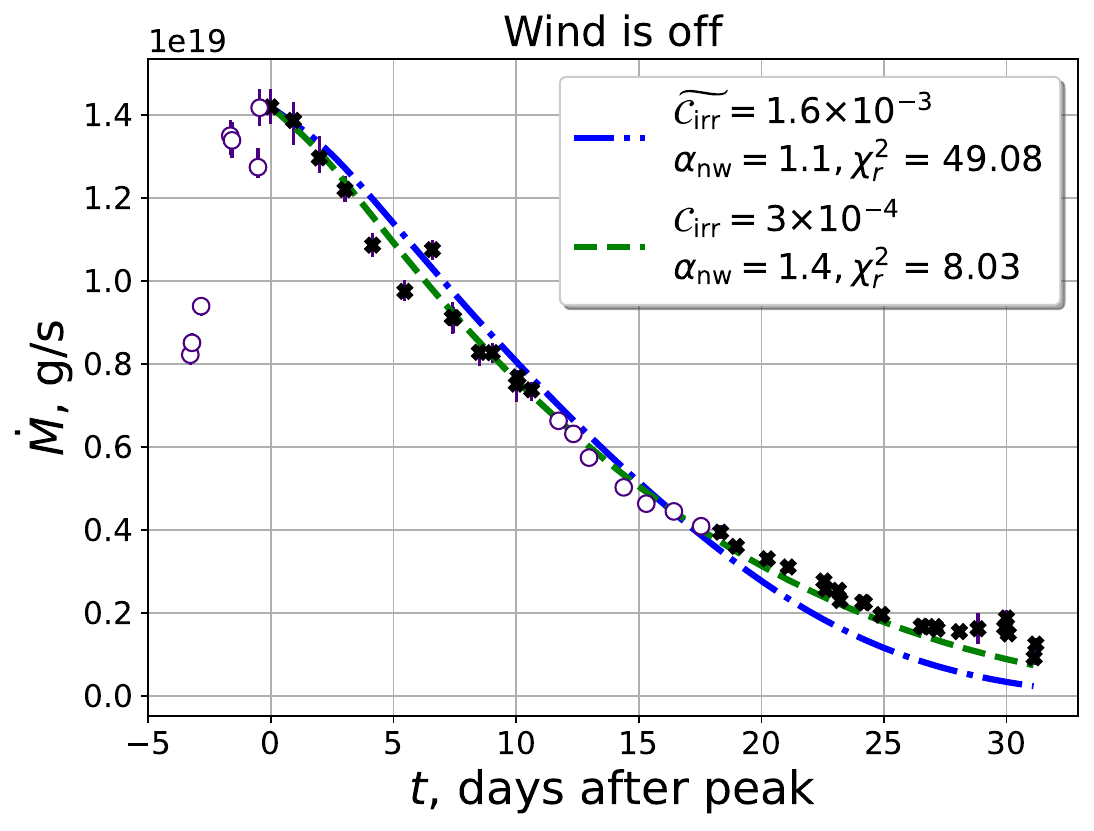}
    \caption{Models and observed $\dot{M}_{\rm acc}(t)$ during the 2002 outburst of~\lup{}. Models care calculated with different prescription for $C_{\rm irr}$: $k=1$ and plane irradiation (the blue line) and $k=0$ and isotropic irradiation (the green line, as LM17 suggests), see Eq.~\eqref{irr_both}.}
    \label{angular}
    \end{center}
    \end{figure}

\begin{figure}
    \begin{center}
    \includegraphics[width=\columnwidth]{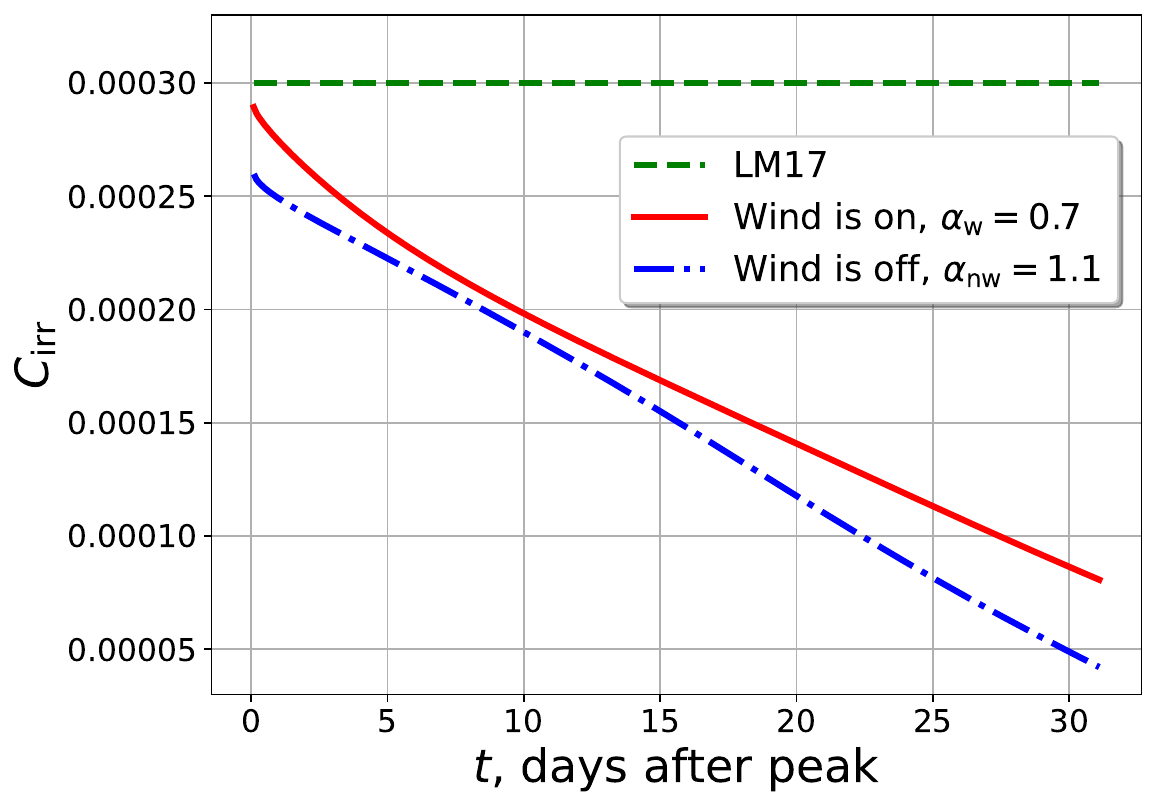}
    \caption{Irradiation parameter of the hot disc vs. time (see Eq.~\eqref{irr_both}) shown for models with and without wind, see Fig.~\ref{mdotvstime}). Horizontal line shows constant $C_{\rm irr}$  from LM17 model of the 2002 outburst of~\lup{}.}
    \label{Cirr_comp}
    \end{center}
    \end{figure}

Consequently, this leads to different resulting estimates of $\alpha$: in LM17 the estimate of the viscosity parameter $ \alpha $ was about $ 1.4 $ for the 2002 X-ray nova outburst of~\lup{} for binary parameters as in Table~\ref{tab:params}. In the present work, $ \alpha_{\rm nw} = 1.1 $ is obtained, when no wind is present~(Fig.~\ref{mdotvstime}). Fig.~\ref{angular} shows two fits of the accretion rate evolution of~\lup{} (2002) for such prescriptions of $C_{\rm irr}$. Fig.~\ref{Cirr_comp} shows corresponding parameters $C_{\rm irr}$: constant, as in the modelling by LM17, and the one changing with $z_0/r$, see Eq.~\eqref{irr_both}.

However, it is important to keep in mind that processes involving scattering in a media above the disc may infringe a simple geometrical approach like \eqref{irr_both}. Also the evolution of the X-ray's albedo is a non-trivial matter~\citep{Tavleev23},
Thus, a decision cannot be made yet, which description of $\mathcal{C}_{\rm irr}$ is more effective for modelling: constant $\mathcal{C}_{\rm irr}$ or thickness-dependent.
In \textsc{freddi} both options are available. Meanwhile, the robust conclusion is that an estimate of $\alpha$ becomes lower if we take into account decreasing of $\mathcal{C}_{\rm irr}$ with time.

\bsp	
\label{lastpage}
\end{document}